\documentclass{pasa}%

\usepackage{graphicx}
\usepackage{mathtools}
\usepackage{longtable}

\title[Hydra]{Hydra II: Characterisation of Aegean, Caesar, ProFound, PyBDSF, and Selavy source finders.}

\author[Boyce et al.]{M. M. Boyce,$^{1}$ A. M. Hopkins,$^{2}$  S. Riggi,$^{3}$  L. Rudnick,$^{4}$ M. Ramsay,$^{1}$ C. L. Hale,$^{5}$  J. Marvil,$^{6}$ M. Whiting,$^{7}$ P. Venkataraman,$^{8}$  C. P. O'Dea,$^{1}$ S. A. Baum,$^{1}$ Y. A. Gordon,$^{1, 9}$ A. N. Vantyghem,$^{1}$ M. Dionyssiou,$^{8}$ H. Andernach,$^{10}$ J. D. Collier,$^{11, 12}$ J. English,$^{1}$ B. S. Koribalski,$^{7, 12}$ D. Leahy,$^{13}$ M. J. Michałowski,$^{14}$ S. Safi-Harb,$^{1}$ M. Vaccari,$^{15, 16}$ 
E. Alexander,$^{17}$ M. Cowley,$^{18,19}$ A. D. Kapinska,$^{6}$ A. S. G. Robotham,$^{20}$ H. Tang.$^{21}$
\affil{$^{1}$Department of Physics and Astronomy, University of Manitoba, 30A Sifton Road, Winnipeg, MB R3T 2N2, Canada}%
\affil{$^{2}$Australian Astronomical Optics, Macquarie University,
105 Delhi Rd, North Ryde, NSW 2113, Australia}
\affil{$^{3}$INAF, Osservatorio Astrofisico di Catania
Via S. Sofia 78, 95123, Catania, Italy}
\affil{$^{4}$Minnesota Institute for Astrophysics, School of Physics and Astronomy, University of Minnesota, 116 Church Street SE, Minneapolis, MN 55455, USA}
\affil{$^{5}$School of Physics and Astronomy, University of Edinburgh, Institute for Astronomy, Royal Observatory, Blackford Hill, Edinburgh EH9 3HJ, UK}
\affil{$^6$National Radio Astronomy Observatory, P.O. Box O, Socorro, NM 87801, USA}
\affil{$^{7}$Australia Telescope National Facility, CSIRO Astronomy and Space Science, PO Box 76, Epping, NSW 1710, Australia}
\affil{$^{8}$Dunlap Institute for Astronomy and Astrophysics, University of Toronto, 50 St. George Street, Toronto, ON M5S 3H4, Canada}
\affil{$^{9}$Department of Physics, University of Wisconsin-Madison, Madison, WI 57306, USA}
\affil{$^{10}$Departamento de Astronom{\'{i}}a, DCNE, Universidad de Guanajuato, Callej\'on de Jalisco s/n, Guanjuato, CP 36023, GTO, Mexico}%
\affil{$^{11}$Inter-University Institute for Data Intensive Astronomy (IDIA), Department of Astronomy, University of Cape Town, Private Bag X3, Rondebosch, 7701, South Africa}
\affil{$^{12}$School of Science, Western Sydney University, Locked Bag 1797, Penrith, NSW 2751, Australia}
\affil{$^{13}$Department of Physics and Astronomy, University of Calgary, 2500 University Dr. NW, Calgary, AB T2N 1N4, Canada}
\affil{$^{14}$Astronomical Observatory Institute, Faculty of Physics, Adam Mickiewicz University, ul.~S{\l}oneczna 36, 60-286 Pozna{\'n}, Poland}
\affil{$^{15}$Inter-University Institute for Data Intensive Astronomy (IDIA), Department of Physics and Astronomy, University of the Western Cape, Robert Sobukwe Road, 7535 Bellville, Cape Town, South Africa}
\affil{$^{16}$INAF - Istituto di Radioastronomia, via Gobetti 101, 40129 Bologna, Italy}
\affil{$^{17}$Jodrell Bank Centre for Astrophysics, Department of Physics and Astronomy, University of Manchester, Manchester, UK}
\affil{$^{18}$School of Chemistry and Physics, Queensland University of Technology, Brisbane, QLD 4000, Australia}
\affil{$^{19}$Centre for Astrophysics, University of Southern Queensland, West Street, Toowoomba, QLD 4350, Australia}
\affil{$^{20}$ICRAR, M468, University of Western Australia, Crawley, WA 6009, Australia}
\affil{$^{21}$Department of Astronomy, Tsinghua University, Beijing 100084, China}
}

\jid{PASA}
\doi{10.1017/pas.\the\year.xxx}
\jyear{\the\year}

\usepackage{aas_macros}
\usepackage{hyperref} 
\hypersetup{colorlinks,citecolor=blue,linkcolor=blue,urlcolor=blue}

\usepackage{soul}

\begin{document}

\begin{frontmatter}
\maketitle

\begin{abstract}
We present a comparison between the performance of a selection of source finders using a new software tool called Hydra. The companion paper, Paper~I, introduced the Hydra tool and demonstrated its performance using simulated data.  Here we apply Hydra to assess the performance of different source finders by analysing real observational data taken from the Evolutionary Map of the Universe (EMU) Pilot Survey. EMU is a wide-field radio continuum survey whose primary goal is to make a deep ($20\mu$Jy/beam RMS noise), intermediate angular resolution ($15^{\prime\prime}$), 1\,GHz survey of the entire sky south of $+30^{\circ}$ declination, and expecting to detect and catalogue up to 40 million sources. With the main EMU survey expected to begin in 2022 it is highly desirable to understand the performance of radio image source finder software and to identify an approach that optimises source detection capabilities. Hydra has been developed to refine this process, as well as to deliver a range of metrics and source finding data products from multiple source finders. We present the performance of the five source finders tested here in terms of their completeness and reliability statistics, their flux density and source size measurements, and an exploration of case studies to highlight finder-specific limitations. 
\end{abstract}

\end{frontmatter}


\section{Introduction}
\label{sec:intro}
We are entering a new era of radio surveys, digging ever deeper and with greater sky coverage, providing catalogues with tens of millions of sources \citep{norris_2017}, with increasing data rates, up to hundreds of gigabytes per second \citep{whiting_2012}. This has created a need for source finder (SF) software tools that are up to the challenge (as discussed in Paper~I). The motivation to assess the performance of and to characterise such tools has led to a number of data challenges in recent years \citep[\textit{e.g.,}][]{hopkins_2015,bonaldi_2021}. 

As a consequence, so-called \textsc{NxGen} SFs, capable of efficiently handling large image tiles through multiprocessing (see Paper~I) have been developed, and include Aegean \citep{hancock_2012,hancock_2018} and Selavy \citep{whiting_2012}, for handling compact or marginally extended sources, and Caesar \citep[Compact And Extended Source Automated Recognition,][]{riggi_2016,riggi_2021}, for handling compact and extended sources with diffuse emission. SFs developed initially for use with optical images have also been explored to understand their performance on radio image data. These include CuTex \citep[Curvature Threshold Extractor,][]{molinari_2010} for its ability to extract compact sources in the presence of intense background fluctuations, SExtractor \citep[Source Extractor,][]{bertin_1996} for its ability to handle extended sources, and ProFound \citep{robotham_2018,hale_2019,boyce_2020} for its ability to handle extended sources with diffuse emission. Traditional Gaussian fitting SFs, capable of handling compact sources, such as APEX \citep[Astronomical Point source Extractor,][]{makovoz_2005}, PyBDSF \citep[Python Blob Detector and Source Finder,][]{mohan_2015}, and SAD \citep[Search and Destroy,][]{condon_1998}, have also been tested in such challenges. These SFs are just the tip of the iceberg, with new tools and approaches continuing to be explored \citep[\textit{e.g.},][]{hopkins_2015,wu_2018,lukic_2019,sadr_2019,koribalski_2020,bonaldi_2021,magro_2022}. This wide variety of tools and techniques makes comparison studies challenging, as it requires expertise spanning an extensive set of quite diverse SF tools, and hence often requires large collaborative efforts \citep[\textit{e.g.,}][]{hopkins_2015,bonaldi_2021}. We have developed a new software tool, Hydra  (see Paper~I), to ease this effort.

Hydra is an extensible multi-SF comparison and cataloguing tool, requiring minimal expert knowledge of the underlying SFs from the user. Hydra is extensible, with the scope for adding new SFs in a modular fashion, using containerisation. Hydra currently incorporates  Aegean, Caesar, ProFound, PyBDSF, and Selavy.\footnote{Hydra is available, along with the data products presented in this paper, by navigating through the CIRADA portal at \url{https://cirada.ca}.} 

Hydra is designed for SFs with RMS and island-like parameters, which are optimised by reducing the False Detection Rate \citep[FDR,][]{whiting_2012b} through a Percentage Real Detections \citep[PRD, \textit{e.g.},][]{williams16,hale_2019} metric (we use a 90\% PRD cutoff, see Paper~I). It also provides optional RMS box parameter pre-optimisation, using \textsc{bane} \citep{hancock_2012} in a process to minimize the background noise (referred to as $\mu$--optimisation, see Paper~I). Hydra is designed to handle simulated images with injected ($\mathcal{J}$) sources, and real (\textit{i.e.}, deep or $\mathcal{D}$) images through comparison of detections in shallow ($\mathcal{S}$) images (\textit{i.e.}, images with $5\sigma$ noise added), for which detections in the $\mathcal{D}$-images are assumed as real. This leads to a rich set of statistics, including completeness ($\mathcal{C}$) and reliability ($\mathcal{R}$) metrics.
We define $\mathcal{C}$ as the ratio of SF detections to real sources, and $\mathcal{R}$ as the ratio of SF detections that are real to detected sources.
In our terminology, for $\mathcal{D}$ or $\mathcal{S}$ images with known input sources $\mathcal{J}$, these are $\mathcal{C_D}$ and $\mathcal{R_D}$, or $\mathcal{C_S}$ and $\mathcal{R_S}$, respectively. For real images, we use $\mathcal{C_{DS}}$ and $\mathcal{R_{DS}}$ describing sources detected in an $\mathcal{S}$ image with respect to sources in $\mathcal{D}$ assumed to be real. For formal definitions see Paper~I. 

A match is defined as the overlap between source components (components or sources, interchangeably, herein), which is achieved through a clustering technique (see Paper~I). The technique is also used to associate SF components, from all SF $\mathcal{D}$ and $\mathcal{S}$ catalogues (including $\mathcal{J}$), spatially together into clumps. 

Hydra produces a cluster catalogue with sets of rows marked by clump ID (\texttt{clump\_id}), defining image depths ($\mathcal{D,\,S}$), spatial positioning, component sizes, flux densities, \textit{etc.}, with links to SF $\mathcal{D}$ and $\mathcal{S}$  (including $\mathcal{J}$) catalogues. It also includes a match ID (\texttt{match\_id}), indicating the closest $\mathcal{D}$ and $\mathcal{S}$ associations. This catalogue is used to create a rich set of diagnostic plots and and cutouts of annotated and unannotated images and residual images, which can be accessed through a local web-browser based Hydra Viewer tool (Paper~I). The user can also mine Hydra's database to produce other diagnostics.

This paper is part two of a two part series. In Paper~I, we introduced the Hydra software suite and evaluated the tool using $2^\circ\times2^\circ$ simulated-compact (CMP) and simulated-extended (EXT) images. In this paper, a comparison study of Aegean, Caesar, ProFound, PyBDSF, and Selavy is performed using a real $2^\circ\times2^\circ$ EMU pilot survey \citep{norris_2021} image sample, along with the previous simulated data. We precede this with a brief overview of relevant radio surveys.

\section{Radio Surveys}
\label{sc:radio_surveys}
Over the last two decades radio surveys have progressed in depth ($\sigma_{rms}$), resolution ($\delta\theta_{res.}$), and sky coverage, due to technological advancements. At the Karl G. Jansky Very Large Array (JVLA, or VLA) in New Mexico, the ongoing VLA Sky Survey \citep[VLASS,][]{lacy_2020,gordon_2020,gordon_2021} has better angular resolution and sensitivity than its predecessors, the National Radio Astronomy Observatory (NRAO) VLA Sky Survey \citep[NVSS,][]{condon_1998} and the Faint Images of the Radio Sky at Twenty-cm survey \citep[FIRST,][]{becker_1995,helfand_2015}. Other surveys with Northern hemisphere telescopes include the Tata Institute of Fundamental Research (TIFR) Giant Metrewave Radio Telescope (GMRT) Sky Survey -- Alternative Data Release \citep[TGSS--ADR,][]{intema_2017} at the GMRT (Khodad India), the Low-Frequency Array (LOFAR) Two-metre Sky Survey \citep[LoTSS,][]{shimwell_2017,shimwell_2019,shimwell_2022} at the LOFAR network \citep[spanning Europe,][]{rottgering_2003,vanhaarlem13} survey, the Westerbork Observations of the Deep APERture Tile In Focus (APERTIF) Northern-Sky \citep[WODAN,][]{rottgering_2010,riggi_2016} at the Westerbork Synthesis Radio Telescope \citep[WSRT; Netherlands,][]{oosterloo_2009,apertif_2016}, and the Allen Telescope Array (ATA) Transients Survey \citep[ATATS,][]{croft_2011} at the ATA (California). The upper partition of Table~\ref{tb:radio} summarises the characteristics of these surveys.

\begin{table}[htb]
   \caption{Radio surveys in the Northern (upper partition) and Southern (lower partition) hemispheres, with columns indicating the telescope (Telescope), radio survey (Survey), percentage sky-coverage (Sky), resolution ($\delta\theta_{res.}$), frequency ($\nu$), and depth ($\sigma_{rms}$).}
   \centering
   {\small
   \begin{tabular}{@{\;}l@{\;\;}l@{\;}c@{\;}r@{\;}c@{\;}c@{\;}}
        \hline\hline
        Telescope & Survey & Sky & \multicolumn{1}{@{}c@{}}{$\delta\theta_{res.}$} & $\nu$ & $\sigma_{rms}$  \\
                  &        &     &                       &       & $\left(\!\frac{\mu\mbox{Jy}}{\mbox{beam}}\!\right)$ \\
        \hline
       VLA   & NVSS      &      82\% & $45^{\prime\prime}$  & 1.4\,GHz &         \;\;\;\;\;\;450 \\
             & FIRST     &      25\% & $5.4^{\prime\prime}$ & 1.4\,GHz &         \;\;\;\;\;\;130 \\
             & VLASS     &      82\% & $2.5^{\prime\prime}$ & 3.0\,GHz &    \;\;\;\;\;\;\;\;70 \\
       GMRT  & TGSS-ADR  &      90\% & $25^{\prime\prime}$  & 150\,MHz & \;\;\;\;\;\;\;\;\;\;5 \\
       LOFAR & LoTSS     &      50\% & $6^{\prime\prime}$   & 144\,MHz &    \;\;\;\;\;\;\;\;70 \\
       WSRT  & WODAN     & $\;\;$1\% & $15^{\prime\prime}$  & 1.3\,GHz &    \;\;\;\;\;\;\;\;10 \\
       ATA   & ATATS     &     1.7\% & $150^{\prime\prime}$ & 1.4\,GHz &         \;\;\;\;\;\;150 \\
        \hline
               ASKAP   & EMU     &      75\% &    $14^{\prime\prime}$ &      940\,MHz &    \;\;\;\;\;\;\;\;15 \\
                & WALLABY &      75\% &    $30^{\prime\prime}$ &      1.4\,GHz &        \;\;\;\;1\,600 \\
                & VAST    & 1.8--24\% &    $10^{\prime\prime}$ & 1.1--1.4\,GHz &               10--500 \\
                & RACS    &      83\% &    $15^{\prime\prime}$ &      890\,MHz &       \;\;\;\;\;\;250 \\
        MWA     & GLEAM   &      60\% &     \;\;$2^{\prime}\;$ &      200\,MHz &           \;\;50\,000 \\
        MeerKAT & MIGHTEE &    0.05\% & \;\;$6^{\prime\prime}$ & 0.9--1.7\,GHz & \;\;\;\;\;\;\;\;\;\;1 \\
        SKAMP   & SUMSS   &      25\% &    $45^{\prime\prime}$ &      840\,MHz & \;\;\;\;\;\;\;\;\;\;8 \\
                & MGPS-2  &      25\% & \;\;$2^{\prime\prime}$ &      840\,MHz & \;\;\;\;\;\;\;\;\;\;8 \\
        ATCA    & SCORPIO &  0.0097\% &    $10^{\prime\prime}$ &      2.1\,GHz &    \;\;\;\;\;\;\;\;30 \\
                & ATLAS   &  0.0090\% & \;\;$6^{\prime\prime}$ &      1.4\,GHz &    \;\;\;\;\;\;\;\;10 \\
        \hline\hline
   \end{tabular}
   }
   \label{tb:radio}
\end{table}

Surveys with Southern hemisphere telescopes include those with ASKAP \citep{johnston_2007,johnston_2008}, at the Murchison Radio Observatory (MRO) in Australia, such as EMU \citep{norris_2011,norris_2021}, Widefield ASKAP L-band Legacy All-sky Blind surveY \citep[WALLABY,][]{koribalski_2020}, Variables and Slow Transients \citep[VAST,][]{banyer_2012,murphy_2013,murphy_2021}, and Rapid ASKAP Continuum Survey \citep[RACS,][]{mcconnell_2020,hale_2021}. MRO is also home to the Murchison Widefield Array \citep[MWA,][]{lonsdale_2009,tingay_2013,wayth_2015} which conducted the GaLactic and Extragalactic All-Sky MWA \citep[GLEAM,][]{wayth_2015,hurley_2017} survey. These facilities along with the Hydrogen Epoch of Reionisation Array \citep[HERA,][]{deboer_2017} and Karoo Array Telescope \citep[MeerKAT,][]{jonas_2009,jonas_2018}, in the Karoo region South Africa, are precursors to the SKA.\footnote{\url{https://www.skatelescope.org/}} MeerKAT is conducting the MeerKAT International GigaHertz Tiered Extragalactic Exploration survey \citep[MIGHTEE,][]{jarvis_2018,heywood22}. The lower partition of Table~\ref{tb:radio} summarises the characteristics of these surveys.

An earlier SKA precursor is the SKA Molonglo Prototype \citep[SKAMP,][]{adams_2004}, in South Eastern Australia, which conducted the Sydney University Molonglo Sky Survey \citep[SUMSS,][]{mauch_2003} and Molonglo Galactic Plane Survey \citep[MGPS-2,][]{murphy_2007} surveys. Other surveys of note include the Stellar Continuum Originating from Radio Physics In Ourgalaxy \citep[SCORPIO,][]{umana_2015} and Australia Telescope Large Area Survey \citep[ATLAS,][]{norris_2006}, conducted with the Australia Telescope Compact Array (ATCA) facility,\footnote{\url{https://www.narrabri.atnf.csiro.au}} in New South Wales, Australia.

As radio telescope technology improves, radio surveys are becoming rich in the number and types of sources detected. EMU is expected to detect up to about 40~million sources \citep{norris_2011,norris_2021}, expanding our knowledge in areas such as galaxy and star formation. VAST, operating at a cadence of $5s$, opens up areas of variable and transient research: \textit{e.g.}, flare stars, intermittent pulsars, X-ray binaries, magnetars, extreme scattering events, interstellar scintillation, radio supernovae, and orphan afterglows of gamma-ray bursts \citep{murphy_2013,murphy_2021}. In short, surveys are approaching a point where the data volumes make transferring and reanalysis challenging at best. This places a strong demand on SF technology, requiring near- or real-time processing strategies.

\section{Analysis}
\label{sc:analysis}
In this section we use Hydra to do a comparison study between the Aegean \citep{hancock_2012,hancock_2018}, Caesar \citep{riggi_2016,riggi_2019}, ProFound \citep{robotham_2018,hale_2019}, PyBDSF \citep{mohan_2015}, and Selavy \citep{whiting_2012} SFs. The image data consists of three $2^\circ\times2^\circ$ images. These are the simulated-compact (CMP) and simulated-extended (EXT) images from Paper~I, and a subregion of a real EMU image (see \S\,\ref{sc:image_data} below). The simulated images are detailed in Paper~I, which also assessed the value of the $\mathcal{C_{DS}}$ and $\mathcal{R_{DS}}$ metrics. The image sizes were restricted to $2^\circ\times2^\circ$ in order to keep processing times manageable through the development of Hydra. 

In the analyses presented here, the total flux densities and component sizes are obtained from the component catalogues generated by each of the SFs. ProFound's total flux densities are computed through pixel sums, whereas the other SFs obtain their total flux densities and component sizes from Gaussian fits (see detailed discussion in Paper~I). ProFound characterises components through flux-weighted measurements of its detected segments (comparable to ``islands''). Accordingly, they are constrained to lie within segment boundaries. This should be kept in mind when comparing total flux densities and source component sizes derived from ProFound with those of the other SFs, as they are measured differently.

\subsection{Image Data}
\label{sc:image_data}
The simulated images are described in Paper~I, and we summarise the key points here for convenience. The simulated beam size is set to 15$^{\prime\prime}$ FWHM at a 4$^{\prime\prime}$/pixel image-scale, with a $20\mu$Jy/beam (RMS) noise floor. The CMP and EXT images consist of 9\,075 sources at 15$^{\prime\prime}$ in size, and 9\,974 sources varying from 15$^{\prime\prime}$ to 45$^{\prime\prime}$ in size (with axis-ratios varying between 0.4 and 1), respectively. The CMP sources have a maximum peak flux density of $1\,$Jy and minimum of $50\,\mu$Jy. The EXT is a combination of simulated compact and extended sources with maxima of $1\,$Jy and $1\,$mJy, respectively. For our CMP and EXT images, the ratios ($\sigma_A$) of the total component area (\textit{i.e.}, $\pi\Sigma_{i=1}^Na_ib_i$ for N components with semi-major axes, $a_i$, and semi-minor axes, $b_i$) to total image area (\textit{i.e.}, $2^\circ\times2^\circ$) are 0.031 and 0.055, respectively, and are thus still above the confusion limit.

\begin{figure}[hbt!]
\centering%
\includegraphics[scale=0.25]{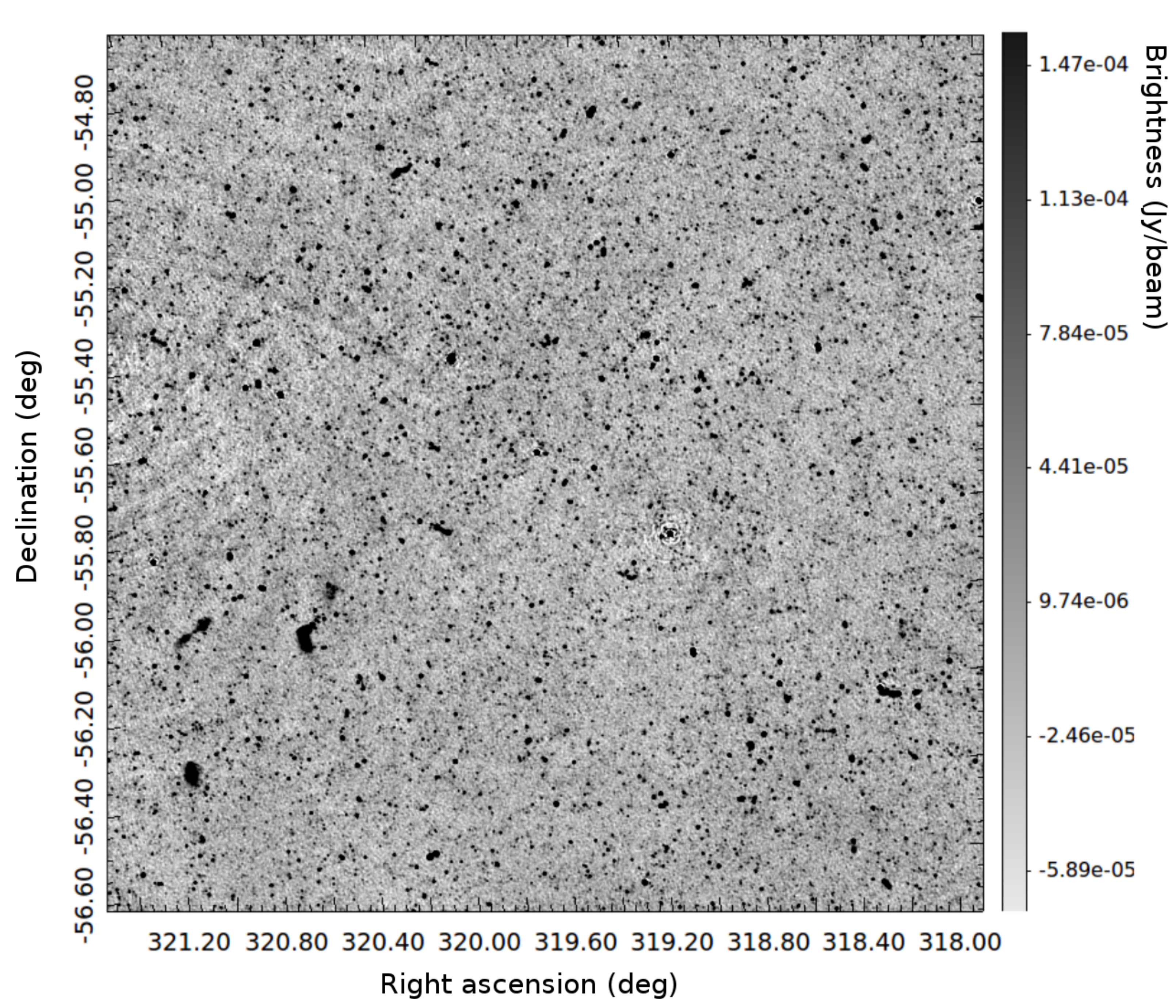}
\caption{$2^\circ\times2^\circ$ central cutout of an EMU pilot tile.}
\label{fg:emupilot}
\end{figure}

Figure~\ref{fg:emupilot} shows a $2^\circ\times2^\circ$ cutout from the centre of an EMU Phase~1 Pilot tile that we use for this analysis. This region provides a good mixture of compact, extended, and complex sources including diffuse extended emission. The simulated images do not include diffuse emission.

\subsection{Processing Speeds}
In the course of our analyses we were able to use Hydra to assess and compare the processing speeds of the different SFs. Table~\ref{tb:sf_speeds} shows order of magnitude CPU times for one PRD calculation step, during the optimisation process (Paper~I).

\begin{table}[hbt!]
\caption{SF order of magnitude PRD CPU times (rounded to one significant figure) for CMP, EXT, and EMU images. The processing was done on a 2GHz 16 core (single threaded) Intel Xeon Processor with 60G of RAM running Ubuntu 20.04.3 LTS.}
\centering
\begin{tabular}{@{\;}llccc@{\;}}
\hline\hline
    &  \multicolumn{3}{c}{PRD CPU Time (s)} \\
SF  & CMP & EXT &  EMU\\
\hline
Aegean   &    \;\;\,300 &    \;\;\,600 &    \;\;\,200 \\
Caesar   &       1\,000 &       4\,000 &    \;\;\,500 \\
ProFound & \;\;\;\;\,90 & \;\;\;\;\,90 & \;\;\;\;\,40 \\
PyBDSF   & \;\;\;\;\,90 & \;\;\;\;\,90 & \;\;\;\;\,40 \\
Selavy   &       1\,000 &       4\,000 &       3\,000 \\
\hline\hline
\end{tabular}
\label{tb:sf_speeds}
\end{table}

\begin{table*}[hbt!]
\caption{Typhon run statistics for $2^\circ\times2^\circ$ CMP and EXT and EMU images, with SF, image depth ($\mbox{D}= \mathcal{D},\,\mathcal{S}$), RMS parameter ($n_{rms}$ [$\sigma$]) island parameter, ($n_{island}$ [$\sigma$]), source numbers (N), residual RMS [$\mu$Jy/beam], and residual MADFM (median absolute deviation from the median) [$\mu$Jy/beam] columns.$^a$ The CMP and EXT results are reproduced from Paper~I.}
\centering
{\small
\begin{tabular}{@{\;}l@{\;\;}l@{\;\,}c@{\;\,}c@{\;\,}r@{\;\,}c@{\;\,}c@{\;\,}c@{\;\,}c@{\;\,}r@{\;\,}c@{\;\,}c@{\;\,}c@{\;\,}c@{\;\,}r@{\;}c@{\;\,}c@{\;\;}}
\hline\hline
SF &  D &
   \multicolumn{5}{@{}c@{}}{CMP Sources} & \multicolumn{5}{@{}c@{}}{EXT Sources} & \multicolumn{5}{@{}c@{}}{EMU Sources}\\
   & &
   \multicolumn{1}{@{}c@{}}{$n_{rms}$} & \multicolumn{1}{@{}c@{}}{$n_{island}$} & \multicolumn{1}{@{}c@{}}{N} & RMS & MADFM &
   \multicolumn{1}{@{}c@{}}{$n_{rms}$} & \multicolumn{1}{@{}c@{}}{$n_{island}$} & \multicolumn{1}{@{}c@{}}{N} & RMS & MADFM &
   \multicolumn{1}{@{}c@{}}{$n_{rms}$} & \multicolumn{1}{@{}c@{}}{$n_{island}$} & \multicolumn{1}{@{}c@{}}{N} & RMS & MADFM \\
\hline
Aegean   & $\mathcal{D}$ & 3.074 & 3.070 & 6\,016 & 20.0 & 19.0 & 3.074 & 3.070 & 4\,112 & 67.0 & 56.0 & 2.676 & 2.674 &  8\,538 & 91.0 & 26.0 \\
Caesar   & $\mathcal{D}$ & 3.074 & 3.000 & 4\,084 & 19.0 & 16.0 & 3.206 & 3.000 & 3\,618 & 54.0 & 44.0 & 2.809 & 2.806 &  7\,838 & 27.0 & 23.0 \\
ProFound & $\mathcal{D}$ & 2.412 & 2.409 & 4\,997 & 18.0 & 16.0 & 2.412 & 2.409 & 3\,730 & 52.0 & 43.0 & 2.279 & 2.277 & 11\,484 & 24.0 & 19.0 \\
PyBDSF   & $\mathcal{D}$ & 2.809 & 2.806 & 5\,991 & 22.0 & 19.0 & 2.809 & 2.806 & 4\,688 & 106 & 54.0 & 2.941 & 2.938 &  8\,292 & 1080 & 26.0 \\
Selavy   & $\mathcal{D}$ & 3.206 & 3.203 & 3\,225 & 45.0 & 20.0 & 3.206 & 3.203 & 2\,544 & 982 & 58.0 & 2.941 & 2.938 &  5\,880 & 566 & 27.0 \\
\hline\hline
Aegean   & $\mathcal{S}$ & 3.868 & 3.864 &    747 & 110  & 110  & 3.603 & 3.599 & 1\,287 & 321  & 317  & 3.603 & 3.599 &     926 & 192  & 169  \\
Caesar   & $\mathcal{S}$ & 4.000 & 3.000 &    657 & 109  & 106  & 3.603 & 3.000 & 1\,297 & 310  & 295  & 3.735 & 3.000 &     885 & 169  & 166  \\
ProFound & $\mathcal{S}$ & 3.074 & 3.070 &    642 & 109  & 107  & 2.941 & 2.938 & 1\,138 & 311  & 298  & 3.735 & 3.732 &     778 & 169  & 165  \\
PyBDSF   & $\mathcal{S}$ & 3.735 & 3.732 &    598 & 111  & 109  & 3.338 & 3.335 & 1\,312 & 316  & 313  & 3.735 & 3.732 &     794 & 409  & 169  \\
Selavy   & $\mathcal{S}$ & 4.000 & 3.996 &    427 & 114  & 110  & 3.735 & 3.732 &    787 & 623  & 320  & 3.603 & 3.599 &     789 & 169  & 169  \\
\hline\hline
\multicolumn{17}{l}{\footnotesize$\!\!^a$The MADFM estimators are normalised by 0.6744888 \citep{whiting_2012b}.}
\end{tabular}
}
\label{tb:hydra_2x2_typhon_stats}
\end{table*}

The wide range in CPU times is due to the background and noise estimators being used. For Caesar and Selavy we are using robust statistics \citep[Paper~I]{whiting_2012,riggi_2016}, which has time complexity \citep{knuth_2000} $\mathcal{O}(n\,\log\,n)$. Aegean also has the same time complexity, however it uses an Inter-Quartile Range (IQR) estimator \citep{hancock_2012}. PyBDSF uses the typical $\mu/\sigma$ estimator \citep{mohan_2015}, which goes as $\mathcal{O}(n)$. ProFound uses $\sigma$-clipping \citep{robotham_2018}, which is also of that order. The CPU times can be seen to vary, sometimes significantly, between images.

\subsection{Typhon Statistics}
\label{sc:typhon_stats}
\subsubsection{Optimisation Run Results}
Hydra's Typhon software module (Paper~I) was used to optimise Aegean, Caesar, ProFound, PyBDSF, and Selavy's RMS and island parameters to a 90\% PRD cutoff (Table~\ref{tb:hydra_2x2_typhon_stats}). Aegean, PyBDSF, and Selavy RMS box parameters were pre-optimised (Table~\ref{tb:hydra_2x2_typhon_rms_box_stats}) using \textsc{bane} \citep{hancock_2018} as part of a process to minimise the mean noise (\textit{i.e.}, $\mu$-optimisation, Paper~I). Here we present the EMU $\mathcal{D/S}$-image results, with relevant CMP and EXT image results from Paper~I incorporated as appropriate for comparison.

\begin{table}[hbt!]
\caption{Hydra $\mu$-optimised Aegean, PyBDSF, and Selavy \texttt{box\_size} and \texttt{step\_size} input parameters,$^a$ for CMP, EXT, and EMU $\mathcal{D}/\mathcal{S}$-images. The CMP and EXT results have been incorporated from Paper~I.}
\centering
\begin{tabular}{@{\;}l@{\;\;}c@{\;\;}c@{\;\;}c@{\;\;}c@{\;}}
\hline\hline
Image & Image &      $\mu$     & \texttt{box\_size}  & \texttt{step\_size} \\
Type  & Depth  & ($\mu$Jy/beam) & ($^{\prime\prime}$) & ($^{\prime\prime}$) \\
\hline
CMP  & $\mathcal{D}$ & 21.81  & 240 & 120 \\
         & $\mathcal{S}$ & 108.2  & 180 &  88 \\
\hline\hline
EXT & $\mathcal{D}$ & 68.01  & 480 & 240 \\
         & $\mathcal{S}$ & 325.3  & 240 & 120 \\
\hline\hline
EMU      & $\mathcal{D}$ & 35.28  & 720 & 270 \\
         & $\mathcal{S}$ & 171.23 & 216 &  80 \\
\hline\hline
\multicolumn{4}{l}{\footnotesize$\!\!^a$Selavy only accepts \texttt{box\_size}.}
\end{tabular}
\label{tb:hydra_2x2_typhon_rms_box_stats}
\end{table}

Table~\ref{tb:hydra_2x2_typhon_stats} summarises the Typhon run statistics for our image data. As noted in Paper I, the RMS and island parameters are similar for CMP and EXT $\mathcal{D}$-images for a given SF; however, they are slightly lower for the real $\mathcal{D}$-image. As for the $\mathcal{S}$-image, the simulated and real parameters are similar.

Table~\ref{tb:hydra_2x2_typhon_rms_box_stats} summarizes the EMU $\mathcal{D/S}$-image optimized RMS box parameters for Aegean, PyBDSF, and Selavy. First, we note that $\mu_{\mathcal{S}}/\mu_{\mathcal{D}}\sim 5$ for all images. This is consistent with the factor of 5 noise increase for the $\mathcal{S}$-image. Secondly, $\mu^{\mbox{\tiny EMU}}_{\mathcal{D}}\sim35\,\mu$Jy/beam falls between $\mu^{\mbox{\tiny CMP}}_{\mathcal{D}}\sim22\,\mu$Jy/beam and $\mu^{\mbox{\tiny EXT}}_{\mathcal{D}}\sim68\,\mu$Jy/beam (and similarly for $\mathcal{S}$-images), suggesting the simulated images are well suited for our analysis.

\begin{figure}[htb!]
\begin{center}
\includegraphics[width=\columnwidth]{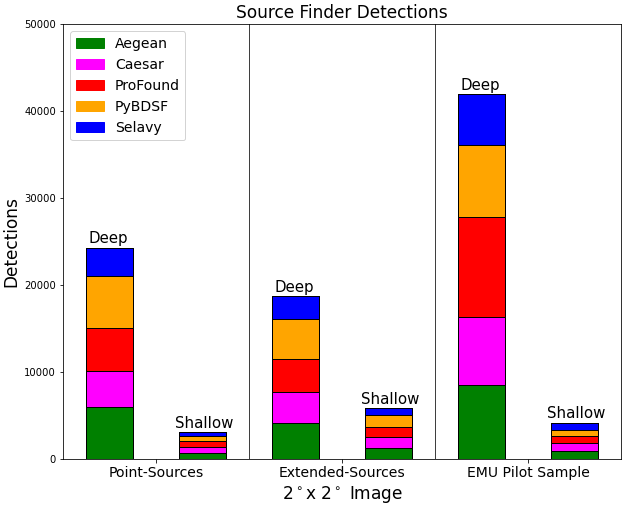}
\caption{SF CMP, EXT, EMU $\mathcal{D/S}$-image detection stacked plots (from the N columns of  Table~\ref{tb:hydra_2x2_typhon_stats}). \label{fg:source_finder_detections}}
\end{center}
\end{figure}

Figure~\ref{fg:source_finder_detections} shows stacked measures of $\mathcal{D/S}$ source detections (N) for the $\mathcal{D/S}$-images, where we have included the simulated data from Paper~I. For the $\mathcal{S}$-images, for a given SF, N is similar between the CMP and EMU sources, but slightly higher for EXT sources. With the exception of Selavy, N is similar between SFs. As for the simulated $\mathcal{D}$-images, the relative proportion of N between SFs is roughly the same for CMP and EXT sources, except for Selavy. In the EMU image there is a significant increase in the relative proportion of N by ProFound. This suggests a possible excess of spurious detections, an increase in real detections missed by other SFs, real detections being split into more components than by other SFs, or a combination of these.

\begin{table}[hbt!]
\caption{$\mathcal{S:D}$ recovery rates.}
\centering
\begin{tabular}{lccc}
\hline\hline
SF & CMP & EXT & EMU\\
\hline   
Aegean   & 12.4\% & 31.3\% &    10.8\% \\
Caesar   & 16.1\% & 35.8\% &    11.3\% \\
ProFound & 12.8\% & 30.5\% & \;\;6.8\% \\
PyBDSF   & 10.0\% & 28.0\% & \;\;9.6\% \\
Selavy   & 13.2\% & 30.9\% &    13.4\% \\
\hline\hline
\end{tabular}
\label{tb:hydra_2x2_typhon_n_stats}
\end{table}

In Paper~I we examined the ratios of deep-to-injected ($\mathcal{D:J}$) and shallow-to-deep ($\mathcal{S:D}$) source detections, or recovery rates. Table~\ref{tb:hydra_2x2_typhon_n_stats} shows the $\mathcal{S:D}$ recovery rates for our image data. The $\mathcal{S:D}$ recovery rates are roughly of the same order between the CMP
and EMU images, indicating the real image has a large population of compact sources (\textit{c.f.} Figure~\ref{fg:emupilot}). The recovery rate is higher for the EXT
image, by a factor around $2-3$, likely a reflection of the different flux density distribution of sources in this image. This would seem to suggest that the CMP image more closely models our real sample. This is also evident when the real image (Figure~\ref{fg:emupilot}) is compared with the simulated images in Paper~I.

\begin{table}[hbt!]
\centering
\caption{$\mathcal{D/S}$ (D) residual (res.) $|\mbox{RMS}-\mbox{MADFM}|/\mbox{MADFM}$ (\%), RMS [$\mu$Jy/beam], and MADFM [$\mu$Jy/beam] statistics extracted from Table~\ref{tb:hydra_2x2_typhon_stats}.}
\begin{tabular}{@{\;}l@{\;\;}c@{\;\;}c@{\;\;}c@{\;\;}c@{\;}}
   \hline\hline
   SF/Res.  & D & \multicolumn{1}{c}{CMP} & \multicolumn{1}{c}{EXT} & \multicolumn{1}{c}{EMU} \\
   \hline
      Aegean & $\mathcal{D}$ & \;\;\;\;5.3\% &  \;\;\;\;19.6\% &     250.0\% \\
     Caesar & $\mathcal{D}$ &    \;\;18.8\% &  \;\;\;\;22.7\% &  \;\;17.4\% \\
   ProFound & $\mathcal{D}$ &    \;\;12.5\% &  \;\;\;\;20.9\% &  \;\;26.3\% \\
     PyBDSF & $\mathcal{D}$ &    \;\;15.8\% &  \;\;\;\;80.4\% &  \;\;58.5\% \\
     Selavy & $\mathcal{D}$ &       125.0\% &  \;\;\;\;69.3\% &     109.6\% \\
   \hline
   RMS      & $\mathcal{D}$ & $24.8\pm10.2$    & $56.4\pm28.2$    & $41.9\pm28.8  $ \\
   MADFM    & $\mathcal{D}$ & $18.0\pm1.7\;\;$ & $51.0\pm6.3\;\;$ & $24.2\pm2.9\;\;$ \\
   \hline\hline
      Aegean & $\mathcal{S}$ & \;\;\;\;0.0\% &  \;\;\;\;1.3\% &    \;\;13.6\% \\
     Caesar & $\mathcal{S}$ & \;\;\;\;2.8\% &  \;\;\;\;5.1\% & \;\;\;\;1.8\% \\
   ProFound & $\mathcal{S}$ & \;\;\;\;1.9\% &  \;\;\;\;4.4\% & \;\;\;\;2.4\% \\
     PyBDSF & $\mathcal{S}$ & \;\;\;\;1.8\% &  \;\;\;\;1.0\% &       142.0\% \\
     Selavy & $\mathcal{S}$ & \;\;\;\;3.6\% &     \;\;94.7\% & \;\;\;\;0.0\% \\
   \hline
   RMS      & $\mathcal{S}$ & $110.6\pm1.9$ & $376.2\pm 123.5$    & $221.6\pm94.1  $ \\
   MADFM    & $\mathcal{S}$ & $108.4\pm1.6$ & $308.6\pm 10.2\;\;$ & $167.6\pm1.7\;\;$ \\
   \hline\hline
\end{tabular}
\label{tb:rms_madfm_prds}
\end{table}

For simulated and real images there is some variability in the residual RMS values, with Selavy being consistently high for $\mathcal{D}$-images and consistently low for $\mathcal{S}$-images. On the whole, however, the residual MADFMs are consistent between all SFs for a given image. Table~\ref{tb:rms_madfm_prds} summarises these results. The $|\mbox{RMS}-\mbox{MADFM}|/\mbox{MADFM}$ percentage differences are a measure of the robustness of the SF residual models (which is also reflection of the Hydra optimisation schema, Paper~I).

\subsubsection{Source Size Distributions}
\begin{figure}[hbt!]
\centering%
\includegraphics[width=\columnwidth]{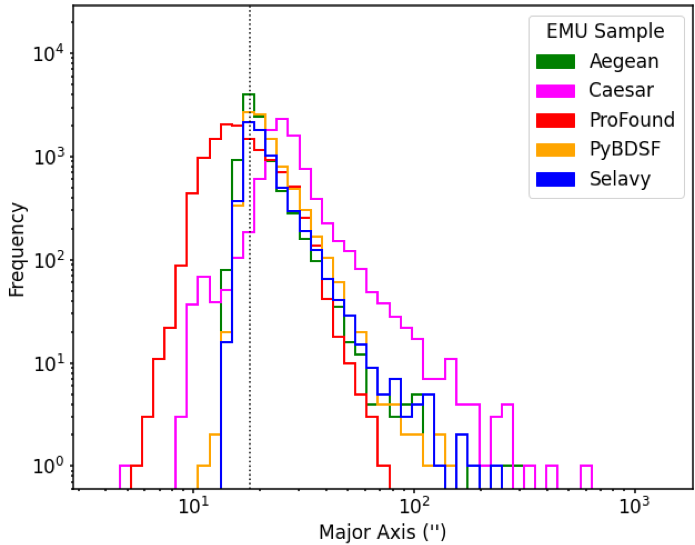}
\caption{Major-axis distributions for EMU sources (showing size distributions from both $\mathcal{D}$ and $\mathcal{S}$ image measurements together). NB: Size estimates between SFs are not necessarily directly comparable as they are estimated using different methods (Paper I).}
\label{fg:major_distributions}
\end{figure}

Figure~\ref{fg:major_distributions} shows the major-axis distribution for the EMU image data. Both the $\mathcal{D}$ and $\mathcal{S}$ source detections are combined, as the $\mathcal{S}$ data provides additional information. Although the underlying sources are in common between $\mathcal{D}$ and $\mathcal{S}$, the higher noise level means that the SFs are operating on quantitatively different pixel values, and may well derive different size estimates. Also, the numbers of detections for $\mathcal{S}$ are relatively small (Figure~\ref{fg:source_finder_detections}), so incorporating them here is unlikely to bias the results. Recall that size estimates for different SFs use different methods and are not necessarily directly comparable.

\begin{figure*}[hbt!]
\centering%
\begin{tabular}{c}
\includegraphics[width=0.95\textwidth]{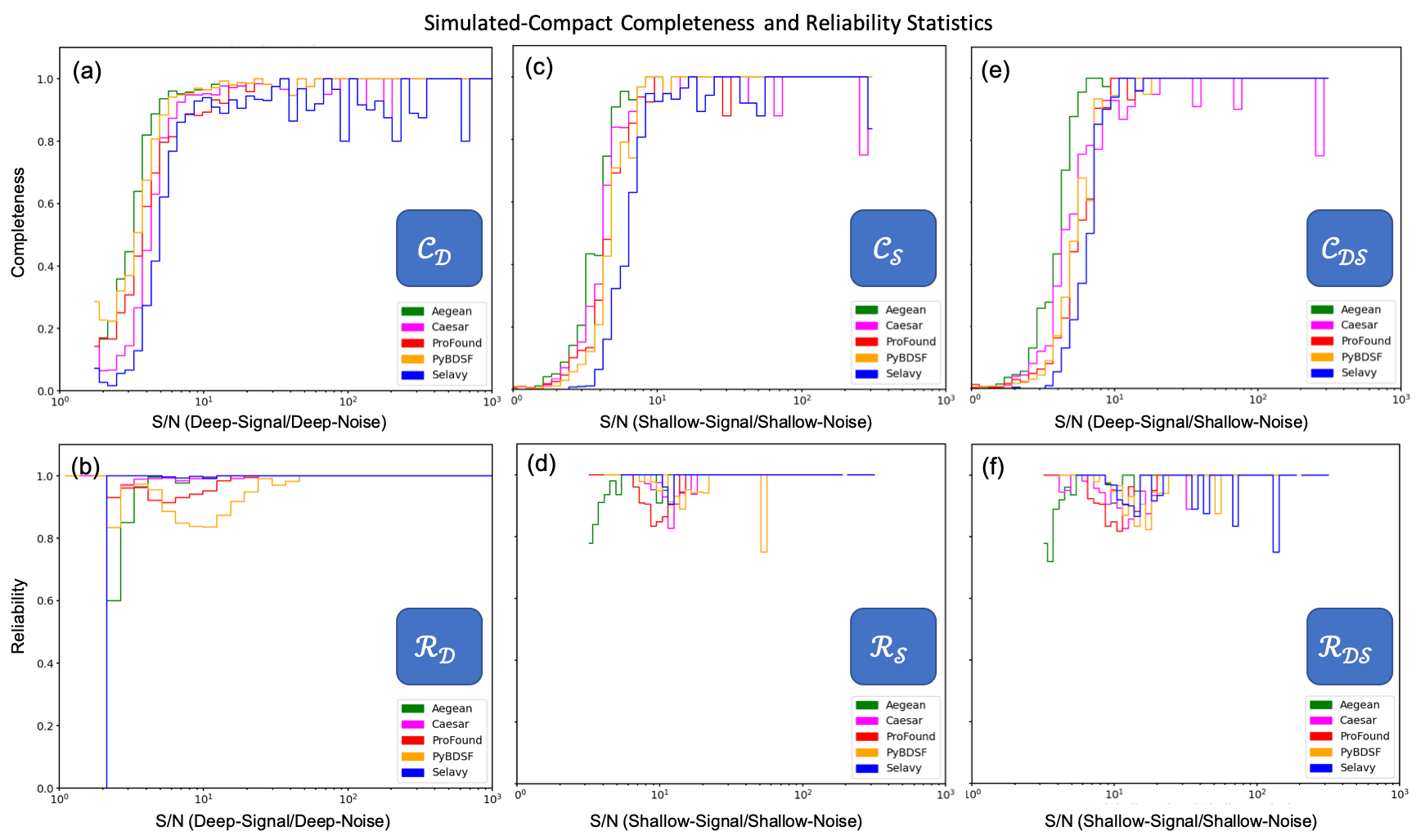}\\
\includegraphics[width=0.95\textwidth]{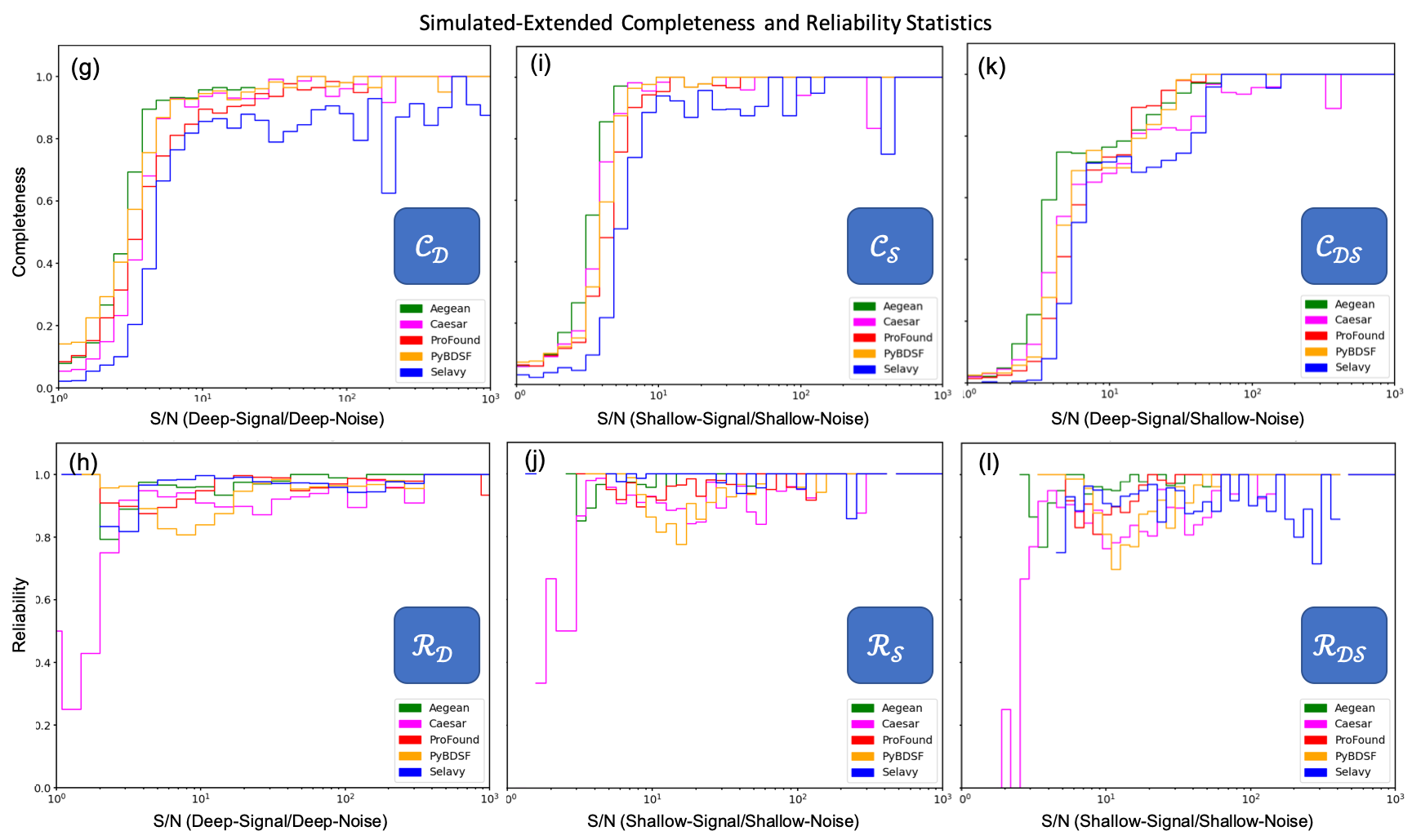}
\end{tabular}
\caption{CMP and EXT $\mathcal{C_D}$ (a and g), $\mathcal{R_D}$ (b and h), $\mathcal{C_S}$ (c and i), $\mathcal{R_S}$ (d and j), $\mathcal{C_{DS}}$ (e and k), and $\mathcal{R_{DS}}$ (f and l) \textit{vs.} S/N. These results are reproduced from Paper~I, Figures~14, 15, and 16.}
\label{fg:cds_rds_cmp_ext}
\end{figure*}

In Figure~\ref{fg:major_distributions}, all but one of the distributions peak at the scale corresponding to the beam size, consistent with point (\textit{i.e.}, compact) source detections. Caesar shows a distribution systematically offset by a factor of about 1.5 toward larger sizes. The results for Aegean, PyBDSF, and Selavy are similar, which is reassuring given that they use similar approaches in fitting elliptical Gaussians to source components (see Paper~I). ProFound shows an excess of very small sources. Some of these will be due to noise spikes, and others to faint sources close to the detection threshold (Table~\ref{tb:hydra_2x2_typhon_stats}) where ProFound only has a few pixels available from which to calculate its flux-weighted sizes, leading to underestimates. Caesar's excess at small sizes is less than that from ProFound, although were the Caesar results to be shifted left by the factor 1.5 noted above, it would be roughly comparable. The origin of small sizes reported by Caesar might be due to deblending issues, as it has a relatively lower source detection number (Figure~\ref{fg:source_finder_detections}). In general, all SFs with detections below the beam size are expected to be contaminated, to some degree, by noise spikes. 

As for EXT images, they have a similar behaviour to the EMU case, except to a lesser degree (Paper~I). In the CMP case, there are significantly fewer detections below the beam size, with some overestimates of source sizes in the case of ProFound and PyBDSF. The overestimates are most likely due to nearby noise peaks being incorporated into a source fit, artificially enlarging the size.

\subsection{Completeness and Reliability}
\label{sc:c_and_r}
\begin{figure}[hbt!]
\centering%
\begin{tabular}{c}
\includegraphics[width=0.95\columnwidth]{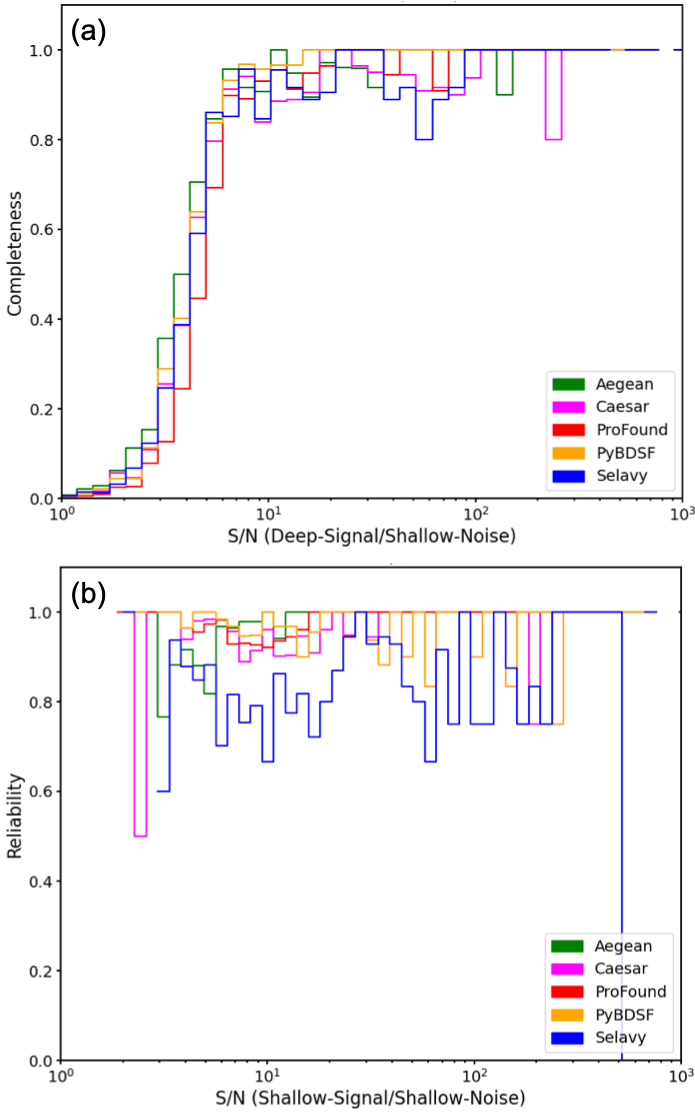}
\end{tabular}
\caption{EMU-image $\mathcal{C_{DS}}$ \textit{vs.} S/N (a) and $\mathcal{R_{DS}}$ (b) \textit{vs.} S/N, with S/N expressed as $\mathcal{D}$-signal/$\mathcal{S}$-noise and $\mathcal{S}$-signal/$\mathcal{S}$-noise, respectively.}
\label{fg:hydra_2x2_all_cr_plots}
\end{figure}

In Paper~I we explored $\mathcal{C_D}$ (and $\mathcal{C_S}$) and $\mathcal{R_D}$ (and $\mathcal{R_S}$) for CMP and EXT sources; for ease of discussion and comparison here, we have duplicated the relevant results in Figure~\ref{fg:cds_rds_cmp_ext}, along with the new measurements from the EMU data in Figure~\ref{fg:hydra_2x2_all_cr_plots}. Aegean was found to have the best $\mathcal{C}$ statistic followed PyBDSF, ProFound, Caesar, and Selavy. Selavy, followed by Caesar, tended to miss bright sources, and were less reliable ($\mathcal{R}$) at high S/N. In general, the statistics for all SFs are poorer for the EXT case, but followed the same performance trends as the CMP case. Some of this can been attributed to confusion.

Figure~\ref{fg:hydra_2x2_all_cr_plots} shows the newly measured $\mathcal{C_{DS}}$ and $\mathcal{R_{DS}}$ metrics for the EMU image. The results are similar to the CMP image case (Figure~\ref{fg:cds_rds_cmp_ext}). Here, the detections appear fairly complete for all SFs, with high completeness ($\mathcal{C_{DS}}\gtrsim 0.9$) above $\mbox{S/N}\gtrsim 10$ dropping to $\mathcal{C_{DS}}\sim 0.5$ by $\mbox{S/N}\sim 5$. Reliability is also generally high ($\mathcal{R_{DS}}\gtrsim 0.9$) for most SFs, although Selavy's performance here is notably poorer ($0.7\lesssim \mathcal{R_{DS}}\lesssim 0.9$) across almost the full range of S/N. All SFs drop in reliability below $\mbox{S/N}\sim 10-20$.

\subsection{Flux Density Ratios and $n\sigma$ Scatter}
\label{sc:flux_ratios}
\begin{figure*}[hbt!]
\centering%
\begingroup
\renewcommand{\arraystretch}{0} 
\begin{tabular}{@{}c@{}}
\includegraphics[width=0.85\textwidth]{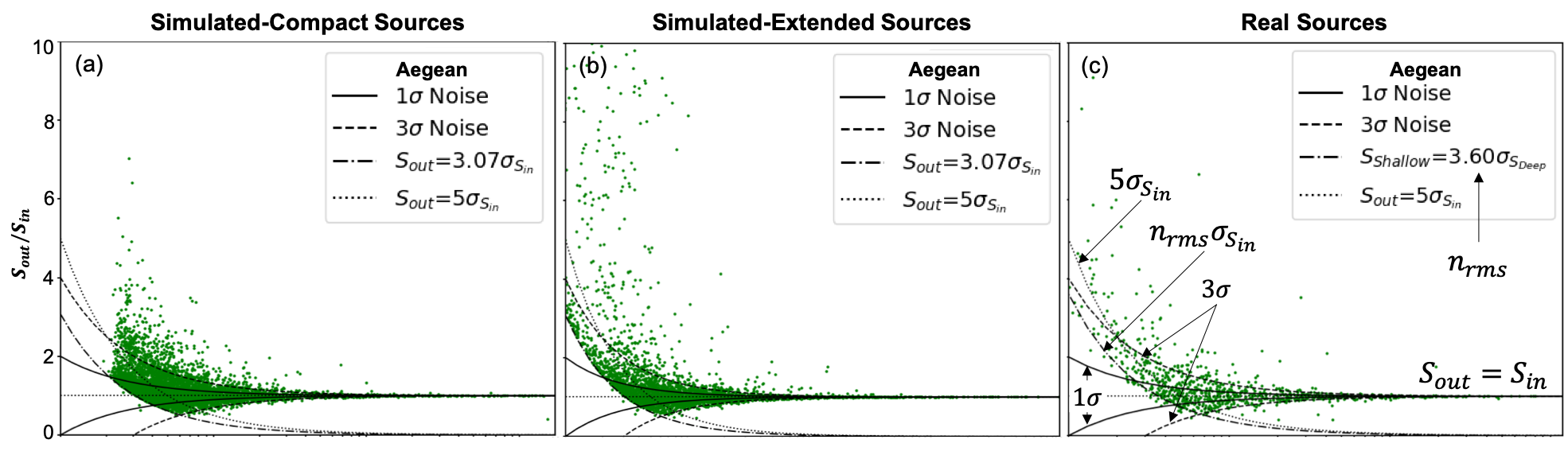}\\
\includegraphics[width=0.85\textwidth]{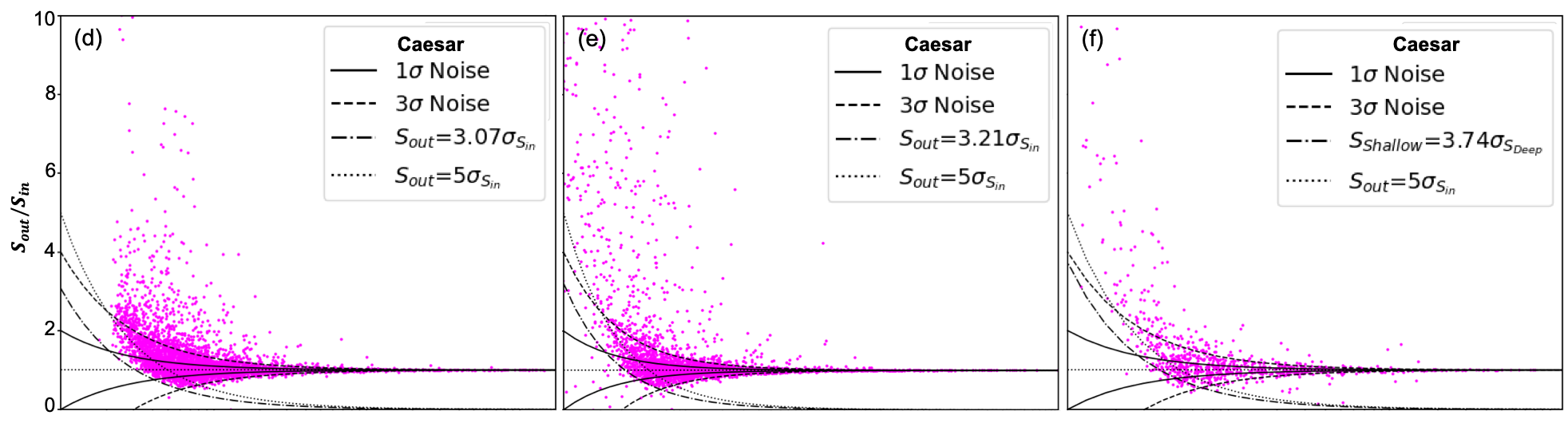}\\
\includegraphics[width=0.85\textwidth]{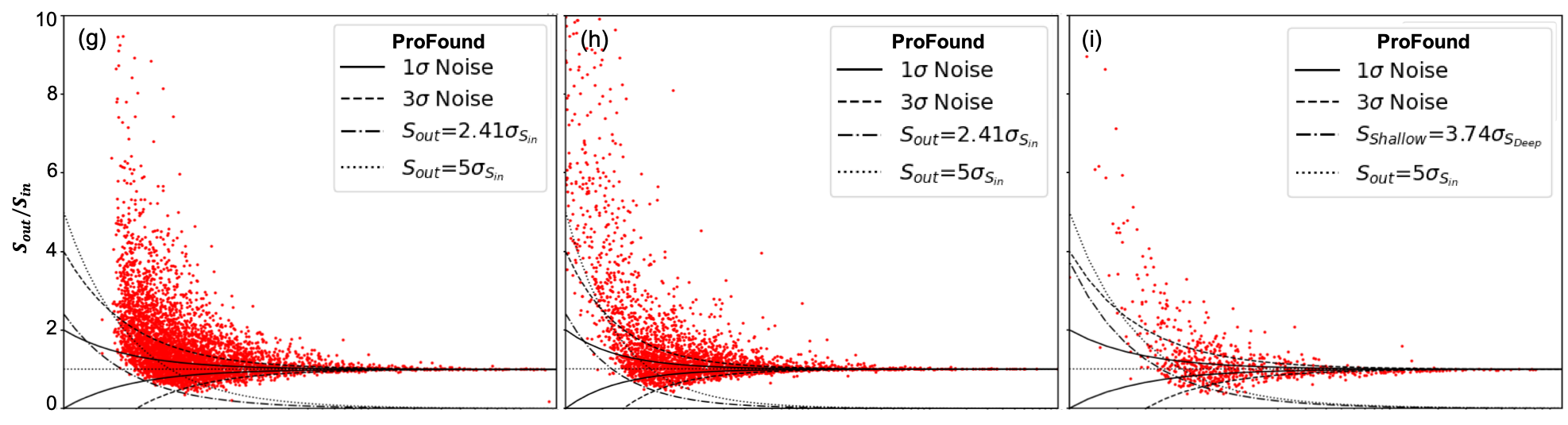}\\
\includegraphics[width=0.85\textwidth]{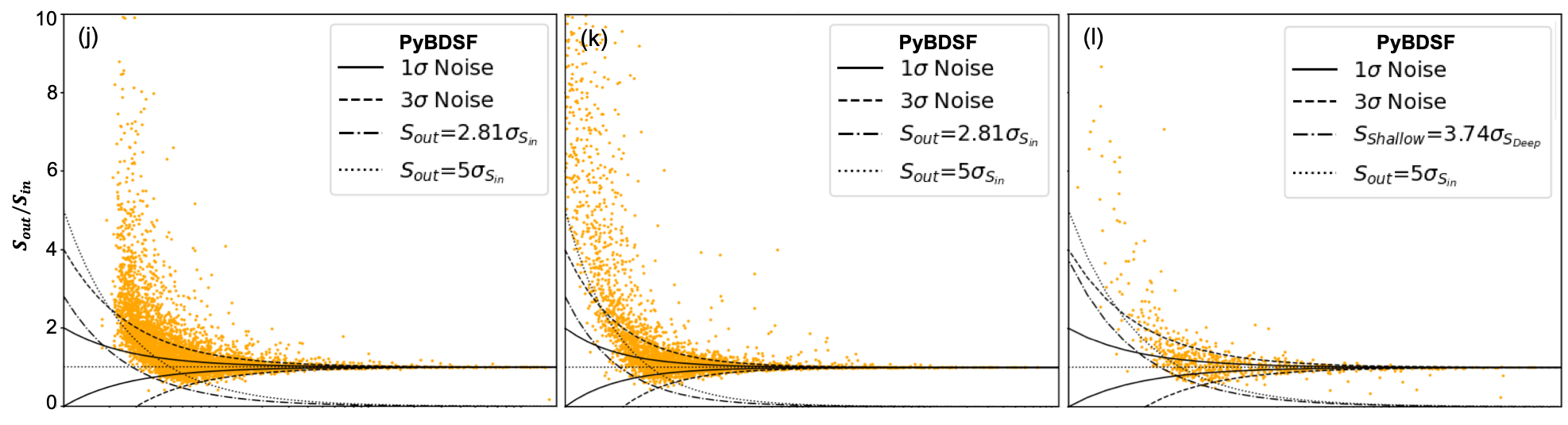}\\
\includegraphics[width=0.85\textwidth]{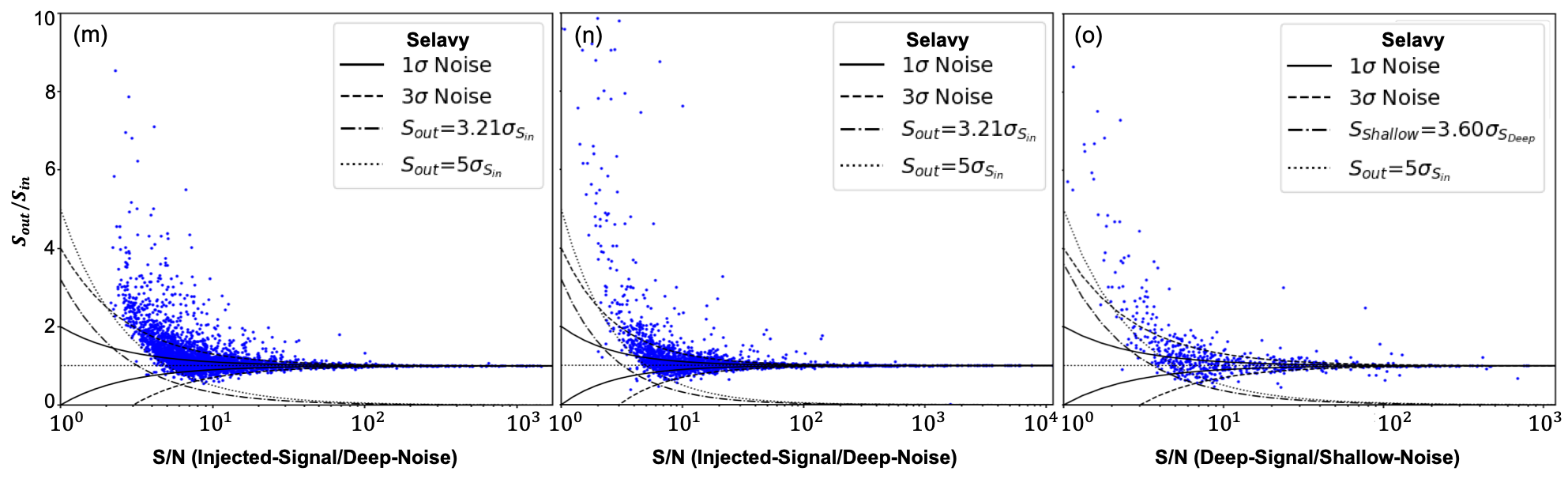}
\end{tabular}
\endgroup
\caption{Flux density ratios ($S_{out}/S_{in}$) for CMP (left), simulated-extended (middle), and real (right) sources for Aegean (a, b, and c), Caesar (c, e, and f), ProFound (g, h, and i), PyBDSF (j, k, and l) and Selavy (m, n, and o) \textit{vs.} S/N, expressed as $\mathcal{J}$-signal/$\mathcal{D}$-noise and $\mathcal{D}$-signal/$\mathcal{S}$-noise for simulated and real sources, respectively.
The $1\sigma$ (solid) and $3\sigma$ (dashed) curves are RMS noise ($\sigma$) deviations from the $\mbox{flux-ratio}=1$ lines (dotted). Also shown are the detection threshold ($n_{rms}$; dot-dashed curves) and nominal $5\sigma$ threshold (dotted curves). These curves are annotated in (c).}
\label{fg:hydra_2x2_deep_flux_ratios}
\end{figure*}

Figure~\ref{fg:hydra_2x2_deep_flux_ratios} shows the flux ratios, $S_{out}/S_{in}$, for each of the SFs. For the simulated images, $S_{in}$ (expressed as S/N) represents the $\mathcal{J}$-signal over the $\mathcal{D}$-noise. For the real images it represents the $\mathcal{D}$-signal over the $\mathcal{S}$-noise.
$S_{out}$ represents the detected measurement. The horizontal (dotted) lines represent $S_{out}=S_{in}$, and the solid and dashed curves about these lines are the $1\sigma$ and $3\sigma$ deviations from $S_{in}$, respectively. The dot-dashed curves are the detection thresholds (\textit{i.e.}, the RMS parameters, $n_{rms}$, in Table~\ref{tb:hydra_2x2_typhon_stats}, corresponding to 90\% PRD), and the dotted curves represent nominal $5\sigma$ thresholds \citep[\textit{c.f.}][]{hopkins_2015}.
The sharp vertical cut off below $\mbox{S/N} \sim 2$ for the CMP images is a consequence of having no artificial input sources fainter than that level.

\begin{figure*}[hbt!]
\centering%
\begingroup
\renewcommand{\arraystretch}{0} 
\begin{tabular}{@{}c@{}}
\includegraphics[width=\textwidth]{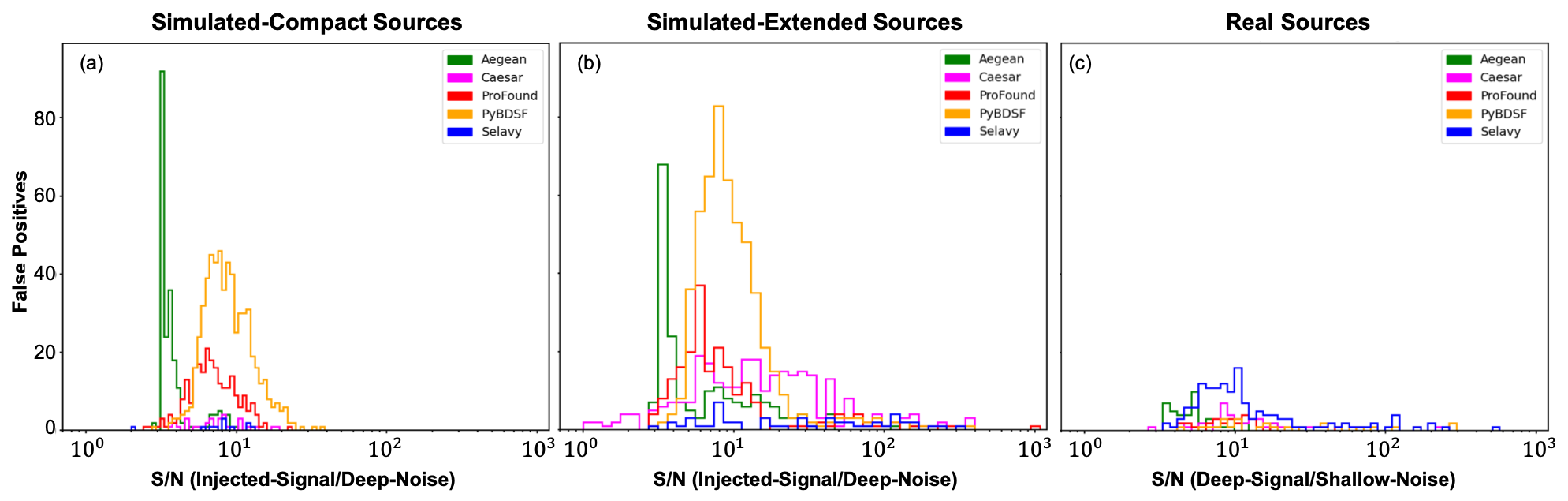}
\end{tabular}
\endgroup
\caption{False positives \textit{vs.} S/N for CMP (a), EXT (b), and real (c) sources, with S/N expressed as $\mathcal{J}$-signal/$\mathcal{S}$-noise and $\mathcal{D}$-signal/$\mathcal{S}$-noise for simulated and real sources, respectively.
}
\label{fg:hydra_2x2_deep_false_positives}
\end{figure*}

\begin{table*}[hbt!]
\caption{$3\sigma$ scatter ($s_{3\sigma}(\hat{S}_i)$, Equation~\ref{eq:scatter}) at $\hat{S}_i=3,\,5,\,10$, for SF CMP, EXT, and EMU source flux density ratios (Figure~\ref{fg:hydra_2x2_deep_flux_ratios}). Also shown are averaged $3\sigma$ scatters.}
\centering
{\small
\begin{tabular}{@{\;}l@{\;}c@{\,}c@{\,}c@{\;}c@{\,}c@{\,}c@{\;}c@{\,}c@{\,}c@{\;}c@{\,}c@{\,}c@{\;}}
\hline\hline
         &  \multicolumn{9}{@{}c@{}}{$3\sigma$ Scatter} & \multicolumn{3}{@{}c@{}}{Averaged $3\sigma$ Scatter}\\
         &  \multicolumn{3}{@{}c@{}}{CMP}  
         &  \multicolumn{3}{@{}c@{}}{EXT}
         &  \multicolumn{3}{@{}c@{}}{EMU}
         &  \multicolumn{3}{@{}c@{}}{$\langle[s_{3\sigma}(3),\,s_{3\sigma}(5),\,s_{3\sigma}(10)]\rangle$}\\
SF         &  $s_{3\sigma}(3)$ &  $s_{3\sigma}(5)$ & $s_{3\sigma}(10)$
         &  $s_{3\sigma}(3)$ &  $s_{3\sigma}(5)$ & $s_{3\sigma}(10)$
         &  $s_{3\sigma}(3)$ &  $s_{3\sigma}(5)$ & $s_{3\sigma}(10)$
         &  \multicolumn{1}{@{}c@{}}{CMP}
         &  \multicolumn{1}{@{}c@{}}{EXT}
         &  \multicolumn{1}{@{}c@{}}{EMU}\\
\hline
Aegean   &  3.49\% &  3.91\% &  4.27\% &  9.20\% &  7.98\% &  7.48\% & 3.40\% & 4.88\% &  5.59\% &  $\;\;$3.89 $\pm$ 0.32\% &  $\;\;$8.22 $\pm$ 0.72\% & $\;\;$4.62 $\pm$ 0.91\% \\
Caesar   &  6.93\% &  7.11\% &  6.65\% & 11.02\% &  9.92\% &  9.08\% & 4.62\% & 4.33\% &  6.76\% &  $\;\;$6.90 $\pm$ 0.19\% & 10.01 $\pm$ 0.79\%       & $\;\;$5.24 $\pm$ 1.08\% \\
ProFound & 14.55\% & 16.70\% & 21.74\% & 14.20\% & 15.93\% & 20.28\% & 6.60\% & 6.94\% & 10.72\% & 17.66 $\pm$ 3.01\%       & 16.80 $\pm$ 2.56\%       & $\;\;$8.09 $\pm$ 1.87\% \\
PyBDSF   &  8.85\% &  6.20\% &  6.64\% & 10.80\% &  9.55\% &  9.57\% & 3.43\% & 3.18\% &  5.88\% &  $\;\;$7.23 $\pm$ 1.16\% &  $\;\;$9.97 $\pm$ 0.58\% & $\;\;$4.16 $\pm$ 1.22\% \\
Selavy   &  6.43\% &  6.92\% &  6.63\% & 10.25\% &  8.91\% &  9.06\% & 3.94\% & 3.77\% &  5.99\% &  $\;\;$6.66 $\pm$ 0.20\% &  $\;\;$9.41 $\pm$ 0.60\% & $\;\;$4.57 $\pm$ 1.01\% \\
\hline\hline
\end{tabular}
}
\label{tb:3sigma_scatter}
\end{table*}

The form of the flux ratio diagrams for the EMU image follow that of the simulated images, however its $\mathcal{DS}$ source numbers are lower due to relying on detections in the $\mathcal{S}$-image. Similar $\mathcal{DS}$ flux ratio statistics can be measured for the simulated cases (not shown). We use $\mathcal{D}$ diagnostics for the simulated images, as they are the most robust, whereas the $\mathcal{S}$ and $\mathcal{DS}$ metrics provide no extra useful information.

For the simulated cases all SFs, with the exception of Selavy, are seen to detect sources down to the specified RMS threshold (dot-dashed line). Selavy was given a $\sim3.2\sigma$ threshold (Table~\ref{tb:hydra_2x2_typhon_stats}), but only appears to recover sources down to a $5\sigma$ level (Figure~\ref{fg:hydra_2x2_deep_flux_ratios} (m) and (n)).
For the real image case, though, Selavy does appear to recover sources down to the nominal S/N threshold.

Figure~\ref{fg:hydra_2x2_deep_false_positives} is a complementary diagnostic to Figure~\ref{fg:hydra_2x2_deep_flux_ratios}, and shows the corresponding false-positive distributions. This emphasises in another way the SF characteristics we have seen in earlier $\mathcal{R}$ distributions (\S~\ref{sc:c_and_r}). Specifically, for CMP sources (\textit{i.e.}, comparing Figures~\ref{fg:cds_rds_cmp_ext}b and~\ref{fg:hydra_2x2_deep_false_positives}a), Aegean picks up false detections predominantly close to the S/N threshold, with ProFound and PyBDSF having false detections peaking between about S/N$\sim 5-10$, arising from blended sources and overestimated source sizes. The false sources seen by each SF correspond to their deficit in $\mathcal{R}$ (Figure~\ref{fg:cds_rds_cmp_ext} and Figure~\ref{fg:hydra_2x2_all_cr_plots}). In particular, Aegean displays higher levels of completeness, but at the expense of reliability at low S/N, with the converse being true for Selavy and Caesar. In this analysis, Aegean shows arguably the best performance, as is evident from its false-positive distribution being limited to low S/N, generally reliable flux densities seen in the limited scattering beyond $S_{out}/S_{in}>3\sigma$, and good completeness to low S/N. This is not surprising, though, as Aegean is well designed for identifying and fitting point sources, and the simulated data used here do not include anything else.

The false-positive distributions for the EXT case (\textit{cf.}, Figures~\ref{fg:cds_rds_cmp_ext}h and~\ref{fg:hydra_2x2_deep_false_positives}b) have similar characteristics to the CMP case (Figures~\ref{fg:cds_rds_cmp_ext}b and~\ref{fg:hydra_2x2_deep_false_positives}a), but are broadened somewhat. As for the real case (Figure~\ref{fg:hydra_2x2_deep_false_positives}c), the distribution for Selavy becomes more prominent, as reflected in its $\mathcal{R_{DS}}$ (Figure~\ref{fg:hydra_2x2_all_cr_plots}b).

Pleasingly, none of the SFs show any systematic overestimate or underestimate of the flux densities (Figure~\ref{fg:hydra_2x2_deep_flux_ratios}). To quantify the reliability of the flux density measurements, we focus on the scatter in the distribution of $S_{out}/S_{in}$ as a function of S/N. We first
define the fraction of sources $r_{n\sigma}(\hat{S}_i)$ above a S/N limit, $\hat{S}_i$, with $S_{out}/S_{in}$ lying between $1-n\sigma$ and $1+n\sigma$:
\begin{equation}
    r_{n\sigma}(\hat{S}_i) = \frac{\displaystyle\left|\left\{\!\frac{S_{out}}{S_{in}}\!\left|1\!\!-\!\!\frac{n}{\hat{S}_{in}}\!\le\!\frac{S_{out}}{S_{in}}\!\le1\!+\!\!\frac{n}{\hat{S}_{in}}\!\ni\!\hat{S}_{in}\!\ge\!\hat{S_i}\right.\!\right\}\right|}{\displaystyle\left|\left\{\!\frac{S_{out}}{S_{in}}\!\left|\hat{S}_{in}\!\ge\!\hat{S_i}\!\right.\right\}\right|}\,,
    \label{eq:flux_ratio_fraction}
\end{equation}
where $S$ represents flux density, and $\hat{S}$ represents S/N.\footnote{So $S_{out}/S_{in}=\hat{S}_{out}/\hat{S}_{in}$.} Using this, we define the scatter in $S_{out}/S_{in}$ outside an $n\sigma$ range as
\begin{equation}
    s_{n\sigma}(\hat{S}_i)=1-r_{n\sigma}(\hat{S}_i)\,.
    \label{eq:scatter}
\end{equation}
Table~\ref{tb:3sigma_scatter} shows the compiled statistics for $3\sigma$ scatter.

For the CMP case, Aegean and Selavy tend to show the least scatter. The scatter for PyBDSF and Selavy is slightly higher than that reported by \cite{hopkins_2015}, who saw scatters above a $5\sigma$ detection threshold of 4.9\% for PyBDSF and 3.3\% for Selavy, compared to the values of $\sim 6\%$ seen here.

\begin{figure*}
\centering
\begin{tabular}{@{}c@{}}
   \includegraphics[width=\textwidth]{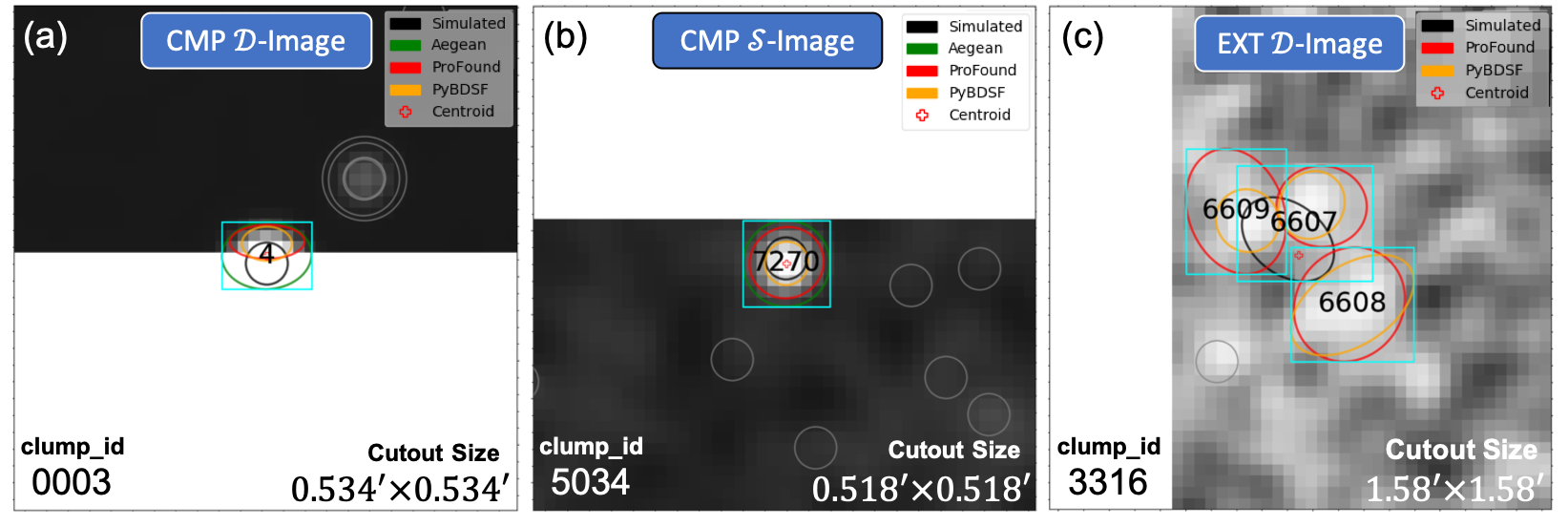}
\end{tabular}
\caption{\textbf{Edge Detection Infographic}: Edge detection example cutouts for (a) CMP $\mathcal{D}$-image (from the $\mbox{S/N}\sim1\,600\pm110$ bin of $\mathcal{C_D}$, Figure~\ref{fg:cds_rds_cmp_ext}a), (b) CMP $\mathcal{S}$-image (from the $\mbox{S/N}\sim42.6\pm2.9$ bin of $\mathcal{C_S}$, Figure~\ref{fg:cds_rds_cmp_ext}c), and (c) EXT $\mathcal{D}$-image (from the $\mbox{S/N}\sim0.318\pm0.036$ bin of $\mathcal{C_D}$, Figure~\ref{fg:cds_rds_cmp_ext}g). In examples (a) and (b), Caesar and Selavy failed to detected the injected sources at \texttt{match\_id}s 4 and 7270, respectively. In example (c), there is an injected source at \texttt{match\_id} 6607 ($\mbox{S/N}\sim0.34$) which is detected by ProFound ($\mbox{S/N}\sim4.82$) and PyBDSF ($\mbox{S/N}\sim6.39$). The remaining detections, by both SFs, at \texttt{match\_id}s 6608 and 6609 are spurious, due to noise fluctuations.}
\label{fg:le_edge}
\end{figure*}

On the whole, the scatter is comparable between the CMP and EMU sources, with the highest scatter for EXT sources. Aegean tends to deliver the least scatter in the flux density estimates, with Caesar, PyBDSF, and Selavy showing similar results at the next level, and ProFound always with the highest scatter. The results in the EMU image appear to show the least scatter overall, but recall that these measurements result from comparing the detections in the $\mathcal{S}$-image against those in the $\mathcal{D}$-image for each SF against itself. These are not directly comparable with simulated image cases where the scatter compares measured against known input values. The implication here is that flux density estimates inferred from $\mathcal{D}/\mathcal{S}$ comparisons may not reflect the full extent of the true underlying uncertainties.

\subsection{Hydra Viewer Cutout Case Studies}
\label{sc:case_studies}
Here we present a gallery of Hydra Viewer cutouts, for the purpose of analysing frequently encountered anomalies and artifacts utilising the metrics explored in the previous subsections. They have been sorted into broad categories: edge detections (sources near image edges), blending/deblending, faint sources, component size errors, bright sources, and diffuse emission. With the exception of diffuse emission (seen only in the real image), all such artefacts are found in the SF catalogues in all of the images, to varying degrees.

\subsubsection{Edge Detections}
Detecting sources at or near the edges of images is problematic, as it can be difficult to estimate the noise and sometimes the source itself is truncated. For sources truncated near an edge, Aegean tends to extrapolate their shape, while ProFound estimates the remaining flux, and PyBDSF treats them as complete sources: \textit{e.g.}, for the CMP source in Figure~\ref{fg:le_edge}a, Caesar and Selavy do not record this source, and hence it contributes to an underestimate of $\mathcal{C_D}$ at $\mbox{S/N}\sim1\,600\pm110$ for those finders (Figure~\ref{fg:cds_rds_cmp_ext}a). Here S/N $\sim$ 1\,519 for the injected source, and 621, 263, and 277 for Aegean, ProFound, and PyBDSF, respectively. So the source flux-density is underestimated. This does not appear as degradation in $\mathcal{C_D}$, but is reflected in the scatter of the flux-ratio (Figure~\ref{fg:hydra_2x2_deep_flux_ratios}).

SF performance is better when sources are just touching an edge: \textit{e.g.}, for the CMP source in Figure~\ref{fg:le_edge}b. Here again Caesar and Selavy do not report this source, which falls in the $\mathcal{C_S}$ $\mbox{S/N}\sim42.6\pm2.9$ bin (Figure~\ref{fg:cds_rds_cmp_ext}a). The other finders detect it and estimate the flux density for this source reliably. The injected source S/N is $\sim$ 41.66, and it is measured at 44.82, 44.92, and 45.19 for Aegean, ProFound, and PyBDSF, respectively.

Figure~\ref{fg:le_edge}c shows an example of noise spikes appearing as low S/N detections, surrounding an $\mbox{S/N}\sim0.34$ EXT source, near an edge. Only ProFound and PyBDSF make detections, and only in the $\mathcal{D}$-image. There is little difference in apparent flux density between the injected source detection at \texttt{match\_id} 6607 and the spurious adjacent detections at \texttt{match\_id}s 6608 and 6609.

A cursory scan suggests Caesar finds sources at image edges less often than other SFs, and Selavy tends not to identify them at all. Most likely, the island is there, but the fit has failed and no source component is recorded.\footnote{NB: Island output information is not completely supported in the current version of Hydra.}  As ProFound is capable of characterising irregularly shaped objects it tends to perform well for such truncated sources.

\begin{figure*}[hbt!]
\centering%
\includegraphics[width=\textwidth]{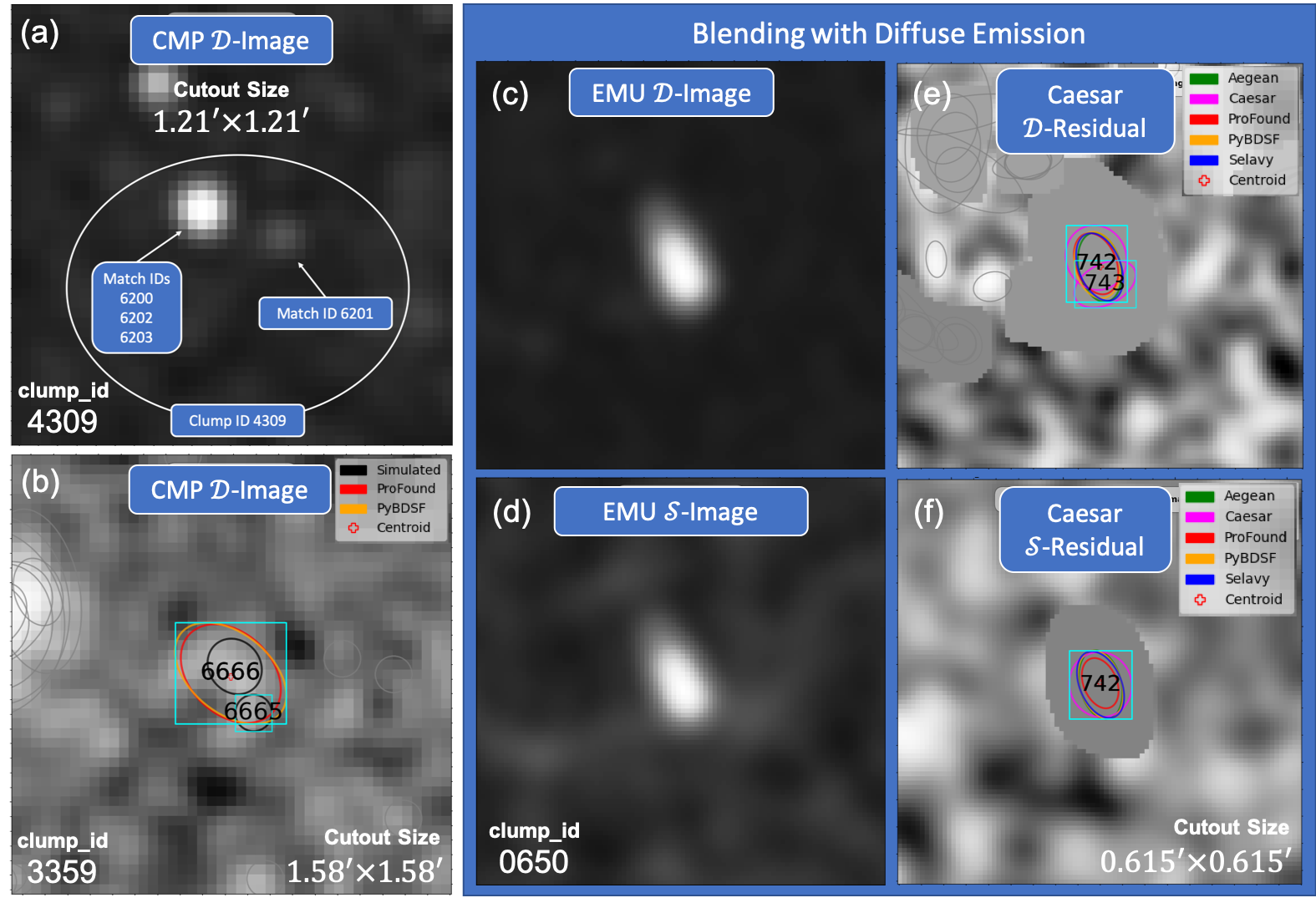}
\caption{\textbf{Blended Sources Infographic}: CMP (a--b) and real (EMU; c--f) $\mathcal{D}$-image examples of blended sources. In example (a) (from the $\mbox{S/N}\sim19.6\pm1.3$ bin of $\mathcal{C_D}$, Figure~\ref{fg:cds_rds_cmp_ext}a), only ProFound missed detection of the isolated source at \texttt{match\_id} 6201. The remaining injected sources overlap to make up a single unresolved compact object, which is identified as \texttt{match\_id} 6203 by all SFs. In example (b) (from the $\mbox{S/N}\sim0.318\pm0.036$ bin of $\mathcal{C_D}$, Figure~\ref{fg:cds_rds_cmp_ext}a), there are two injected sources at \texttt{match\_id}s 6665 ($\mbox{S/N}\sim0.070$) and 6666 ($\mbox{S/N}\sim0.336$). Only \texttt{match\_id} 6666 is detected by ProFound ($\mbox{S/N}\sim7.468$) and PyBDSF  ($\mbox{S/N}\sim12.446$). Examples (c) and (d) show a $\mathcal{D}$-image and $\mathcal{S}$-image cutouts, respectively, of a compact object with a diffuse tail. Only Caesar is able to resolve the $\mathcal{D}$-image (into two components; top), but not the $\mathcal{S}$-image (bottom) as the diffuse emission is washed out. Examples (e) and (f) are the corresponding a $\mathcal{D}$-residual and $\mathcal{S}$-residual image cutouts in the previous example, respectively, for Caesar. All SFs make $\mathcal{D}$ and $\mathcal{S}$ detections at \texttt{match\_id} 742. Caesar separately detects the bright peak at \texttt{match\_id} 743 in the $\mathcal{D}$-image, but with no corresponding $\mathcal{S}$ match. This contributes to a reduction in the inferred $\mathcal{C_{DS}}$ for Caesar at $\mbox{S/N}\sim3.4$ in Figure~\ref{fg:hydra_2x2_all_cr_plots}a.}
\label{fg:le_blender}
\end{figure*}

\subsubsection{Blended Sources}
For the simulated images there are example cases of unresolved compact sources: \textit{e.g.}, Figure~\ref{fg:le_blender}a. Here \texttt{match\_id}s 6200, 6202, and 6203, with injected flux densities of 0.0529, 0.4032, and 0.6928 mJy, respectively, appear as a single compact unresolved source. All SFs detect it, with measured flux densities of 1.1118, 1.1113, 1.2481, 1.1296, and 1.1230 mJy, for Aegean, Caesar, ProFound, PyBDSF, and Selavy, respectively. These values are consistent with the total injected flux $\sim1.1489\,$mJy. So the total flux is accounted for. Regardless, this causes a degradation in the $\mathcal{C_D}$; in particular, the $\mbox{S/N}\sim20$ bin of Figure~\ref{fg:cds_rds_cmp_ext}a.

Blending is also encountered with overlapping point and extended sources, with the former having slightly lower S/N. Figure~\ref{fg:le_blender}b shows an example of a very low S/N detection being influenced by an even fainter adjacent source.
Here, only ProFound and PyBDSF, which have the lowest detection thresholds (Table~\ref{tb:hydra_2x2_typhon_stats}), make detections, and only in the $\mathcal{D}$-image. Both SFs treat the two adjacent injected sources as a single source. This leads to a degradation in $\mathcal{C_D}$, but not $\mathcal{R_S}$ (Figure~\ref{fg:cds_rds_cmp_ext}a--b).

For simple cases of real sources with diffuse emission, such as a compact object with a diffuse tail, Aegean, ProFound, PyBDSF, and Selavy tend to characterise them with a single flux-weighted component position, whereas Caesar tends to resolve them in $\mathcal{D}$-images, but not $\mathcal{S}$-images where the diffuse emission is diminished: \textit{e.g.}, Figure~\ref{fg:le_blender}c--f. Here Caesar decomposes this object into two sources in the $\mathcal{D}$-image (e), but not in the $\mathcal{S}$-image (f). The remaining SFs detect this as a single source in both the $\mathcal{D}$ and $\mathcal{S}$ images. The diffuse emission is washed out in the $\mathcal{S}$-image, leading to a slight ($\sim3$\%) drop in the overall S/N.

\begin{figure*}[hbt!]
\centering%
\includegraphics[width=\textwidth]{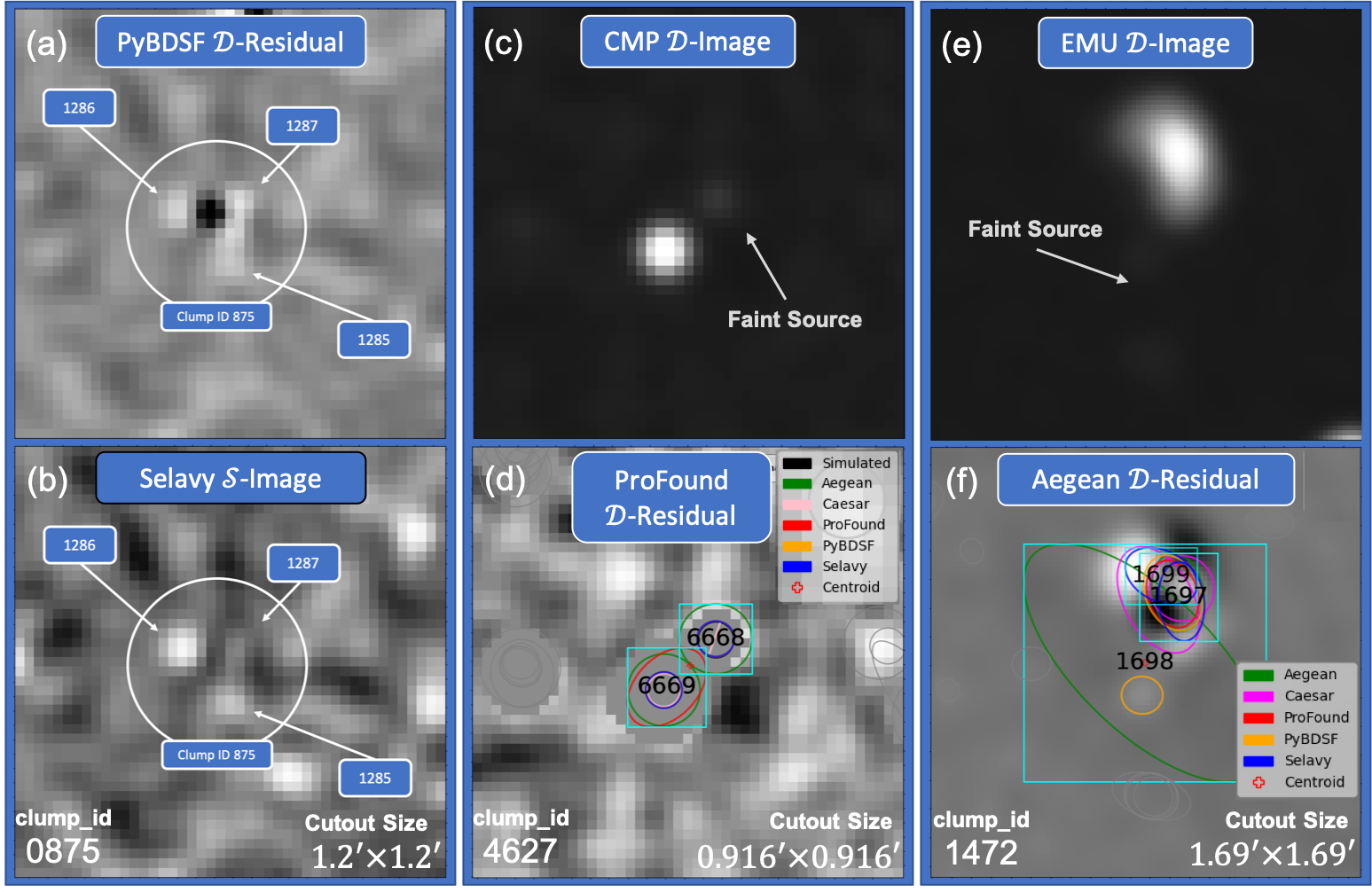}
\caption{\textbf{Deblending Issues Infographic}: CMP PyBDSF $\mathcal{D}$-residual-image (a) with Selavy $\mathcal{S}$-residual-image (b) (from \texttt{clump\_id} 875), CMP $\mathcal{D}$-image (c) with ProFound $\mathcal{D}$-residual-image (d) (from \texttt{clump\_id} 4672), and real (EMU) $\mathcal{D}$-image (e) with Aegean $\mathcal{D}$-residual-image (f)  (from \texttt{clump\_id} 875) cutout examples of deblending issues. In example (a), only PyBDSF makes a detection at \texttt{match\_id} 1287. In example (b), only Selavy makes a detection at \texttt{match\_id} 1287. In example (c--d), ProFound provides a single flux-weighted component that blends these two adjacent sources, and which is best matched to \texttt{match\_id} 6669. This leads to the result that \texttt{match\_id} 6668 is deemed undetected in the $\mathcal{C_D}$ statistics (Figure~\ref{fg:cds_rds_cmp_ext}a). In example (e--f), Aegean overestimates the extent of the vary faint (diffuse) source at \texttt{match\_id} 1698 (bottom image, middle): \textit{i.e.}, $(a,b,\theta)\sim(130^{\prime\prime},\,57.6^{\prime\prime},\,45.8^\circ)$, with major, minor, and  position-angle components, respectively. There is no corresponding $\mathcal{S}$ match. This leads to a degradation in the $\mathcal{C_{DS}}$ S/N $\sim$ 25 bin of Figure~\ref{fg:hydra_2x2_all_cr_plots}a.}
\label{fg:le_delending}
\end{figure*}

\subsubsection{Deblending Issues}
Deblending issues can occur in systems having a relatively low S/N neighbour to a brighter object, causing a positional fit which is offset from the true centre of the bright object: \textit{e.g.}, Figure~\ref{fg:le_delending}a. This leads to a degradation in $\mathcal{R_{D}}$ (and $\mathcal{R_{DS}}$). In such cases, PyBDSF tends to bias its position estimates towards flux-weighted centers. The other SFs tend to be less biased in such a fashion: \textit{e.g.}, Figure~\ref{fg:le_delending}b. ProFound tends to extend its component footprint to include faint sources: \textit{e.g.}, Figure~\ref{fg:le_delending}d. 

\begin{figure}[hbt!]
\centering%
\includegraphics[width=\columnwidth]{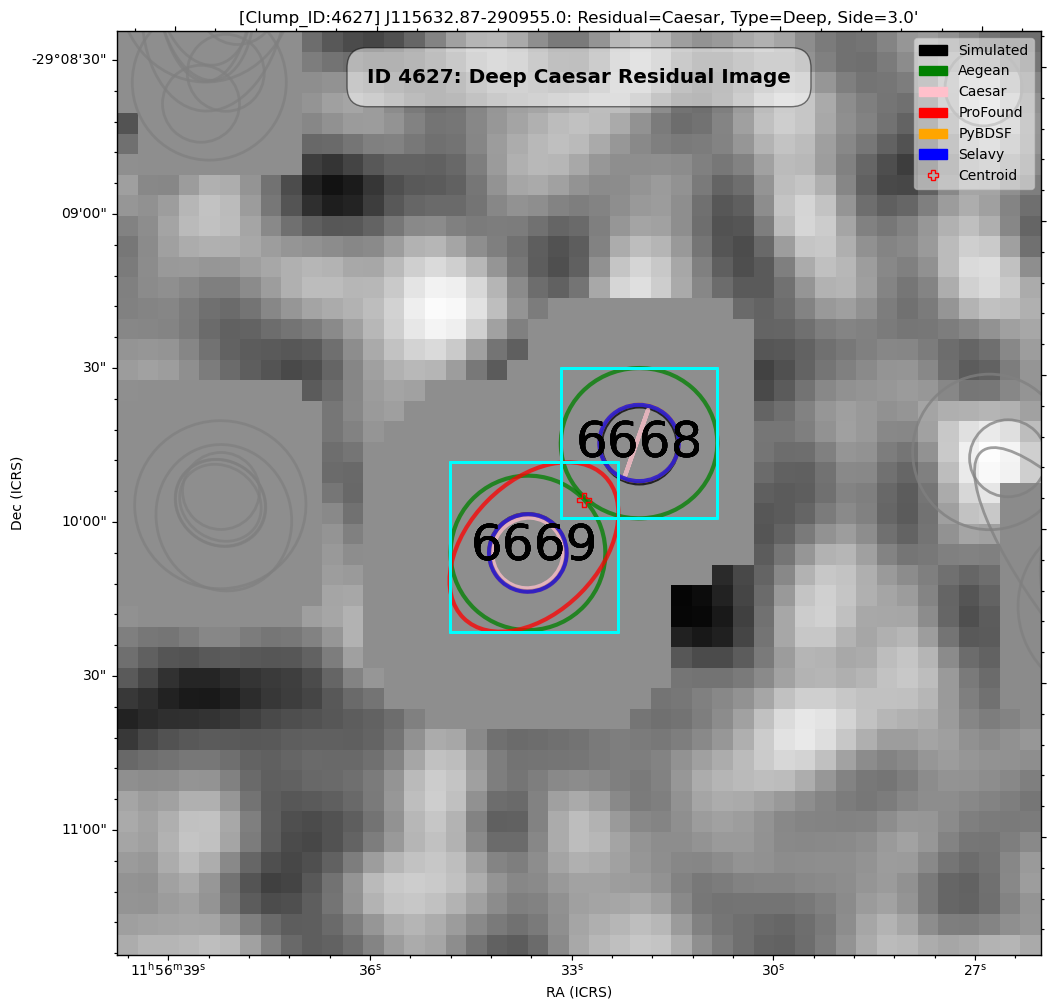}
\caption{Caesar \texttt{clump\_id} 4627 $\mathcal{D}$ $0.916^\prime\times0.916^\prime$ residual-image cutout for simulate-compact sources. Caesar underestimates the flux density for \texttt{match\_id} 6668 ($\sim0.664\,\mu$Jy, compared to $\sim291\,\mu$Jy for the injected source), resulting in an artifact in the calculated reliability, placing this source at an artificially low S/N.  This information was extracted from the $\mathcal{R_D}$ $\mbox{S/N}\sim0.0331\pm0.0036$ bin  (Figure~\ref{fg:cds_rds_cmp_ext}b).}
\label{fg:hydra_2x2_rd_cid_4627}
\end{figure}

Figure~\ref{fg:hydra_2x2_rd_cid_4627} shows an example of an injected source with a flux density of $\sim291\,\mu$Jy which is erroneously measured by Caesar as $\sim0.664\,\mu$Jy. This is likely to be a consequence of a poor deblending of the two adjacent sources, initially identified by Caesar as a single island. The position is accurate, so it is counted as a correct match, but because $\mathcal{R_D}$ (Figure~\ref{fg:cds_rds_cmp_ext}b) is calculated using the measured S/N, this source contributes to $\mathcal{R_D}$ at the unreasonably low value of $\mbox{S/N}\sim0.033$. For this source, Aegean, PyBDSF, and Selavy correctly estimate the flux density as $281.9^{+4.2}_{-8.4}\,\mu$Jy, and it contributes to $\mathcal{R_D}$ for those SFs at $\mbox{S/N}\sim12$. It is worth noting here that requiring flux-density matching in the reliability metric calculation would lead (for Caesar) to this source being deemed a false detection, although it has been detected at the correct location. ProFound does not separately detect \texttt{match\_id} 6668 (Figure~\ref{fg:le_delending}d) as it has blended it with \texttt{match\_id} 6669. Comparing Figure~\ref{fg:hydra_2x2_rd_cid_4627} and~\ref{fg:le_delending}d we can see the similar islands detected by ProFound and Caesar, although ProFound's boundary is tighter, which provides a clear indication of why only a single match was identified.

\begin{figure}[hbt!]
\centering%
\includegraphics[width=\columnwidth]{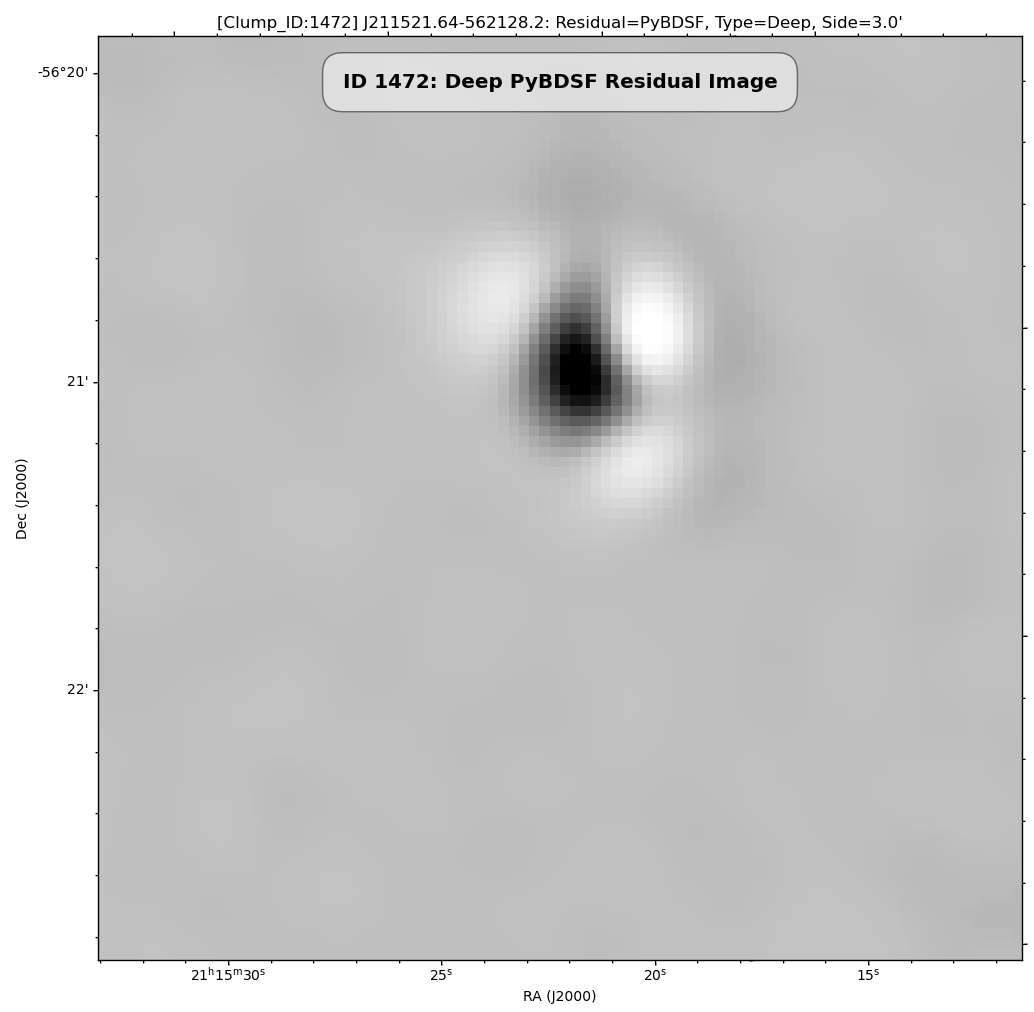}
\includegraphics[width=\columnwidth]{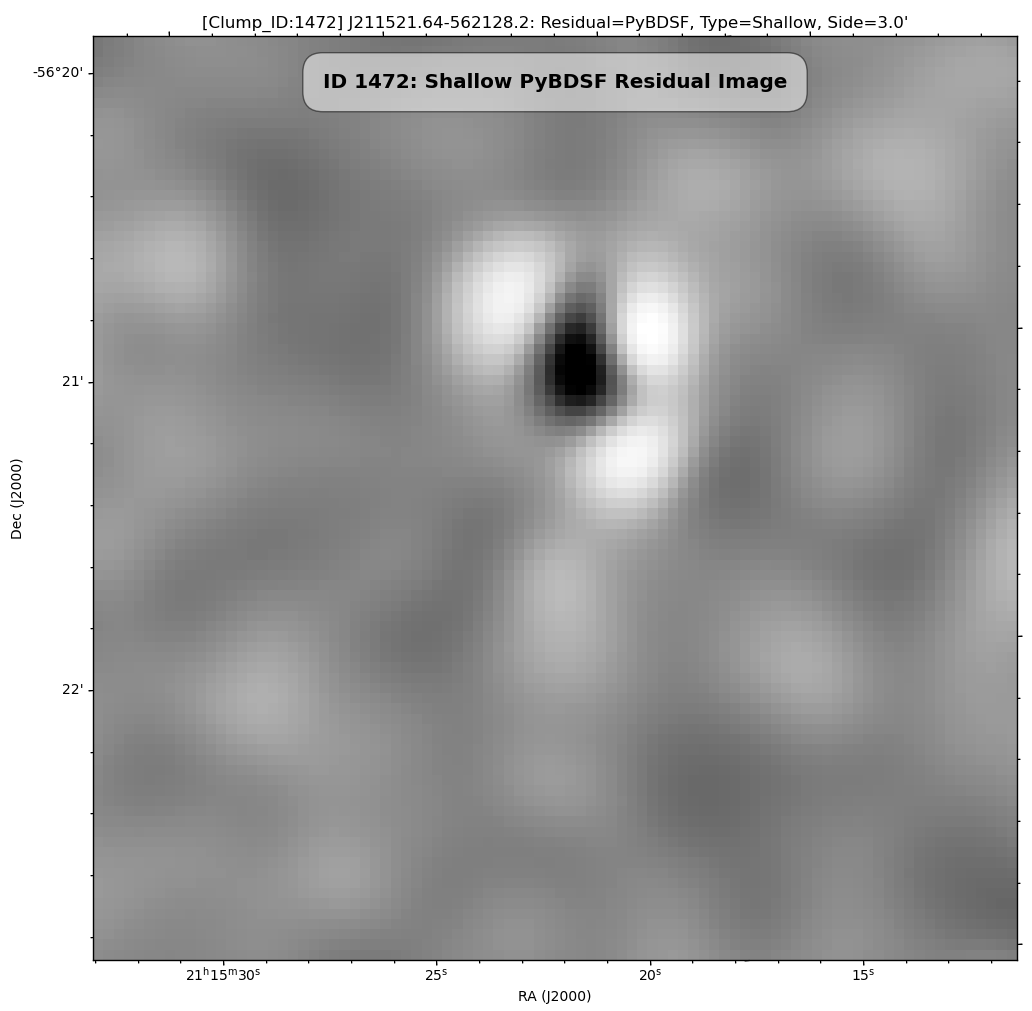}
\caption{PyBDSF \texttt{clump\_id} 1472 $\mathcal{D}$ (top) and $\mathcal{S}$ (bottom) $1.69^\prime\times1.69^\prime$ residual-image cutouts for real (EMU) sources.  PyBDSF fits the core of the source shown in Figure~\ref{fg:le_delending}a, while leaving out the diffuse emission (\textit{i.e.}, it over-subtracts), in both $\mathcal{D}$ and $\mathcal{S}$ images. This contributes to the $\mathcal{C_{DS}}$ S/N $\sim$ 152 bin (Figure~\ref{fg:cds_rds_cmp_ext}a).}
\label{fg:hydra_2x2_emu_pydbsf_d_img_cid_1472}
\end{figure}

Figure~\ref{fg:le_delending}e--f shows an example of a compact system with two roughly opposing diffuse tails, along with a faint diffuse neighbour to the south. All SFs detect the core at \texttt{match\_id} 1697, in both the $\mathcal{D}$ and $\mathcal{S}$ images: \textit{e.g.}, Aegean in Figure~\ref{fg:le_delending}f, and PyBDSF in Figure~\ref{fg:hydra_2x2_emu_pydbsf_d_img_cid_1472}. Caesar and Selavy further detect aspects of the diffuse emission, with Caesar fitting it as an extended halo (with its fit slightly biased towards the faint neighbour at \texttt{match\_id} 1698) and Selavy as a tail (\texttt{match\_id} 1699), but only in the $\mathcal{D}$-image. As for the faint neighbouring source, only Aegean and PyBDSF detect it (\texttt{match\_id} 1698), and only in the $\mathcal{D}$-image (Figure~\ref{fg:le_delending}f). Aegean overestimates its extent (leading to an overestimated $\mbox{S/N} \sim 25$), likely due to inclusion of emission from the diffuse tails in its fit. As there are no corresponding $\mathcal{S}$ detections for \texttt{match\_id}s 1698 and 1699, they contribute to a degradation in $\mathcal{C_{DS}}$. ProFound, on the other hand, includes both the bright source and the faint southern neighbour together in the $\mathcal{D}$-image (with its fit slightly biased towards the faint source, S/N $\sim$ 154) but detects only the bright source core in the $\mathcal{S}$-image (S/N $\sim$ 145). This is a further example of the impact on $\mathcal{C_{DS}}$ of extended or faint diffuse emission, characterised differently by different SFs in the $\mathcal{D}$-image but absent in the $\mathcal{S}$-image, resulting in mismatches between the reported sets of sources.

\subsubsection{Noise Spike Detection}
\begin{figure*}[hbt!]
\centering%
\includegraphics[width=\textwidth]{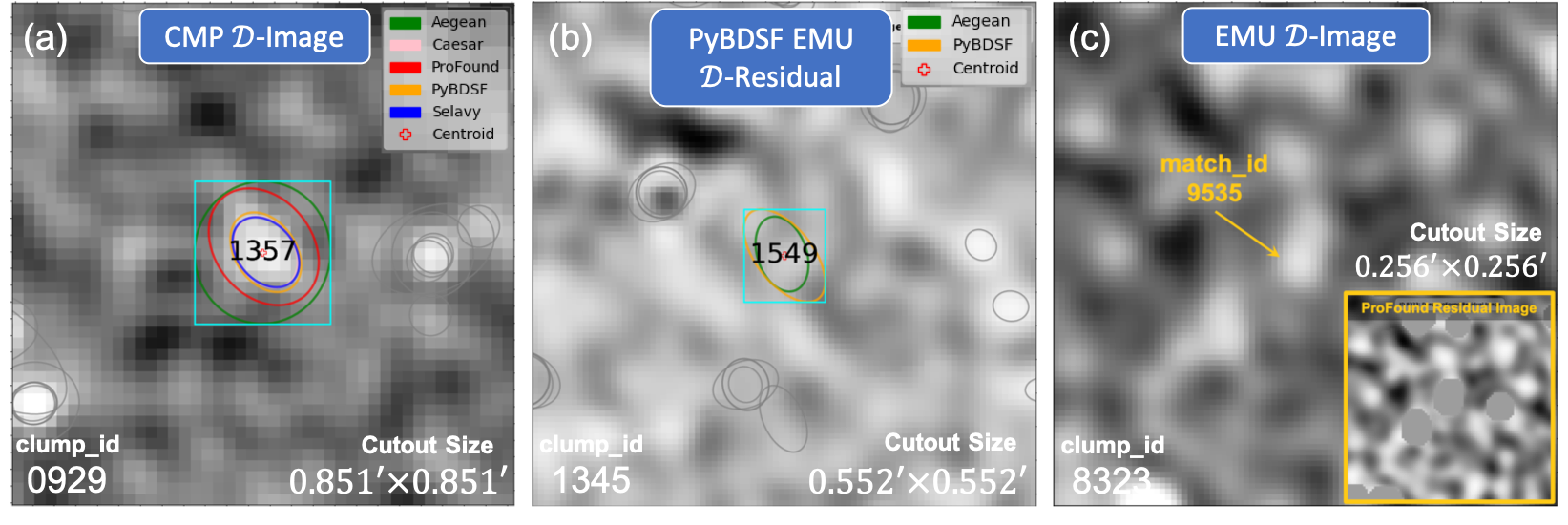}
\caption{\textbf{Noise Spike Detection Infographic}: CMP $\mathcal{D}$-image (a), EMU PyBDSF $\mathcal{D}$-residual-image (b), and EMU $\mathcal{D}$-image (/w ProFound residual-image inset) (c) cutout examples of noise spike detection. In example (a) (from the $\mbox{S/N}\sim12.4\pm1.4$ bin of $\mathcal{R_D}$, Figure~\ref{fg:cds_rds_cmp_ext}b), the detections within this clump are anomalous, as there is no injected source. In example (b), Aegean and PyBDSF detect a $\mathcal{D}$ source (\texttt{match\_id} 1549) here, although visually this object is consistent with a noise spike. With no corresponding $\mathcal{S}$ detection, such results contribute to the inferred  $\mathcal{C_D}$ at low S/N (Figure~\ref{fg:hydra_2x2_all_cr_plots}a). Example (c): At first glance, \texttt{match\_id} 9535 appears to be part of a faint ring structure, something for which ProFound is uniquely suited. The size of the emission is on the order of common EMU's beam size ($18^{\prime\prime}$), making it consistent with either a noise spike or a faint compact source.}
\label{fg:le_spike}
\end{figure*}

All of the SFs detect noise spikes at low S/N that are impossible to distinguish from true sources, \textit{e.g.}, Figure~\ref{fg:le_spike}. Such cases must be handled statistically, with a knowledge of the reliability of a given SF as a function of S/N, in order to treat the properties of such catalogues robustly. As image (and survey) sizes become larger
, such false sources will be detected in higher numbers, as the number simply scales with the number of resolution elements sampled \citep{hopkins_2002}.

\subsubsection{Bright, High S/N Sources Missed}
\begin{figure}[hbt!]
\centering%
\includegraphics[width=\columnwidth]{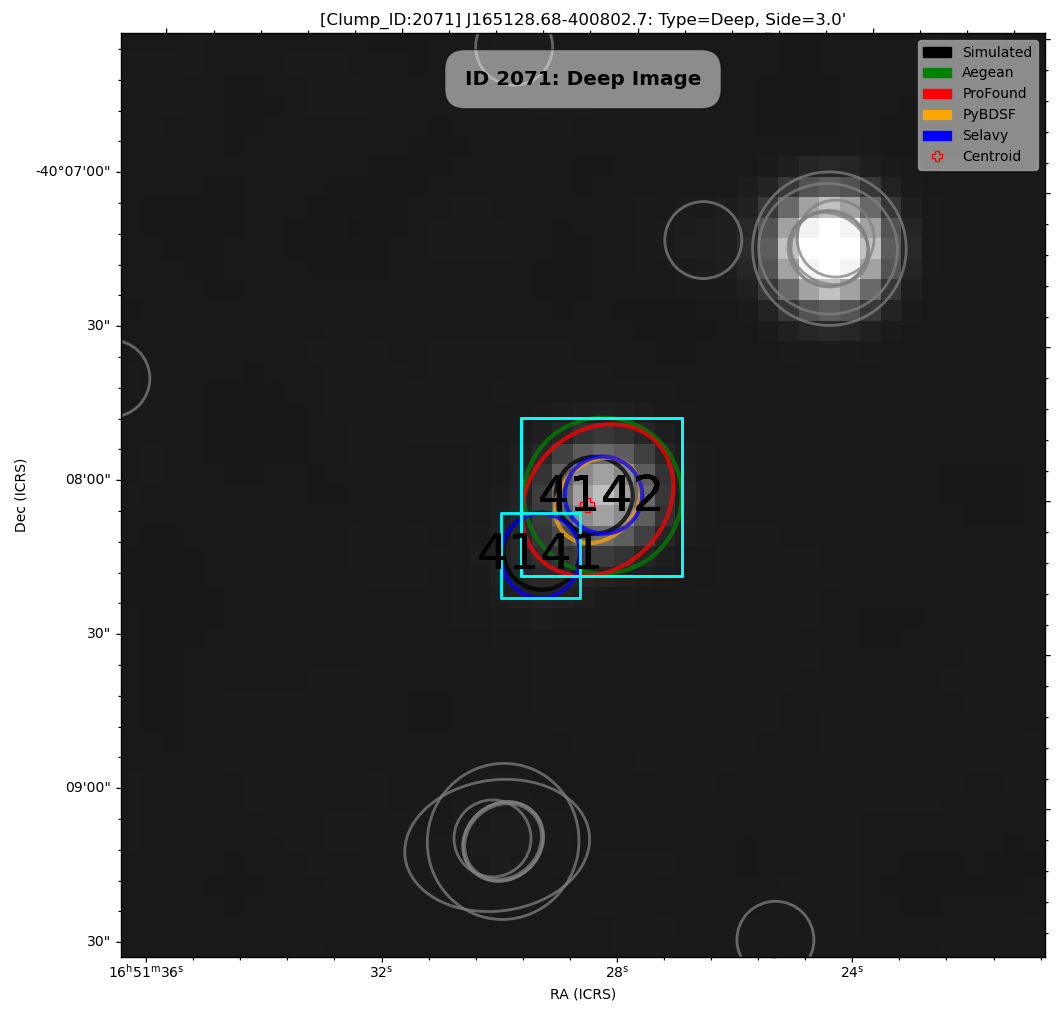}
\caption{$3^\prime\times3^\prime$ image cutout of \texttt{clump\_id} 2071 of the EXT $\mathcal{D}$-image. All SFs detected the injected source ($\mbox{S/N}\sim226$) at \texttt{match\_id} 4142, except for Caesar. Only Selavy detects the adjacent fainter source, \texttt{match\_id} 4141, where $(\mbox{S/N})_{injected}\sim19$. This information was extracted from the $\mathcal{C_D}$ $\mbox{S/N}\sim220\pm25$ bin of Figure~\ref{fg:cds_rds_cmp_ext}g.}
\label{fg:hydra_ext2x2_cd_cid_2071_caesar}
\end{figure}

Figure~\ref{fg:hydra_ext2x2_cd_cid_2071_caesar} shows an EXT $\mathcal{D}$-image example of a bright source (\texttt{clump\_id} 4142 /w $\mbox{S/N}\sim226$) near a fainter source to the south (\texttt{clump\_id} 4141 /w $\mbox{S/N}\sim19$), representing a high $\mathcal{C_D}$ $\mbox{S/N}\sim220$ bin (Figure~\ref{fg:cds_rds_cmp_ext}g) failure mode. All SFs detected the bright source, except for Caesar. Only Selavy detects the fainter source.

Even at the brightest end, sources can be missed by SFs. An example is Selavy failing to detect one of two injected sources in the $C_D$ $\mbox{S/N}\sim4130\pm460$ bin (Figure~\ref{fg:cds_rds_cmp_ext}g): \textit{i.e.}, it fails to detect the first of $\mbox{(S/N})_{\mathcal{D}}^{\mbox{\tiny Injected}}$ $\sim$ 4320 and 4540 at \texttt{clump\_id}s 2363 and 270, respectively. Oddly, Selavy does detect this source in the $\mathcal{S}$-image (matching  $\mbox{(S/N})_{\mathcal{S}}^{\mbox{\tiny Injected}}\sim1365$ with $\sim1364$), suggesting the failure in the $\mathcal{D}$-image may be related to its background estimation in this case.

\subsubsection{Simulated Anomalies}
\begin{figure}[hbt!]
\centering%
\includegraphics[width=\columnwidth]{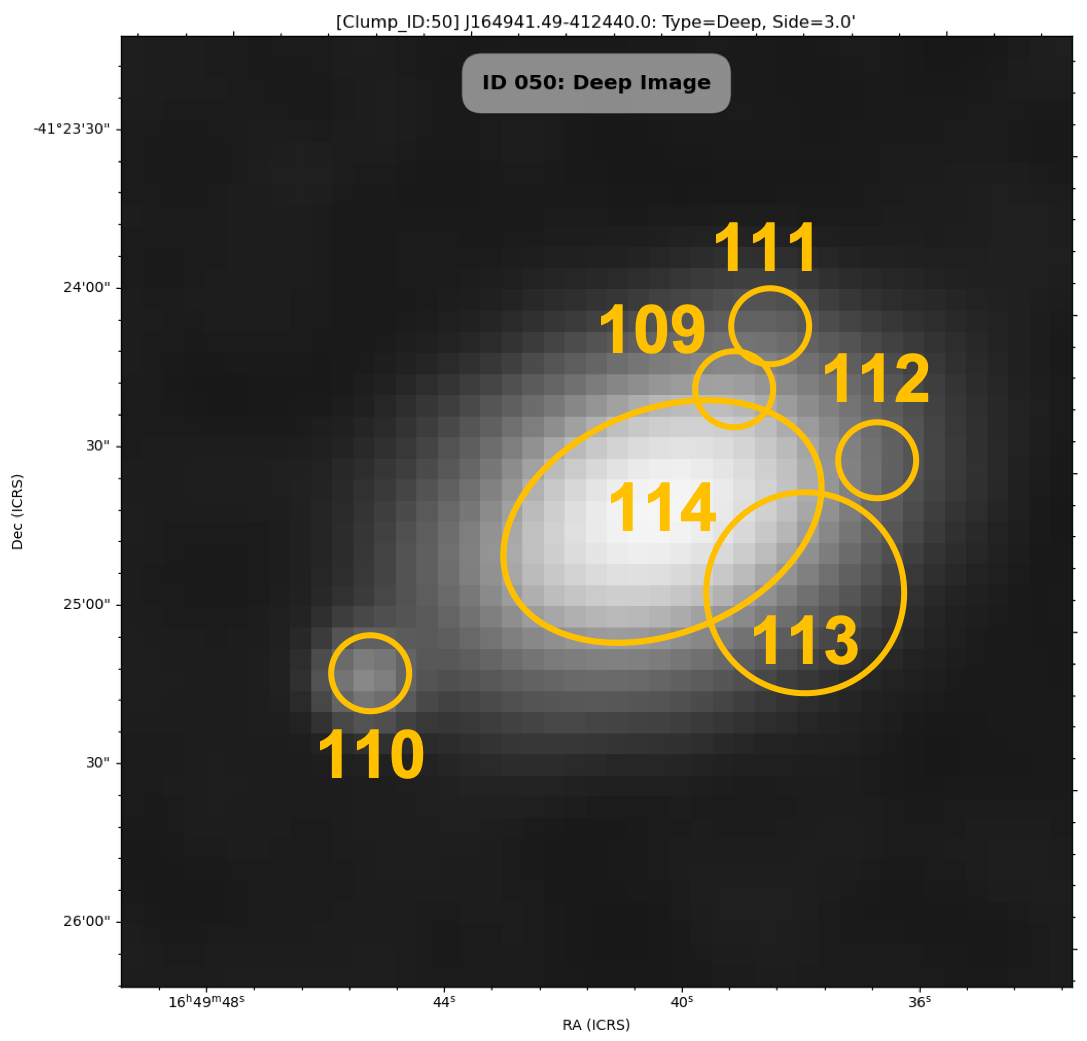}
\caption{$3^\prime\times3^\prime$ cutout of \texttt{clump\_id} 50 of the EXT $\mathcal{D}$-image. The orange ellipses indicate injected sources, summarised in Table~\ref{tb:hydra_ext2x2_cd_cid_50_matrix}. The associated SF detections are shown in Figure~\ref{fg:hydra_ext2x2_cd_cid_50_matrix}. These sources span $0.28<\mbox{S/N}<80$ in $\mathcal{C_{D}}$ (Figure~\ref{fg:cds_rds_cmp_ext}h).
}
\label{fg:hydra_ext2x2_cd_cid_50}
\end{figure}

\begin{figure*}[hbt!]
\centering%
\includegraphics[width=\textwidth]{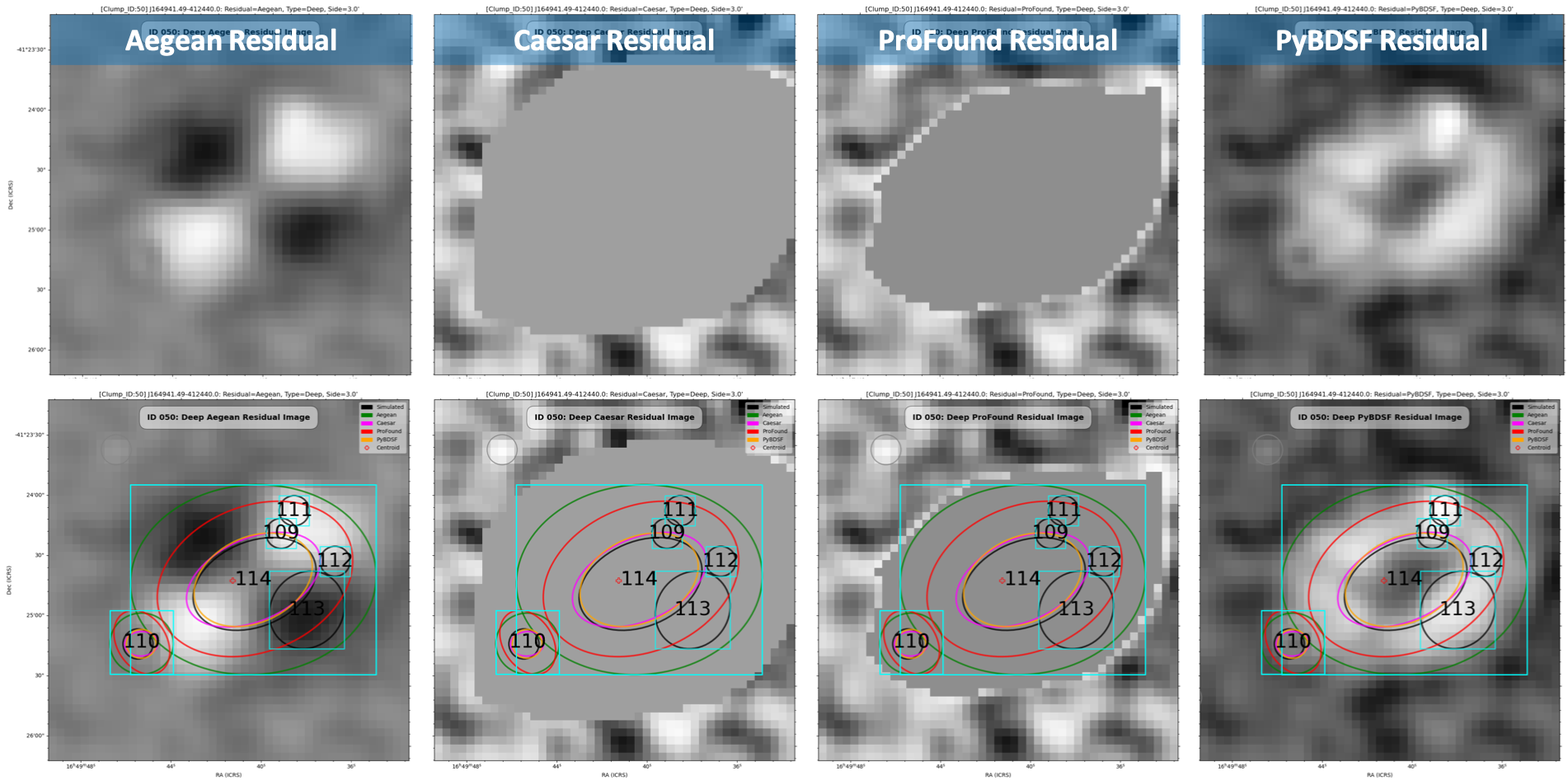}
\caption{$2.51^\prime\times2.51^\prime$ cutout, without (top) and with (bottom) annotations, of \texttt{clump\_id} 50 of the EXT $\mathcal{D}$-image. The corresponding injected sources are shown in Figure~\ref{fg:hydra_ext2x2_cd_cid_50}. Aegean (green), Caesar (cyan), ProFound (red), and PyBDSF (gold) make $\mathcal{D}$ and $\mathcal{S}$ detections at \texttt{match\_id}s 110 and 114 only, except for Selavy which only finds them in the $\mathcal{S}$-image (not shown).}
\label{fg:hydra_ext2x2_cd_cid_50_matrix}
\end{figure*}

Figure~\ref{fg:hydra_ext2x2_cd_cid_50} shows an example of injected sources spanning $0.28<\mbox{S/N}<80$ in \texttt{clump\_id} 50, with details in Table~\ref{tb:hydra_ext2x2_cd_cid_50_matrix}, for our EXT $\mathcal{D}$-image. This is an informative clump, as it encapsulates a moderately rich environment in which to explore detection anomalies. The $\mathcal{D}$-residual-image cutouts from the four SFs that make detections are shown in Figure~\ref{fg:hydra_ext2x2_cd_cid_50_matrix}. All SFs make detections in the $\mathcal{D}$ and $\mathcal{S}$ images, except for Selavy which only finds sources in the $\mathcal{S}$-image. This may arise from Selavy's underlying Duchamp-style thresholding \citep{whiting_2012b}.

\begin{table}[hbt!]
\caption{Summary of injected EXT-sources in Figure~\ref{fg:hydra_ext2x2_cd_cid_50}. The Detected column indicates if at least one SF has detected an injected source.}
\label{tb:hydra_ext2x2_cd_cid_50_matrix}
\centering
{\small
\begin{tabular}{@{\;}c@{\;\,}c@{\;\,}c@{\;\,}c@{\;\,}c@{\;}}
\hline\hline
$\mathcal{C_{D}}$ S/N Bins & \multicolumn{1}{c}{Injected} & RMS & \texttt{match\_id} & Detected \\
(Fig.~\ref{fg:cds_rds_cmp_ext}g) & S/N & (mJy/beam) & (Fig.~\ref{fg:hydra_ext2x2_cd_cid_50}) & (Fig.~\ref{fg:hydra_ext2x2_cd_cid_50_matrix}) \\
\hline
$0.318\pm0.036$ & 0.287 & 0.130 & 113 & False \\
$0.784\pm0.088$ & 0.810 & 0.129 & 112 & False \\
$2.42\pm0.27$   & 2.30  & 0.115 & 109 & False \\
$3.80\pm0.43$   & 4.01  & 0.114 & 111 & False \\
$36.2\pm4.0$    & 37.4  & 0.0949 & 110 & True \\
$71.2\pm8.0$

& 69.2  & 0.114 & 114 & True \\
\hline\hline
\end{tabular}}
\end{table}

For simulated images, based on the flux ratios in Figure~\ref{fg:hydra_2x2_deep_flux_ratios}, Selavy appears to produce a catalogue with an effective detection threshold of $5\sigma$, regardless of its RMS parameter setting.  Recalling that the EXT image contains extended-component elements with a maximum peak flux density of $1\,$mJy, as opposed to $1\,$Jy for its compact-component elements  (\S~\ref{sc:image_data}), this suggests the issue may be related to the extended low surface brightness emission around such objects. There is a roughly 10\% degradation in $\mathcal{C_{D}}$, comparing the results for CMP sources alone (Figure~\ref{fg:cds_rds_cmp_ext}a) to the EXT sources (Figure~\ref{fg:cds_rds_cmp_ext}g). While this degradation approximates the fraction of injected EXT sources, not all the missed sources are extended. It does seem likely, though, that it is the EXT sources that contribute disproportionately to the shortfall in $\mathcal{C_{D}}$.

Comparing $\mathcal{C_{S}}$ (Figure~\ref{fg:cds_rds_cmp_ext}i) and $\mathcal{C_{D}}$ (Figure~\ref{fg:cds_rds_cmp_ext}g) for EXT sources, the distributions are consistent with the addition of the 5$\sigma$ noise level (shifting the S/N axis by 5 units), for most SFs ({\em e.g.}, the value of $\mathcal{C_S}$ at $\mbox{S/N}=10$ is similar to the value of $\mathcal{C_D}$ at $\mbox{S/N}=5$). Selavy shows a different behaviour in that its performance in the $\mathcal{S}$-image appears to be better than in the $\mathcal{D}$-image. At the lowest S/N end this is related to faint low surface brightness objects lying below the detection threshold in the $\mathcal{S}$-image, and not contributing to the $\mathcal{C_S}$ and $\mathcal{C_D}$ distributions. At moderate and high S/N, though, the reason for the difference is less clear. It may be that faint extended emission around bright objects misleads Selavy in the $\mathcal{D}$-image, by being pushed below the noise level in the $\mathcal{S}$-image it makes the bright central region of those sources more easily detectable.

It is clear from the details in Table~\ref{tb:hydra_ext2x2_cd_cid_50_matrix} that the sources not explicitly detected by any SF are at low S/N. Their detection is also confounded by the overlap with the brightest source (\texttt{match\_id} 114). Aegean and ProFound have handled this by extending the size of the detection associated with \texttt{match\_id} 114, influenced by the additional low S/N emission. This leads to over-estimation by Aegean of the source flux density and mischaracterisation of its orientation (evidenced by the oversubtraction in its residual). ProFound simply associates all the flux with the single object, which leads to a flux density overestimate compared to the input flux density of \texttt{match\_id} 114. Caesar searches for blobs nested inside large islands, allowing a more accurate characterisation of such overlapping sources. These residual statistics are demonstrated in Table~\ref{tb:hydra_2x2_typhon_stats}. Here the residual statistic values are significantly lower for ProFound and Caesar than the other SFs.

\begin{figure*}[hbt!]
\centering%
\includegraphics[width=\textwidth]{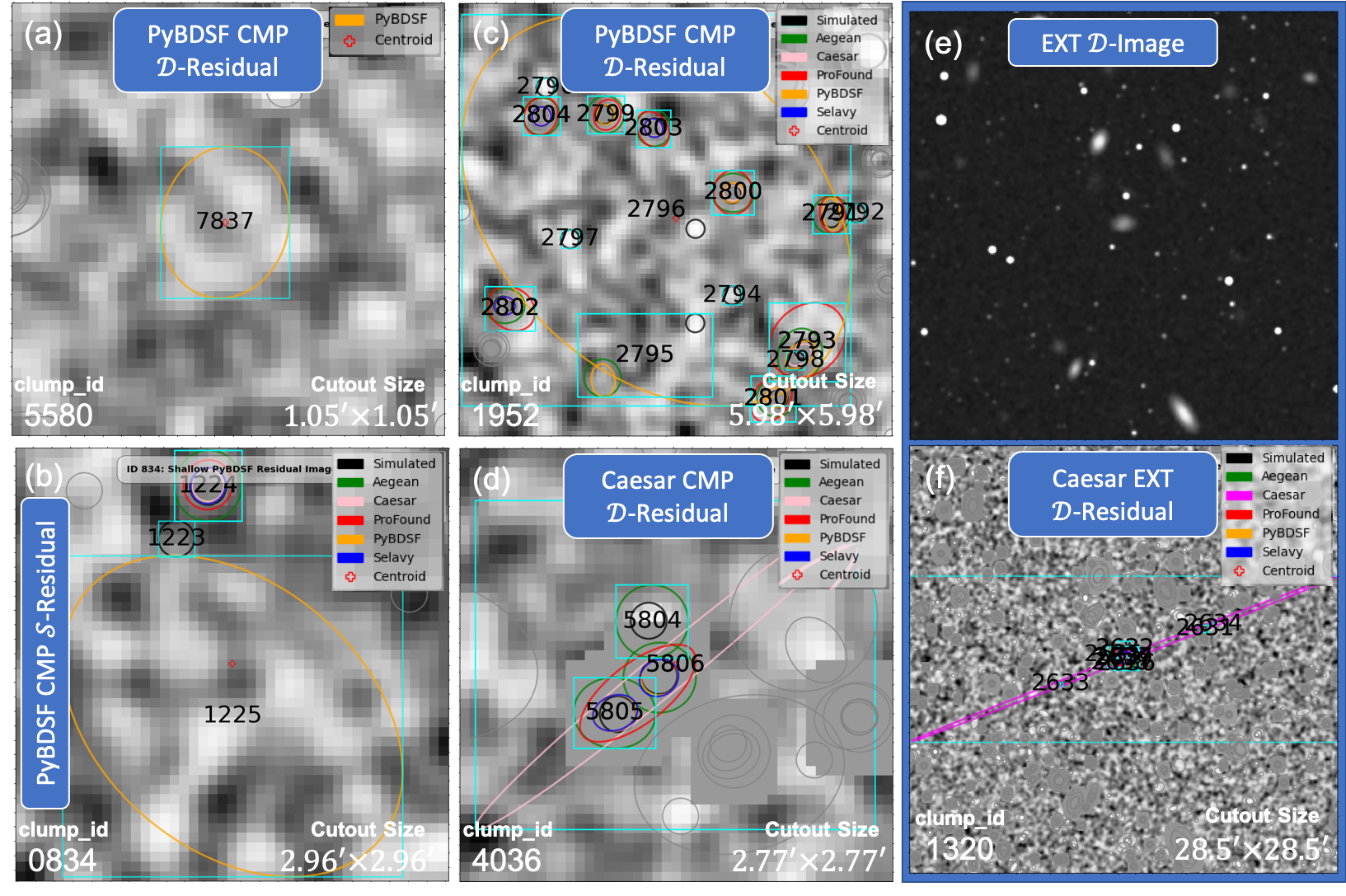}
\caption{\textbf{Oversized Components Infographic}: CMP PyBDSF $\mathcal{D}$ ((a) and (c)) and $\mathcal{S}$ (b) residual-image, Caesar $\mathcal{D}$ (d) residual-image, and EXT $\mathcal{D}$-image (e) and Caesar $\mathcal{D}$-residual-image (f) cutout examples of oversized components. Example (a) (from the $\mbox{S/N}\sim12.4\pm1.4$ bin of $\mathcal{R_D}$, Figure~\ref{fg:cds_rds_cmp_ext}b) is a spurious detection as there is no injected source. In example (b) (from the $\mbox{S/N}\sim12.4\pm1.4$ bin of $\mathcal{R_D}$, Figure~\ref{fg:cds_rds_cmp_ext}b), the detection by PyBDSF at \texttt{match\_id} 1225 is spurious, as there is no injected source at this location. In example (c), the large-footprint detection by PyBDSF at \texttt{match\_id} 2796 demonstrates one of its most frequent failure modes. Its flux density is 1.84$\,$mJy, as compared to 0.0798$\,$mJy for the injected source. In example (d) (from the $\mbox{S/N}\sim20.7\pm1.4$ bin of $\mathcal{C_{DS}}$, Figure~\ref{fg:cds_rds_cmp_ext}a), Caesar's Gaussian fit at \texttt{match\_id} 5806 extends well beyond its island. Its flux density estimate is 2.18$\,$mJy compared to 0.14$\,$mJy for the injected source. (For \texttt{clump\_id}s 5804 and 5805 the injected flux densities are 0.07$\,$mJy and 0.12$\,$mJy, respectively.) In examples (e--f) (from the $\mbox{S/N}\sim259\pm39$ bin of $\mathcal{C_S}$, Figure~\ref{fg:cds_rds_cmp_ext}a), Caesar overestimates the flux density at \texttt{match\_id} 2639, with $18\,$mJy, compared to $0.074\,$mJy for the injected source. Its respective semi-major and semi-minor axes are $920^{\prime\prime}$ and $11^{\prime\prime}$, compared to $20^{\prime\prime}$ and $10^{\prime\prime}$ for the injected source.} 
\label{fg:le_oversized}
\end{figure*}

\subsubsection{Oversized Components}
For CMP sources, PyBDSF sees the largest number of false detections within the $\mathcal{R_D}$ $\mbox{S/N}\sim12$ bin of Figure~\ref{fg:cds_rds_cmp_ext}b (a false/true ratio of 86/522). Figure~\ref{fg:le_oversized}a shows an example of one of these, illustrating a major failure mode for PyBDSF. It tends to overestimate source sizes due to inclusion of nearby noise spikes. At low or modest S/N even small noise fluctuations can lead it to expand its island in fitting to the source. Figure~\ref{fg:le_oversized}b shows another example, here with an excessively large size ($\sim67^{\prime\prime}$), associated with the dip in the $\mathcal{R_S}$ $\mbox{S/N}\sim12$ bin of Figure~\ref{fg:cds_rds_cmp_ext}d.  Figure~\ref{fg:le_oversized}c shows an example where a real injected source is the centre of a 10-component PyBDSF clump, spanning $\sim 156^{\prime\prime}$.

Turning to $\mathcal{R_{D}}$ (Figure~\ref{fg:cds_rds_cmp_ext}b) and $\mathcal{R_{S}}$ (Figure~\ref{fg:cds_rds_cmp_ext}d), at the lowest S/N levels ($\mbox{S/N}\lesssim 3$) Caesar appears to be reporting large numbers of false sources, leading to poor reliability. All of these artifacts are of the same nature as shown in Figure~\ref{fg:le_oversized}d. They spatially coincide with the injected sources, but overestimate the flux density due to fitting an extreme ellipse with artificially large major axis. A catastrophic example of this effect is found in the $\mathcal{C_S}$ $\mbox{S/N}\sim 260$ bin of Figure~\ref{fg:cds_rds_cmp_ext}g for EXT sources, shown in Figure~\ref{fg:le_oversized} (e--f).

\subsubsection{Oddities}
There are some oddities that occur less frequently, such as Caesar underestimating flux (Figure~\ref{fg:le_delending}d) or Selavy detecting bright sources in the $\mathcal{S}$, but not $\mathcal{D}$ (Figure~\ref{fg:hydra_ext2x2_cd_cid_50_matrix}), images, where all the other SFs succeed. In the former case, it is a deblending issue with mixtures of low and high S/N sources in close proximity. The latter case is not easily explained, but may be related to poor background estimation.

\subsection{Diffuse Emission Case Study}
\label{sc:le_bent_tail}
\begin{figure}[hbt!]
\centering%
\includegraphics[width=\columnwidth]{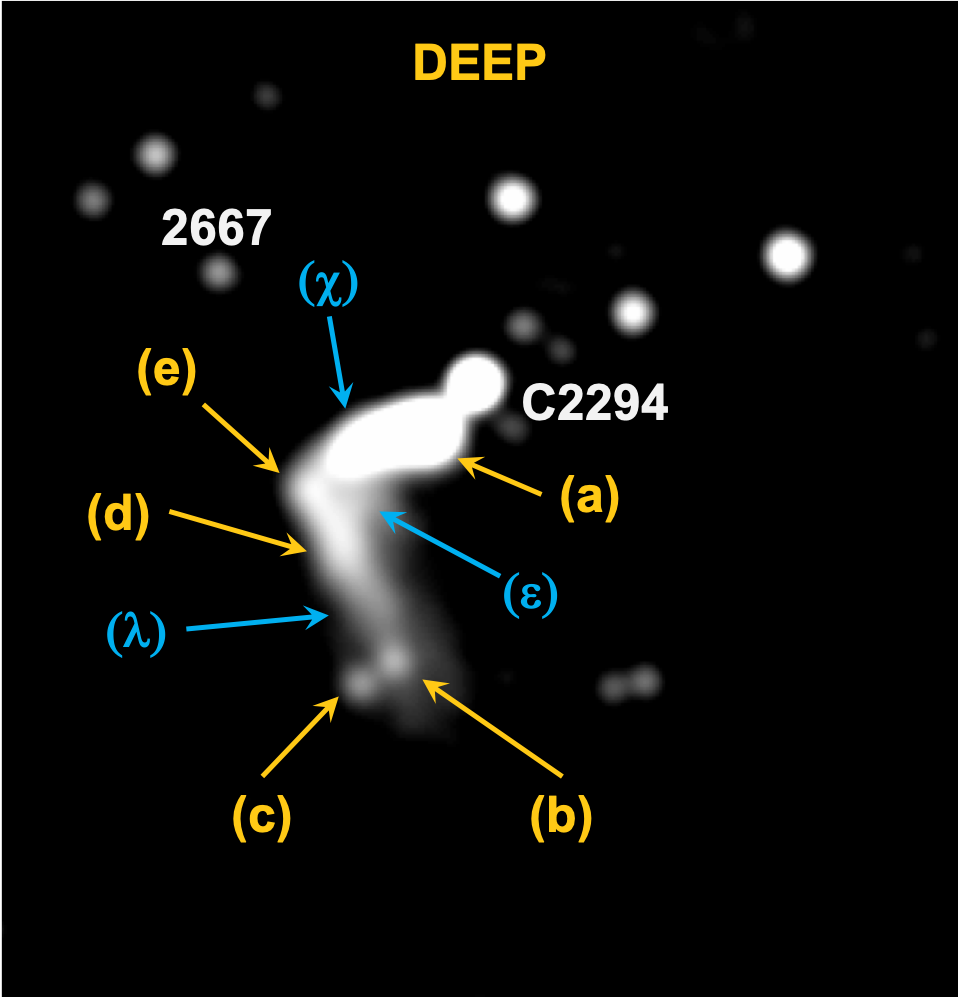}
\includegraphics[width=\columnwidth]{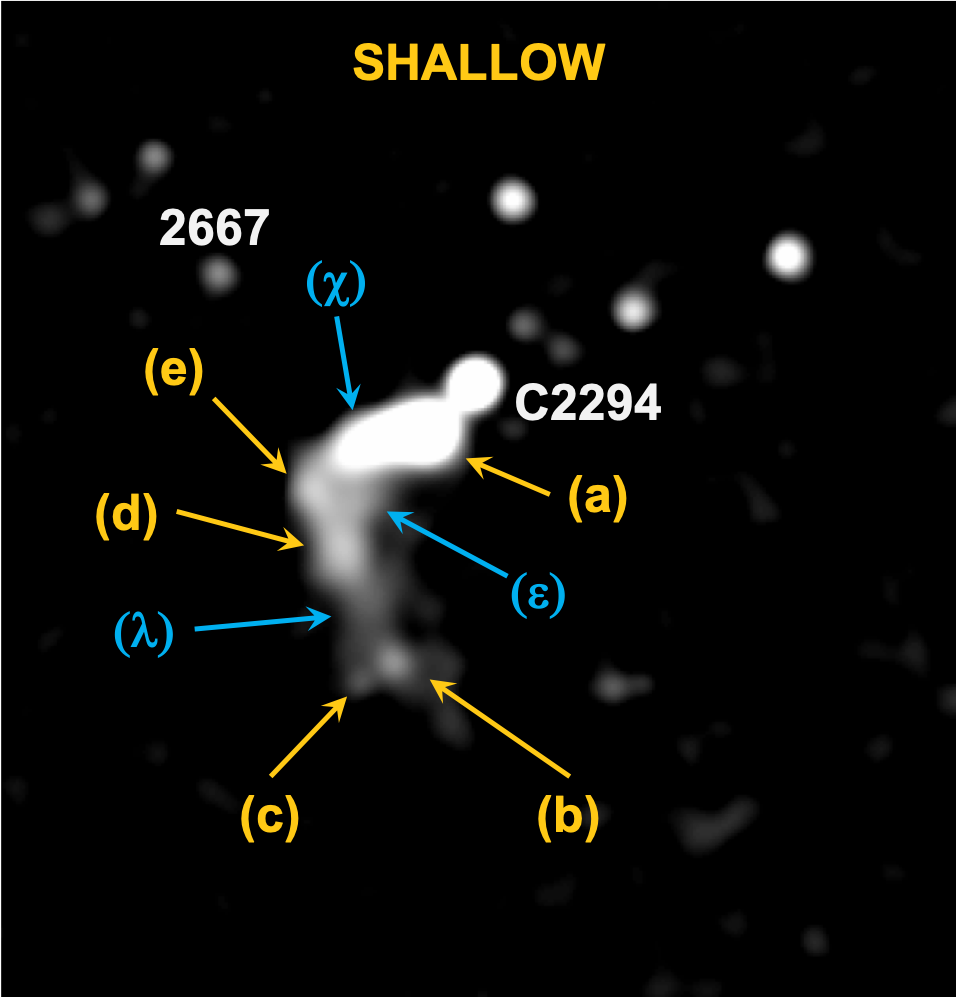}
\caption{$\mathcal{D}$ (top) and $\mathcal{S}$ (bottom) real image (EMU) cutouts of \texttt{clump\_id} 2293 (see also Figure~\ref{fg:emu_diffuse_emission} and Table~\ref{tb:emu_2x2_clump_id_2293}). The elements of the main clump (referred to as B2293) are labelled (a) through (e) for bright compact emission, and ($\chi$), ($\epsilon$), and ($\lambda$) for diffuse emission. Also shown is \texttt{match\_id} 2667, which is identified as part of this clump, although not physically associated. While highly likely to be related to the emission from B2293, the clump labelled C2294 is separated in our analysis. Its compact nature leads the SFs to characterise it with ellipses that do not overlap with any ellipse in B2293, resulting in these being labelled as independent clumps.}
\label{fg:emu_ds_clump_2293_image_cutout}
\end{figure}

We now consider an example of a source with a combination of complex structures. The system is introduced in Figure~\ref{fg:emu_ds_clump_2293_image_cutout}, with bright compact features labelled (a) through (e), and diffuse extended emission as ($\chi$), ($\epsilon$), and ($\lambda$). We distinguish the main complex, broadly referred to herein as a bent-tail source, although the details of its nature are likely much more complex, from the bright adjacent component. This main complex is given the label B2293 (not marked in the figure, with the prefix ``B'' for ``bent-tail''). The bright adjacent component (labelled C2294, with the prefix ``C'' for ``compact'') is identified here as a separate clump. While the emission in C2294 seems highly likely to be related to the emission from B2293, our approach to associating detections into clumps relies on their component footprints touching or overlapping. The compact nature of C2294 leads the SFs to characterise it with ellipses that do not overlap with any ellipse in B2293, and these are consequently marked as independent clumps. The source at \texttt{match\_id} 2667, indicated in the figure, does end up being associated with clump B2293, although it is not physically associated with the complex system (B2293+C2294). Details are given in Table~\ref{tb:emu_2x2_clump_id_2293}, where we have chosen to list the MADFM as a representative statistic to characterise residuals \citep[e.g.,][]{riggi_2019}, although the other statistics are still computed, and give similar results. This complex of emission provides a good illustration of how the different SFs perform with multiple overlapping, extended and diffuse structures.

\begin{figure*}[hbt!]
\centering%
\begin{tabular}{@{}c@{}}
\includegraphics[width=0.95\textwidth]{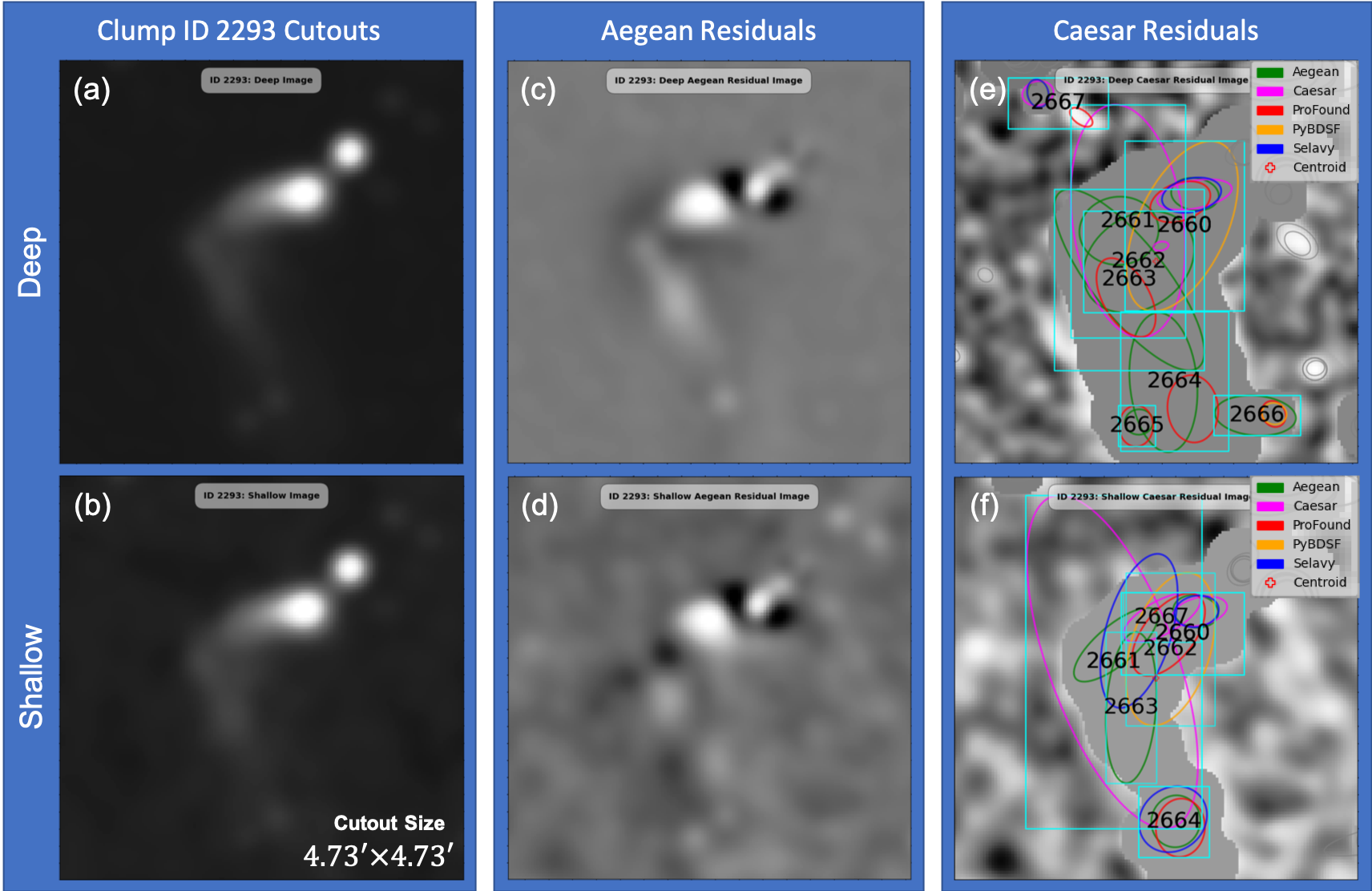}\\
\includegraphics[width=0.95\textwidth]{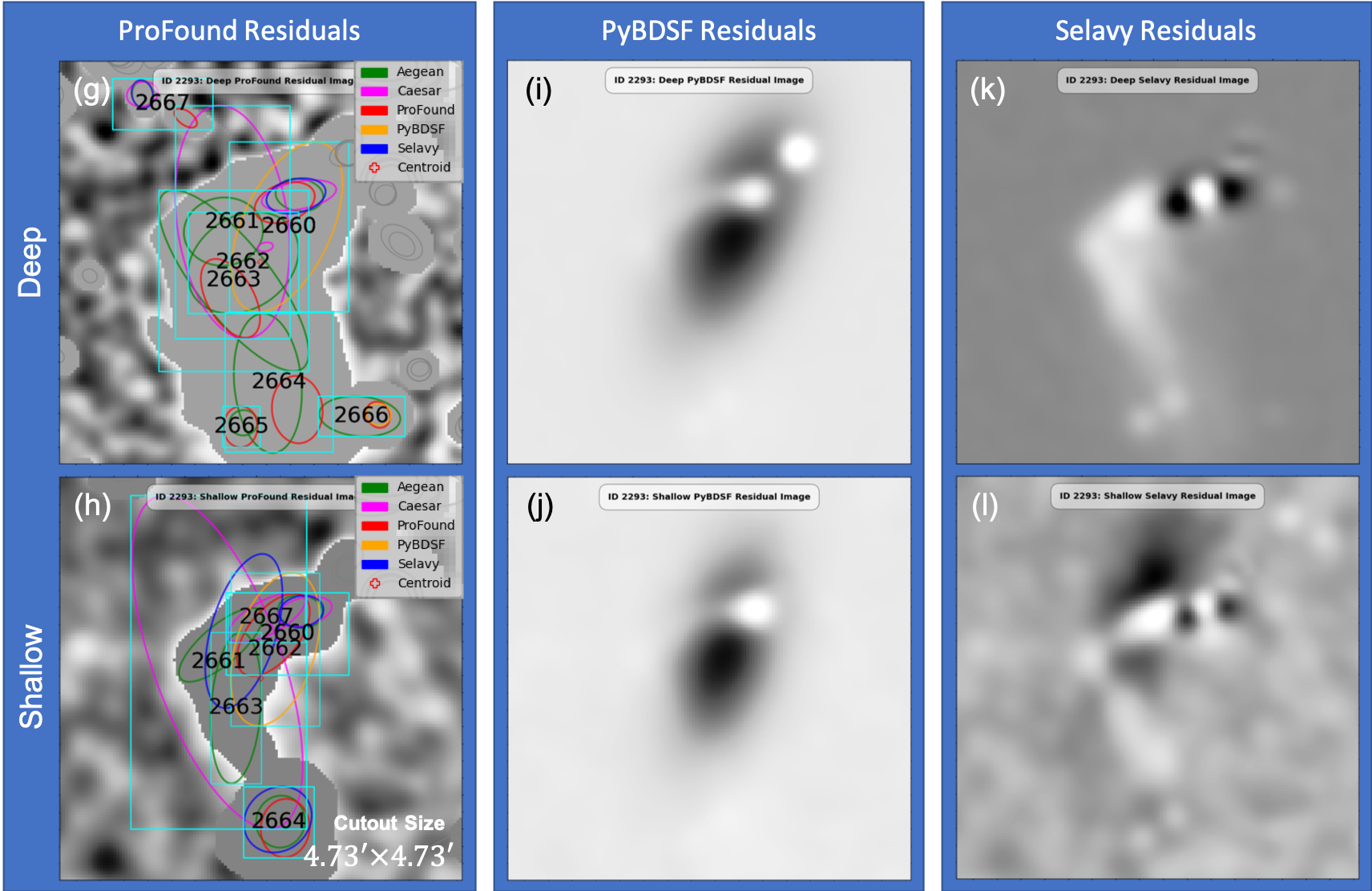}\\
\end{tabular}
\caption{$4.73^\prime\times4.73^\prime$ $\mathcal{D}$ and $\mathcal{S}$ real-image (a--b), and Aegean (c--d), Caesar (e--f), ProFound (g--h), PyBDSF (i--j), and Caesar (k--l) residual-image cutouts for \texttt{clump\_id} 2293 of the B2293+C2294 bent-tail system. Only the Caesar and ProFound cutouts are annotated with \texttt{match\_id} information (Table~\ref{tb:emu_2x2_clump_id_2293}) for clarity.}
\label{fg:emu_diffuse_emission}
\end{figure*}

\begin{table*}[hbt!]
\caption{Cluster table information for \texttt{clump\_id}s 2293 (upper partition) and 2294 (lower partition). The cutouts shown in Figure~\ref{fg:emu_diffuse_emission} are for the upper partition. The source component footprint parameters, $a$, $b$, and $\theta$, correspond to the major axis, minor axis, and position angle, respectively. The S/N is calculated using the noise in the $\mathcal{S}$-image, estimated using \textsc{bane}. The MADFMs are computed within the source component footprints and normalised with respect to their areas for the residual images. The values for ProFound are its flux-weighted estimates (Paper I).}
\centering
\begin{tabular}{@{\;}c@{\;\;\;}l@{\;\;\;}l@{\;\;\;}r@{\;\;\;}r@{\;\;\;}r@{\;\;\;}r@{\;\;\;}r@{\;\;\;}r@{\;\;\;}r@{\;\;\;}c@{\;}}
\hline\hline
\multicolumn{1}{@{}c@{}}{Match} & \multicolumn{1}{@{}l@{}}{SF} & \multicolumn{1}{@{}l@{}}{Image} & \multicolumn{1}{@{}c@{}}{RA} & \multicolumn{1}{@{}c@{}}{Dec} & \multicolumn{1}{@{}c@{}}{$a$} & \multicolumn{1}{@{}c@{}}{$b$} & \multicolumn{1}{@{}c@{}}{$\theta$} & \multicolumn{1}{@{}c@{}}{$S$} & \multicolumn{1}{@{}c@{}}{S/N} & \multicolumn{1}{@{}c@{}}{MADFM}\\
\multicolumn{1}{@{}c@{}}{ID} & \multicolumn{1}{@{}l@{}}{} & \multicolumn{1}{@{}l@{}}{Depth} & \multicolumn{1}{@{}c@{}}{($^\circ$)} & \multicolumn{1}{@{}c@{}}{($^\circ$)} & \multicolumn{1}{@{}c@{}}{($^{\prime\prime}$)} & \multicolumn{1}{@{}c@{}}{($^{\prime\prime}$)} & \multicolumn{1}{@{}c@{}}{($^\circ$)} & \multicolumn{1}{@{}c@{}}{(mJy)} & \multicolumn{1}{@{}c@{}}{} & \multicolumn{1}{@{}c@{}}{$\mbox{mJy}/(\prime^2\mbox{beam})$}\\
\hline
2660 & Aegean & Deep & 320.614 & -56.011 & 33.75 & 22.85 & 82.447 & 51.0417 & 105.4360 & 2.4513$\times10^{3}\;\;$\\
2660 & Aegean & Shallow & 320.614 & -56.012 & 31.88 & 22.87 & 82.612 & 49.0351 & 101.2908 & 1.0319$\times10^{2}\;\;$\\
2660 & Caesar & Deep & 320.614 & -56.011 & 53.35 & 19.46 & -79.362 & 52.5567 & 108.5653 & 5.5643$\times10^{4}\;\;$\\
2660 & Caesar & Shallow & 320.612 & -56.012 & 39.18 & 19.61 & -76.716 & 37.7645 & 79.6115 & 5.0744$\times10^{3}\;\;$\\
2660 & ProFound & Deep & 320.619 & -56.013 & 43.97 & 26.57 & 109.235 & 69.8364 & 136.9028 & 6.1224$\times10^{5}\;\;$\\
2660 & ProFound & Shallow & 320.624 & -56.016 & 68.82 & 35.74 & 141.154 & 85.2258 & 159.2764 & 5.6682$\times10^{3}\;\;$\\
2660 & PyBDSF & Deep & 320.619 & -56.017 & 128.89 & 62.74 & 155.166 & 126.1139 & 246.0558 & 1.2288$\times10^{2}\;\;$\\
2660 & Selavy & Deep & 320.615 & -56.012 & 42.40 & 23.71 & 102.080 & 61.3140 & 125.3935 & 2.6793$\times10^{3}\;\;$\\
2660 & Selavy & Shallow & 320.613 & -56.012 & 31.47 & 22.05 & 100.170 & 46.5150 & 97.0618 & 1.0414$\times10^{2}\;\;$\\
2661 & Aegean & Deep & 320.642 & -56.017 & 55.90 & 48.05 & -84.490 & 17.5477 & 37.3929 & 2.4513$\times10^{3}\;\;$\\
2661 & Aegean & Shallow & 320.642 & -56.017 & 73.95 & 27.27 & -49.783 & 17.7016 & 38.2606 & 1.0319$\times10^{2}\;\;$\\
2661 & Caesar & Deep & 320.638 & -56.016 & 165.70 & 78.08 & 8.087 & 73.9215 & 151.3647 & 5.5643$\times10^{4}\;\;$\\
2661 & Caesar & Shallow & 320.644 & -56.020 & 248.30 & 90.18 & 20.210 & 90.2743 & 195.1955 & 5.0744$\times10^{3}\;\;$\\
2661 & Selavy & Shallow & 320.634 & -56.014 & 112.44 & 46.23 & 163.650 & 60.2680 & 118.3390 & 1.0414$\times10^{2}\;\;$\\
2662 & Aegean & Deep & 320.636 & -56.024 & 83.96 & 64.66 & -55.037 & 17.7797 & 34.2307 & 2.4513$\times10^{3}\;\;$\\
2662 & Caesar & Deep & 320.627 & -56.021 & 10.93 & 6.24 & -78.426 & 0.1349 & 0.2410 & 5.5643$\times10^{4}\;\;$\\
2662 & PyBDSF & Shallow & 320.624 & -56.018 & 111.50 & 56.04 & 162.988 & 156.9845 & 293.7207 & 1.7932$\times10^{2}\;\;$\\
2663 & Aegean & Deep & 320.639 & -56.027 & 157.56 & 52.04 & 38.318 & 3.2044 & 6.4656 & 2.4513$\times10^{3}\;\;$\\
2663 & Aegean & Shallow & 320.639 & -56.029 & 106.85 & 35.06 & -4.002 & 21.3041 & 44.3798 & 1.0319$\times10^{2}\;\;$\\
2663 & ProFound & Deep & 320.641 & -56.030 & 61.32 & 32.77 & 29.961 & 21.4079 & 46.5896 & 6.1224$\times10^{5}\;\;$\\
2664 & Aegean & Deep & 320.630 & -56.048 & 98.68 & 47.38 & 5.068 & 14.5981 & 43.9117 & 2.4513$\times10^{3}\;\;$\\
2664 & Aegean & Shallow & 320.627 & -56.052 & 36.73 & 34.34 & -12.335 & 6.0130 & 21.4578 & 1.0319$\times10^{2}\;\;$\\
2664 & ProFound & Deep & 320.621 & -56.053 & 47.60 & 35.67 & 5.535 & 7.5893 & 28.1344 & 6.1224$\times10^{5}\;\;$\\
2664 & ProFound & Shallow & 320.625 & -56.054 & 41.23 & 34.49 & 173.862 & 10.5205 & 39.3395 & 5.6682$\times10^{3}\;\;$\\
2664 & Selavy & Shallow & 320.628 & -56.052 & 50.22 & 43.97 & 128.550 & 10.8400 & 37.7821 & 1.0414$\times10^{2}\;\;$\\
2665 & Aegean & Deep & 320.640 & -56.055 & 19.85 & 17.86 & -84.110 & 0.8030 & 3.3544 & 2.4513$\times10^{3}\;\;$\\
2665 & ProFound & Deep & 320.641 & -56.056 & 29.00 & 23.98 & 178.631 & 2.4034 & 10.3152 & 6.1224$\times10^{5}\;\;$\\
2666 & Aegean & Deep & 320.599 & -56.056 & 57.28 & 28.06 & 85.426 & 0.9795 & 4.0272 & 2.4513$\times10^{3}\;\;$\\
2666 & ProFound & Deep & 320.593 & -56.056 & 19.18 & 16.12 & 31.535 & 0.3273 & 1.3524 & 6.1224$\times10^{5}\;\;$\\
2666 & PyBDSF & Deep & 320.593 & -56.056 & 17.38 & 15.17 & 80.152 & 0.1721 & 0.7110 & 1.2288$\times10^{2}\;\;$\\
2667 & Aegean & Deep & 320.666 & -55.989 & 18.68 & 14.42 & 2.981 & 0.1361 & 0.5255 & 2.4513$\times10^{3}\;\;$\\
2667 & Caesar & Deep & 320.666 & -55.989 & 22.20 & 17.07 & -76.756 & 0.1552 & 0.5994 & 5.5643$\times10^{4}\;\;$\\
2667 & Caesar & Shallow & 320.626 & -56.012 & 59.98 & 23.76 & -62.079 & 22.7529 & 42.1717 & 5.0744$\times10^{3}\;\;$\\
2667 & ProFound & Deep & 320.651 & -55.995 & 18.49 & 9.79 & 54.245 & 0.1416 & 0.3971 & 6.1224$\times10^{5}\;\;$\\
2667 & PyBDSF & Deep & 320.666 & -55.989 & 19.98 & 14.83 & 7.938 & 0.1606 & 0.6202 & 1.2288$\times10^{2}\;\;$\\
2667 & Selavy & Deep & 320.666 & -55.989 & 19.92 & 14.88 & 12.690 & 0.1610 & 0.6216 & 2.6793$\times10^{3}\;\;$\\
\hline\hline
2668 & Aegean & Deep & 320.595 & -56.004 & 18.82 & 18.40 & -87.136 & 20.9655 & 53.8435 & 1.3433$\;\;\;\;\;\;\;\;\;\;$\\
2668 & Aegean & Shallow & 320.595 & -56.004 & 18.65 & 18.41 & 87.954 & 20.9354 & 53.7664 & 1.1211$\;\;\;\;\;\;\;\;\;\;$\\
2668 & Caesar & Deep & 320.595 & -56.004 & 27.80 & 16.61 & -68.410 & 20.6564 & 53.0499 & 0.0000$\;\;\;\;\;\;\;\;\;\;$\\
2668 & Caesar & Shallow & 320.595 & -56.004 & 25.56 & 20.72 & -48.870 & 21.5519 & 55.3497 & 0.0000$\;\;\;\;\;\;\;\;\;\;$\\
2668 & ProFound & Deep & 320.596 & -56.004 & 16.68 & 16.01 & 141.799 & 21.8717 & 55.5098 & 0.0000$\;\;\;\;\;\;\;\;\;\;$\\
2668 & ProFound & Shallow & 320.596 & -56.004 & 16.05 & 14.91 & 132.551 & 21.5683 & 54.7398 & 0.0000$\;\;\;\;\;\;\;\;\;\;$\\
2668 & PyBDSF & Shallow & 320.595 & -56.004 & 19.06 & 18.04 & 123.511 & 20.8894 & 53.6482 & 7.8357$\;\;\;\;\;\;\;\;\;\;$\\
2668 & Selavy & Deep & 320.595 & -56.004 & 18.53 & 17.57 & 99.270 & 19.9640 & 51.2715 & 1.1786$\;\;\;\;\;\;\;\;\;\;$\\
2668 & Selavy & Shallow & 320.595 & -56.004 & 19.04 & 18.21 & 123.250 & 21.3080 & 54.7232 & 4.7455$\times10^{-1}$\\
\hline\hline
\end{tabular}
\label{tb:emu_2x2_clump_id_2293}
\end{table*}

C2294 corresponds to \texttt{match\_id} 2668 of \texttt{clump\_id} 2294. The clump information for this object is given in the lower partition of Table~\ref{tb:emu_2x2_clump_id_2293}. We have not included the cutout images for this system (whose SF components are greyed out in Figure~\ref{fg:emu_diffuse_emission}e--h), as, in general, we are focusing our attention on the complex source (B2293). The following discussion arises from consideration of the detections and residuals shown in Figures~\ref{fg:emu_ds_clump_2293_image_cutout} and~\ref{fg:emu_diffuse_emission}, and Table~\ref{tb:emu_2x2_clump_id_2293}, with explicit reference where clarity is required.

The interpretation of $(\varepsilon)$ and $(\lambda)$ as purely diffuse emission is subjective, based only on their appearance in this radio image, without consideration of possible optical or infrared counterparts. Similarly, the component labelled $(\chi)$ appears subjectively to be a bright extension from $(a)$. The annotations $(a)$ through $(e)$ are likewise subjectively identified as compact components. This choice is adequate for the current discussion given that this is the same information being used by the SFs.

Aegean cleanly identifies C2294 (\texttt{match\_id} 2668) and the component at $(a)$ (\texttt{match\_id} 2660) in the $\mathcal{D}$ and $\mathcal{S}$ images. While it identifies several additional components (\texttt{match\_id}s 2661, 2662, 2663, 2664, 2665, 2666) to characterise the complex of compact and diffuse emission, these poorly model the emission itself as evidenced in the residuals, for both $\mathcal{D}$ and $\mathcal{S}$ images. It is clear from the annotations, too, that the components identified in the $\mathcal{S}$-image (\texttt{match\_id}s 2661, 2663, 2664 only) are markedly different from their counterparts in the $\mathcal{D}$-image. The component sizes tend to overestimate the extent of the true emission and are often poorly aligned with the structure. Such mismatches, as discussed above, lead to a reduction in $\mathcal{C_{DS}}$ and $\mathcal{R_{DS}}$.
For these complex structures, Aegean generally identifies source positions biased towards flux-weighted centers, with fitted component sizes compensating for the total flux within. This affects the position and size accuracy of the region it is trying to characterise.

Caesar robustly identifies C2294 (\texttt{match\_id} 2668) and $(a)$  (\texttt{match\_id} 2660) in the $\mathcal{D}$ and $\mathcal{S}$ images. In the $\mathcal{D}$-image, $(\chi)$ is incorporated into its estimate of the source $(a)$. In the $\mathcal{S}$-image, it resolves the two separately into $(a)$ and $(\chi)$ (\texttt{match\_id} 2667). Recall that Caesar is designed to be sensitive to diffuse emission \citep{riggi_2016}, which is evident in the extent of its residual-image footprint (representing the parent-blob, Figure~\ref{fg:emu_diffuse_emission}). It correctly characterises the extent and overall shape of the emission. The individual components (child-blobs) are determined in its second iteration. Refinement of the parameters for that step may lead to improvements in the way the various components are characterised. This step was not deemed to be practical for the current implementation of Hydra, and it is also unclear whether fine-tuning for such an individual source would have unintended effects for rest of the image. This is an aspect that can be explored in future developments of the tool.
Having acknowledged this point, in this implementation Caesar does poorly in terms of characterising most of the components, with the exception of detecting $(\varepsilon)$ in the $\mathcal{D}$-image (\texttt{match\_id} 2662 with S/N $\sim$ $0.2\,$mJy). It attempts to fit everything by a single component (\texttt{match\_id} 2661), both in the $\mathcal{D}$ ($a\sim166^{\prime\prime}$, $b\sim78.1^{\prime\prime}$, and $\theta\sim8.09^\circ$ with $S\sim73.9\,$mJy) and $\mathcal{S}$ ($a\sim248^{\prime\prime}$, $b\sim90.2^{\prime\prime}$, and $\theta\sim20.2^\circ$ with $S\sim90.3\,$mJy) images. Here $a$, $b$ and $\theta$ are the fitted major and minor axes and position angle respectively. These correspond to S/N bins 152 and 181, respectively. The mismatch between these components in the $\mathcal{D}$ and $\mathcal{S}$ images contribute to a degradation in $\mathcal{C_{DS}}$ (Figure~\ref{fg:hydra_2x2_all_cr_plots}a).

ProFound does a good job of characterising the features of B2293+C2294 in the $\mathcal{D}$-image, but less so in the $\mathcal{S}$-image. As with Caesar, ProFound accurately characterises the extent and overall shape of the emission in general, as seen in the residual images for both $\mathcal{D}$ and $\mathcal{S}$.  It tends to merge brighter peaks with adjacent diffuse emission. The combinations $(a)+(\chi)$, $(e)+(\varepsilon)+(d)+(\lambda)$, $(b)$, and $(c)$ are identified in \texttt{match\_id}s 2660, 2663, 2664, and 2665, respectively. This is consistent with ProFound's design, identifying diffuse islands and characterising them through such flux-weighted components. In the $\mathcal{S}$-image, ProFound does not do as well because much of the diffuse emission is washed out. This becomes clear by comparing the $\mathcal{D}$ and $\mathcal{S}$ residual images in Figure~\ref{fg:emu_diffuse_emission} through the reduction in its island size (this is also apparent with Caesar, but to a lesser degree). In the $\mathcal{S}$-image ProFound breaks B2293 into two main components, everything along the chain from $(a)-(\lambda)$ in \texttt{match\_id} 2660, and $(b)+(c)$ in \texttt{match\_id} 2664. This again results in a mismatch between the components in the $\mathcal{D}$ and $\mathcal{S}$ images.

PyBDSF does not perform well for extended sources with diffuse emission. It tends to merge things together and attempt to fit complex regions with a single Gaussian. In the $\mathcal{D}$-image it merges the B2294+C2294 system into one component (\texttt{match\_id} 2660), whereas in the $\mathcal{S}$-image they are separated into two components, B2263 (\texttt{match\_id} 2662) and C2294 (\texttt{match\_id} 2668). Clearly it is representing the system largely as a single island in the $\mathcal{D}$-image, due to the presence of the extended diffuse emission. In the $\mathcal{S}$-image it is resolved into two, as the diffuse emission linking B2294 and C2294 is masked by the higher noise level. With our settings (Paper I), PyBDSF uses the number of peaks found within the island as its initial guess for the number of Gaussians to fit, iteratively reducing them until a good fit is achieved \citep{mohan_2015}. Here, presumably, it is the diffuse emission that is affecting the fit quality, as can be seen in the residual images. This is also consistent with the failure modes seen in our simulated point source image, (Figures~\ref{fg:le_oversized}a--c). Figure~\ref{fg:le_oversized}c may be a good analogy, where PyBDSF fits a single component to a large region encompassing multiple injected sources presumably influenced also by peaks in the background noise, mimicking the diffuse emission in this case.

Selavy accurately characterises C2294 (\texttt{match\_id} 2668) in the $\mathcal{D}$ and $\mathcal{S}$ images, but does poorly with B2294. B2294 is characterised by a single component in the $\mathcal{D}$-image, at \texttt{match\_id} 2660, and by three components in the $\mathcal{S}$-image, at \texttt{match\_id}s 2660, 2661, and 2664. It appropriately characterises $(a)+(\chi)$ in the $\mathcal{D}$ and $(a)$ in the $\mathcal{S}$, at \texttt{match\_id} 2660. This is not unexpected, given the diffuse emission is reduced in the latter. Perhaps surprisingly, though, it misses the remainder of the emission beyond $(\chi)$ in the $\mathcal{D}$-image. In the $\mathcal{S}$-image it detects two components here, linking the emission spanning $(\chi)-(\lambda)$ (\texttt{match\_id} 2661) and $(b)+(c)$ (\texttt{match\_id} 2664). Selavy's approach also includes a stage of rejecting poorly fit components \citep{whiting_2012}, and that may be the issue with the lack of components reported for the emission in the $\mathcal{D}$-image. This may have been slightly more successful in the $\mathcal{S}$-image (due to a reduction in the diffuse emission). This contributes to a reduction in $\mathcal{R_{DS}}$ (Figure~\ref{fg:hydra_2x2_all_cr_plots}b), as the $\mathcal{S}$ detection at \texttt{match\_id} 2661 has no $\mathcal{D}$ counterpart, seen in the degradation in the S/N $\sim$ 125 bin. This effect occurs elsewhere at similarly high S/N from other sources.

\subsection{Discussion}
The Hydra software was used to compare the Aegean, Caesar, ProFound, PyBDSF, and Selavy SFs by first minimising the FDR based on a 90\% PRD cutoff, through Typhon. Aegean, PyBDSF, and Selavy RMS box parameters were $\mu$-optimised prior to the main run. The process was done for both $\mathcal{D}$ and $\mathcal{S}$ images for $2^\circ\times2^\circ$ CMP, EXT, and real (EMU pilot) images.

\begin{figure*}[htb!]
\begin{center}
\includegraphics[width=\textwidth]{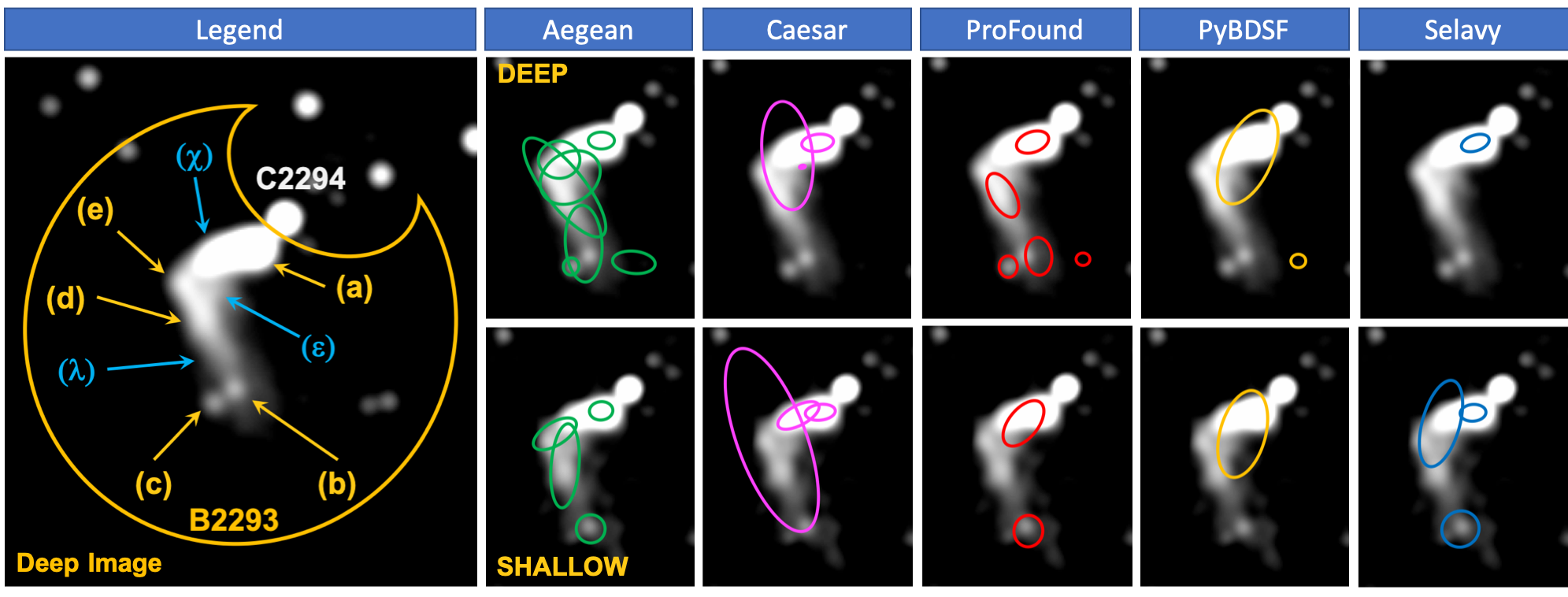}
\caption{\textbf{Diffuse Emission Case Study Summary}: Summary of bent-tail system (B2293+C2294) diffuse emission case study (see \S~\ref{sc:le_bent_tail}, Figures~\ref{fg:emu_ds_clump_2293_image_cutout} and~\ref{fg:emu_diffuse_emission}, and Table~\ref{tb:emu_2x2_clump_id_2293}).} 
\label{fg:le_bent_tail}
\end{center}
\end{figure*}

The RMS box optimisation produces a $\mathcal{S}$-to-$\mathcal{D}$ noise ratio of 5 for all images (Table~\ref{tb:hydra_2x2_typhon_rms_box_stats}), validating the routine. In addition, the baseline statistics show that the CMP source distribution is closer to the real (EMU) image than the EXT results. This is also reflected in the $\mathcal{C_{DS}}$ and $\mathcal{R_{DS}}$ plots, for CMP (Figure~\ref{fg:cds_rds_cmp_ext} (e-f)) and EXT (Figure~\ref{fg:cds_rds_cmp_ext} (k-l)) sources, when compared with the real image (Figure~\ref{fg:hydra_2x2_all_cr_plots}). For the simulated $\mathcal{D/S}$-images, the source detection numbers (N, Table~\ref{tb:hydra_2x2_typhon_n_stats}) are consistent between SFs, with the exception of Selavy being unusually low. As for the real image, the number of sources detected by Aegean, Caesar, and PyBDSF are comparable, while ProFound is significantly high and Selavy is low. In the case of ProFound, this is attributed to noise, or potentially to splitting sources more than done by other SFs, whereas for Selavy it is not as clear. In general, the residual RMS values vary between SFs for each image case, whereas the residual MADFMs are consistent (Table~\ref{tb:rms_madfm_prds}). This should not be surprising as MADFM (or robust) statistics tend to miss outliers, such as bright sources \citep{whiting_2012}. This could provide a viable explanation as to Selavy's behaviour, in this regard, as it has a tendency to miss bright sources in $\mathcal{D}$-images more often than $\mathcal{S}$-images, and we have its robust statistics flag set (Paper~I). This line of reasoning is consistent with \cite{whiting_2012b}, and so warrants further investigation.

In our analysis we examined major-axes, completeness, reliability, flux-ratio, and false-positive statistics (Figures~\ref{fg:major_distributions} through~\ref{fg:hydra_2x2_deep_false_positives}). This was later framed in the context of residual/image cutouts case studies, in which we investigated anomalies regarding edge detection, blending/deblending, component size estimates, bright sources, diffuse emission, \textit{etc.} (Figures~\ref{fg:le_edge} through~\ref{fg:le_oversized}): Aegean, ProFound, and PyBDSF are good at detecting sources at the edge of images, while it is more problematic for Caesar and Selavy (Figure~\ref{fg:le_edge}). As expected, all SFs detect tightly overlapped sources as single unresolved compact sources (Figure~\ref{fg:le_blender}), at the expense of a degradation in completeness. This is of course somewhat artificial, since with real images this situation cannot be distinguished. For $\mathcal{D}$ compact sources with diffuse emission, Aegean, ProFound, PyBDSF, and Selavy tend to flux-weight (blend) their components position and sizes (Figure~\ref{fg:le_blender}c), characterising the core, whereas Caesar and Selavy tend to characterise (deblend) the core and diffuse halo separately (Figure~\ref{fg:le_delending}f). For the $\mathcal{S}$ case (Figure~\ref{fg:le_blender}d), the diffuse emission tends to be somewhat washed out, so that the SFs see predominantly compact sources (Figure~\ref{fg:le_blender}f). Similar issues can occur for systems with a bright source among low S/N sources, contributing to a degradation in reliability (Figure~\ref{fg:le_delending}a--b). For low S/N thresholds, all SFs are susceptible to noise spike detection, particularly so for ProFound (Figure~\ref{fg:le_spike}c); as can be seen from its scatter, Table~\ref{tb:3sigma_scatter}. Noise spikes also affect PyBDSF, in that it tends to collect them together, sometimes with true sources, producing oversized components (Figure~\ref{fg:le_oversized}a-c).  Caesar also overestimates component size on occasion, however this is related to deblending issues (Figure~\ref{fg:le_delending}d). It also underestimates on occasion, leading to a degradation in reliability (Figure~\ref{fg:hydra_2x2_rd_cid_4627}).

A detailed diffuse emission case study was also explored (Figures~\ref{fg:emu_ds_clump_2293_image_cutout} and~\ref{fg:emu_diffuse_emission}) with Figure~\ref{fg:le_bent_tail} summarising the results. Aegean characterises B2293 by fitting its bright spots and diffuse emission as separate components, but ignores the diffuse emission in the $\mathcal{S}$-image. Caesar detects $(a)$ in both the $\mathcal{D/S}$-images, but only identifies $(\chi)$ in the $\mathcal{S}$-image; the remaining information, is approximated by a single component. ProFound correctly characterises the shape of B2293 in the $\mathcal{D}$-image, however its components only highlight the flux-weighted centers of each segment (``island''); for the $\mathcal{S}$-image, only  $(a)+(\chi)$ and $(b)$ are detected. PyBDSF treats the complex largely as a single source (it also picks up the faint isolated SW $\mathcal{D}$ source). Finally, for the $\mathcal{D}$-image, Selavy treats B2293 as a single system centered at (a), whereas it seems more successful at characterising the system in the $\mathcal{S}$-image.

\section{Future Prospects}
\subsection{Detection Confidence}
\begin{figure}[hbt!]
\centering%
\includegraphics[width=\columnwidth]{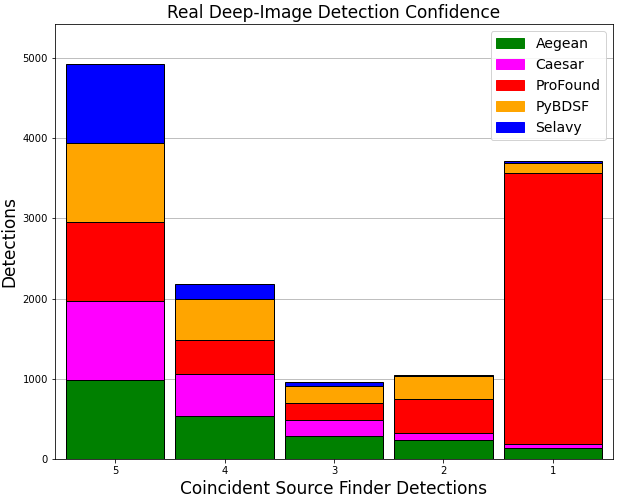}
\includegraphics[width=\columnwidth]{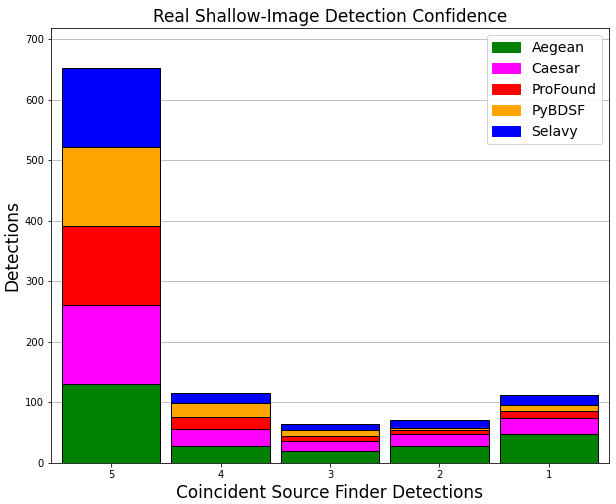}
\caption{Real image $\mathcal{D}$ (top) and $\mathcal{S}$ (bottom) detection confidence charts. The stacked bars indicate agreement between SF detections. From left to right, 5 SFs agree, 4 SFs agree, \textit{etc.}}
\label{fg:hydra_emu2x2_detection_confidence_charts}
\end{figure}

Given Hydra uses multiple SFs, it leads to the possibility of exploring cross-comparison metrics. For example,
Figure~\ref{fg:hydra_emu2x2_detection_confidence_charts} shows detection confidence charts, indicating coincident detections between SFs. The bar on the far left shows the number of sources detected in common between all 5 SFs. The next bar shows the numbers of sources from each SF detected in common by 4 SFs, and so on. This metric uses a ``majority rules'' process to help determine whether a detection is more likely to be real. If a source is only picked up by a single SF, chances are that detection is spurious (\textit{e.g.}, Figure~\ref{fg:le_spike}c). The more SFs that agree on a source, the more likely it is to be real. This could be used to constrain metrics, such as completeness, reliability, flux-ratios, \textit{etc}. This could be particularly useful in refining such statistics for real images, where the true underlying source population is not known a priori.

\begin{figure}[htb!]
\begin{center}
\includegraphics[width=\columnwidth]{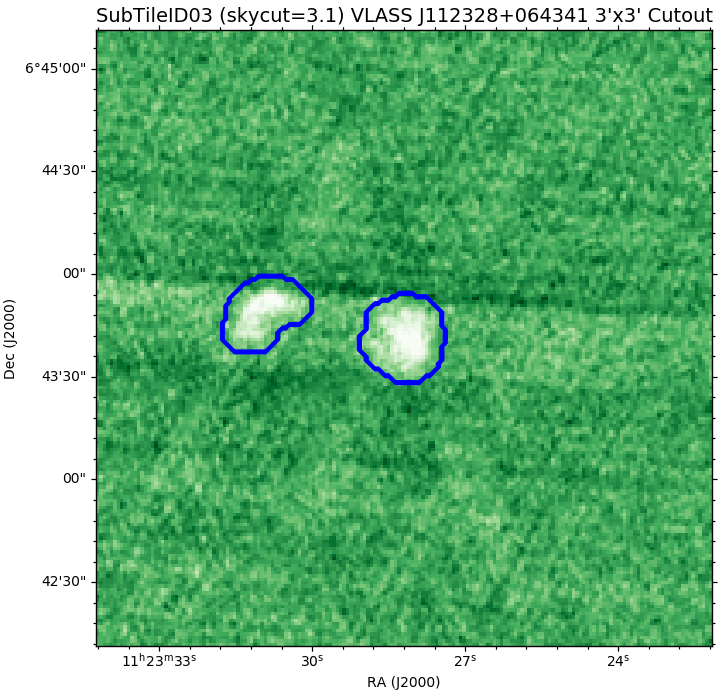}
\caption{Example of a diffuse VLASS source, uniquely detected by ProFound in a comparison study with PyBDSF \citep{boyce_2020}. The source was found in a $3^{\prime}\times3^{\prime}$ region centered at J112328+064341. The cutout was extracted from a QL image tile, available at NRAO (\url{https://science.nrao.edu}), and then processed using ProFound in R Studio. \label{fg:vlass_profound_gem}}
\end{center}
\end{figure}

One should be cautious, however, not to take this as \textit{ipso facto} true. For example, a detection by all SFs could be a bright artifact or noise spike that mimics a source. At the other extreme one could end up excluding real sources only detected by finders well-adapted to recognising them (\textit{e.g.}, Figure~\ref{fg:vlass_profound_gem}). The middle ground may indicate the nature of detection, such as compact and/or extended sources with or without diffuse emission, depending on a SF's strengths (Paper~I).

The number of possible cross-source-SF diagnostics that could be developed is substantial. Consequently we have only briefly touched on this subject here, leaving further details for future investigation.

\subsection{Processing Residual Images}
\label{sc:residuallity}
The existence of initially undetected sources in the residual images suggests that running a second iteration of the SFs on such residual images may potentially improve on the completeness of the delivered catalogues. 
This may not be practical for Caesar or ProFound, however, as they compute residuals by subtracting out all of the flux within an island; although, one could create residual images from their island components. Aegean, PyBDSF, and Selavy are perhaps more suitable, as they are Gaussian-fit based. There is, however, a challenge that arises from over-subtraction, where the initial source list includes overestimated sources (\textit{e.g.}, Figure~\ref{fg:le_delending}a). Here the residuals reflect a poor initial fit, rather than true sources remaining to be detected in such a second pass.

The PyBDSF residual $\mathcal{D}$-image in Figure~\ref{fg:hydra_ext2x2_cd_cid_50_matrix}, which emphasises the undetected sources at \texttt{match\_id}s 109, 111, 112, and 113, suggests that a second detection pass on such residual images may have merit. How best to implement such an approach remains a challenge, as it may introduce new false detections if residuals such as seen in Figure~\ref{fg:hydra_ext2x2_cd_cid_50_matrix} with Aegean dominate, and likewise for more complex objects, containing diffuse emission.

In order to explore this further, let us consider the residual statistics of our case-study system (\textit{i.e.}, \textit{clump\_id} 2293 in Table~
\ref{tb:emu_2x2_clump_id_2293}) and the full EMU image (Table~\ref{tb:hydra_2x2_typhon_stats}), summarized in Table~\ref{tb:residuallity}. Three potential approaches for reprocessing residual images are:
\begin{enumerate}
    \item process all $2^\circ\times2^\circ$ residual images,\label{en:residual_proc_img_opt_1}
    \item process the $2^\circ\times2^\circ$ residual image with the best MADFM,\label{en:residual_proc_img_opt_2}
    \item process an aggregate clump-based residual-image consisting of the best MADFM's on a clump-by-clump basis.\label{en:residual_proc_img_opt_3}
\end{enumerate}
Approach~\ref{en:residual_proc_img_opt_1} is ruled out as being computationally impractical. To explore the other options we consider the MADFMs for CMP and EXT sources in Table~\ref{tb:hydra_2x2_typhon_stats}, and real sources in Table~\ref{tb:residuallity}.

\begin{table}[hbt!]
\caption{Residual image statistics for the full $2^{\circ}\times 2^{\circ}$ EMU pilot sample image and for \texttt{clump\_id} 2293 drawn from it (extracted from Tables~\ref{tb:hydra_2x2_typhon_stats} and~\ref{tb:emu_2x2_clump_id_2293}, respectively). The MADFMs are normalised by cutout area, and are in units of mJy arcmins$^{-2}$\,beam$^{-1}$). N is the component count.}
\centering
\begin{tabular}{@{\;}l@{\;\;}l@{\;\,}r@{\;\;}c@{\;\,}r@{\;\;}c@{\;}}
\hline\hline
\multicolumn{2}{l}{Residual Image}  & \multicolumn{2}{c}{EMU Sample} & \multicolumn{2}{c}{Cutout 2293}\\ \hline
Source & Image & \multicolumn{2}{c}{$2^\circ\times2^\circ$} & \multicolumn{2}{c}{$4.73^{\prime}\times4.73^{\prime}$}\\
Finder & Depth &  \multicolumn{1}{c}{N} & MADFM  &  \multicolumn{1}{r}{N$\,$} & MADFM \\
\hline
Aegean   & Deep    &  8,538 & $2.60\!\!\times\!\!10^{-5}$ & 8 & $2.45\!\!\times\!\!10^{-3}$ \\
Caesar   & Deep    &  7,838 & $2.30\!\!\times\!\!10^{-5}$ & 4 & $5.56\!\!\times\!\!10^{-4}$ \\
ProFound & Deep    & 11,484 & $1.90\!\!\times\!\!10^{-5}$ & 6 & $6.10\!\!\times\!\!10^{-5}$ \\
PyBDSF   & Deep    &  8,292 & $2.60\!\!\times\!\!10^{-5}$ & 3 & $1.23\!\!\times\!\!10^{-2}$ \\
Selavy   & Deep    &  5,800 & $2.70\!\!\times\!\!10^{-5}$ & 2 & $2.68\!\!\times\!\!10^{-3}$ \\
\hline\hline                                                                       
Aegean   & Shallow &    926 & $1.69\!\!\times\!\!10^{-4}$ & 4 & $1.03\!\!\times\!\!10^{-2}$ \\
Caesar   & Shallow &    885 & $1.66\!\!\times\!\!10^{-4}$ & 3 & $5.07\!\!\times\!\!10^{-3}$ \\
ProFound & Shallow &    778 & $1.65\!\!\times\!\!10^{-4}$ & 2 & $5.67\!\!\times\!\!10^{-3}$ \\
PyBDSF   & Shallow &    794 & $1.69\!\!\times\!\!10^{-4}$ & 1 & $5.67\!\!\times\!\!10^{-3}$ \\
Selavy   & Shallow &    789 & $1.69\!\!\times\!\!10^{-4}$ & 3 & $1.04\!\!\times\!\!10^{-2}$ \\
\hline\hline
\end{tabular}
\label{tb:residuallity}
\end{table}

Given that ProFound and Caesar assign all flux in an island to a source, their residuals are zero by default at the location of sources, leading to their MADFM statistics understandably being the lowest. In order to compare them fairly to the other SFs, residual images created instead from their inferred Gaussian components would be necessary. The analysis of the complex system B2293+C2294 (Figure~\ref{fg:emu_ds_clump_2293_image_cutout}) suggests that ProFound would likely produce residual MADFMs similar to the other SFs, and potentially perform better in the case of complex extended sources with diffuse emission. Caesar would probably not fare as well, given that Hydra is not implementing its full capability for decomposing complex structures. For simplicity in the remainder of this discussion we focus on the other three SFs as potential starting places for image reprocessing.

Considering approach~\ref{en:residual_proc_img_opt_2}, we see that Aegean and PyBDSF have identical MADFMs in the full image. It would be easy to choose one at random, or invoke the other residual image metrics as a further discriminator. Regardless, the next challenge will be understanding and accounting for the artifacts associated with the residuals of the chosen SF. Discriminating in a second SF run between a truly overlooked component and a residual peak left by a poor first-pass subtraction seems on its face to be intractable. One approach may be not to treat each iterative pass as independent and concatenate the output catalogues, but instead to first combine the components together where they overlap in order to better reconstruct the true underlying flux distribution for each complex.

\begin{figure}[hbt!]
\centering%
\includegraphics[width=\columnwidth]{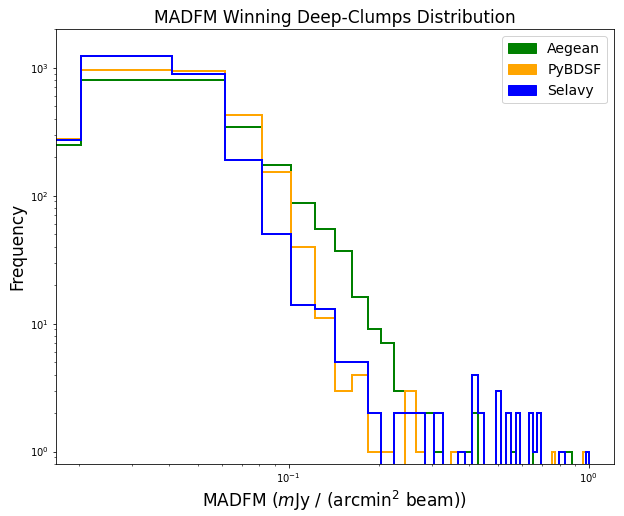}
\caption{Distribution of clumps with the lowest MADFM wrt Aegean, PyBDSF, and Selavy.}
\label{fg:hydra_2x2_emu_deep_clump_madfm_wins_distribution_gaussian_sf_only}
\end{figure}

Approach~\ref{en:residual_proc_img_opt_3} is perhaps more unconventional, as it draws on the aggregate of results from the different SFs. By using the cutouts with the best MADFM in each case, though, it is likely to contain fewer artifacts than the monolithic approach~\ref{en:residual_proc_img_opt_2} above. Figure~\ref{fg:hydra_2x2_emu_deep_clump_madfm_wins_distribution_gaussian_sf_only} shows a distribution of MADFMs for Aegean, PyBDSF, and Selavy. Here we only include a MADFM if it is the lowest for each clump. In other words, each clump is represented only once, and the best MADFM for it contributes to the distribution for that SF. At low MADFM values each SF appears to contribute roughly equally to the set of best residual cutouts. The corresponding aggregate residual image would be roughly a homogeneous mixture of results from the three SFs. For increasing MADFM, Selavy contributes fewer of the best residuals, followed by PyBDSF, with Aegean contributing the most at the higher values of MADFM. For our sample complex system, the aggregate would include 8 components within \texttt{clump\_id} 2293 (which contains B2293) arising from Aegean (Table~\ref{tb:residuallity}) and one component within \texttt{clump\_id} 2294 (which only contains C2294) from Selavy (Table~\ref{tb:emu_2x2_clump_id_2293} bottom partition). This approach may have some merit, but clearly also adds a layer of complexity that may prove more challenging than desired.

An alternative approach would be to first identify compact components only and produce a residual image by subtracting those, presuming that all finders perform (reasonably) well on such compact emission. This would leave a residual image consisting only of emission with more complex structure. This would then need to be processed using a SF that has demonstrated good performance with complex structures, established with another run of Hydra. This may provide the opportunity to better characterise extended emission in the vicinity of bright compact emission. If this model of multi-pass source-finding is promising, it could be further implemented iteratively, perhaps with S/N limits imposed on the initial pass(es) to first characterise and remove the bright sources, and progress to the faint population in later iterations. Either of the approaches~\ref{en:residual_proc_img_opt_2} or~\ref{en:residual_proc_img_opt_3} could be considered in these scenarios. The results from each pass in such approaches would then have to be incorporated with the existing cluster catalogue and appropriately tagged. While this iterative approach was one of the motivations for developing Hydra, it is clear that a considerable amount of further analysis would be required to demonstrate its value.

\section{Summary and Conclusions}
\label{sc:summary}
In Paper~I we introduced the Hydra tool, and detailed its implementation, demonstrating its use with simulated image data.
In this second paper of the two part series we use Hydra to characterise the performance of five commonly used SFs.

\subsection{Source finder comparison and performance}
Hydra was used to explore the performance of SFs through $\mathcal{C}$ and $\mathcal{R}$ statistics, and their size and flux-density measurements, as well as through targeted case studies (\S~\ref{sc:case_studies}). The most significant differences appeared when encountering real sources with diffuse emission. Other frequently encountered anomalies were with sources at image edges, blending/deblending, component size errors, and poorly characterised bright sources.

In terms of characterising sources, all SFs seem to handle compact objects well. With the exception of Selavy, this generally applies to extended objects as well. In the case of compact objects with diffuse emission, Aegean, ProFound and PyBDSF tend to characterise them as point sources, whereas Caesar deblends them into cores with diffuse halos, and Selavy as cores with jet-like structures. This is all consistent with how the SFs are designed (Paper~I). For more complex sources with diffuse emission, Aegean performs best at characterising their complexity through a combination of elliptical Gaussian components, with PyBDSF and Selavy overestimating component sizes, missing regions of flux, or fitting components misled by adjacent flux peaks. Caesar and ProFound characterise extended islands well, but their corresponding components are not always robust representations of the whole.

In moving from simulated to real images, we find the major distinction comes when considering diffuse sources. ProFound performs the best when it comes to complex sources with diffuse emission, characterising them as islands of flux, but without resolving them further \citep[by design,][]{robotham_2018}. Caesar in general performs similarly, although it is not implemented optimally in the current version of Hydra. Caesar can implement different RMS and island parameters for the parent and child segments in order to improve the deblending of complex structures \citep{riggi_2016,riggi_2019}. These are currently defined by Hydra to have the same settings, though, due to its present implementation requiring a single pair of RMS and island parameters for each SF (Paper~I). In short, given the SCORPIO survey analysis by \cite{riggi_2016}, Caesar is expected to outperform ProFound in terms of its deblending approach, although it may still not characterise diffuse emission to the same degree (\textit{e.g.}, compare Caesar and ProFound footprints in Figure~\ref{fg:emu_diffuse_emission}e--f and g--h, respectively). PyBDSF and Selavy have similar performance issues. They both tend to merge together sources embedded in regions of diffuse emission, leading to different characterisations in the $\mathcal{D}$ and $\mathcal{S}$ images depending on the degree of diffuse emission present, with a corresponding impact on $\mathcal{C_{DS}}$ and $\mathcal{R_{DS}}$. Aegean is similar, but performs slightly better in $\mathcal{C_{DS}}$ and $\mathcal{R_{DS}}$ perhaps due to its approach of using a curvature map to tie its Gaussian fits to peaks of negative curvature, which may reduce the degree to which the fitting differs in the $\mathcal{D}$ and $\mathcal{S}$ images.

Figure~\ref{fg:major_distributions} shows that all but one of the SFs have source size distributions peaking at the scale corresponding to the beam size, consistent with point source detections. Caesar's distribution is systematically offset by a factor of about 1.5 toward larger sizes, consistent with its performance seen in the examples so far, where its attempts to capture diffuse emission can lead to overestimates of source sizes. The distribution for ProFound has fewer very extended sources, suggesting the largest sources identified by Aegean, PyBDSF, and Selavy are likely to be poor fits with overestimated sizes. ProFound also shows an excess of very small sources, likely to be noise spikes or faint sources where the true extent of the flux distribution is masked by the noise. This is a consequence of the size metric given by ProFound which is linked to the size of the island containing only those pixels lying above the detection threshold, in contrast to the fitted Gaussian sizes from the other SFs. It is natural that these sizes will be smaller than the beam size for faint objects. Caesar shows similar results to ProFound at the small scales albeit offset to larger sizes by the factor 1.5 already noted. These outcomes are in line with the specific examples described above for the B2293+C2294 complex structure. Comparing Figure~\ref{fg:major_distributions} with the distributions for EXT sources in Paper~I, the distributions show largely the same characteristics, in particular with ProFound fitting fewer extreme extended sources, and Caesar performing similarly at those large sizes to the remaining SFs. The main difference in the real image appears to be the excess of artificially small source sizes fit by ProFound and Caesar, and Caesar's systematic overestimate of sizes.

\subsection{Processing of Residual Images}
Hydra produces an aggregated clump catalogue consisting of one row per clump ID, from the cluster catalogue of the SFs with best residual RMS, MADFM, and $\Sigma I^2$ metrics, normalised by clump cutout area. Hydra also has summary information of residual RMS, MADFM, and $\Sigma I^2$ statistics for the entire $\mathcal{D}$ and $\mathcal{S}$ images, and independent $\mathcal{D}$ and $\mathcal{S}$ catalogues for each SF. There are cases where taking a catalogue generated from the SF that provides the ``best'' fit on a source-by-source basis (a heterogeneous catalogue) may be advantageous, and Hydra provides this option. This is not possible with individual SFs alone. The more traditional homogeneous catalogue, separated by SF, is also available.

This leads to the concept of creating residual images from either heterogeneous or homogeneous catalogues. In either case, the residual image would be created based on either of the residual RMS, MADFM, and $\Sigma I^2$ metrics, or some other metric. The simplest concept would be to create the residual $\mathcal{D}$-images by subtracting out only compact sources (\textit{i.e.}, clumps consisting only of single components from any of the SFs). Running Hydra again on the residual may deliver better results for extended emission in the vicinity of bright compact emission. This approach is promising, as it is very similar to the operations of Caesar, which reprocesses its blobs after subtracting their compact components \citep{riggi_2016}. This is also planned for future versions of Hydra.

\subsection{Final thoughts}
\label{sc:conclusions}
The past two decades have seen an explosion in technologies, providing radio telescope facilities with the ability to perform deep large sky-coverage surveys \citep{lacy_2020,norris_2011,norris_2021}, or transient surveys with modest sky coverage at high cadence \citep{banyer_2012,murphy_2013}. This has led to researching and developing SF technologies capable of handling such data volumes and rates \citep{hopkins_2015,riggi_2016,riggi_2019,robotham_2018,hale_2019,bonaldi_2021}. These case studies involved the fine tuning of parameters by the SF experts in order to get the best performance. The advent of large scale surveys such as EMU \citep{norris_2021}, aiming to detect up to 40 million sources, makes such fine-tuning difficult at best.

We developed Hydra to automate the process of SF optimisation and to provide data products and diagnostics to allow for comparison studies between different SFs. Hydra is designed to be extensible and user friendly. Each SF is containerised in modules with standardised interfaces, allowing for optimisation through RMS and island parameters, which are common to the traditional class of SFs. The parameters are optimised by constraining the false detection rate \citep[\textit{e.g.,}][]{williams16,hale_2019}. New modules can be added to Hydra by following a standardised set of rules.

Future improvements to Hydra include adding the island catalogues where provided by SFs, improvements to optimisation schemes through using parameters more finely tuned for different SFs, development of completeness and reliability metrics for handling complex sources with diffuse emission, and methods of flux recovery through processing residual images with compact components removed. Catalogue post-processing is also an option. Once the false-detection fraction has been established for a given SF, a user-specified limit to false detections can be translated to an effective S/N threshold and only sources detected above that threshold will be provided to the user. In such a scenario the full set of detected sources would still be retained in Hydra's \texttt{tar} archives for subsequent exploration as needed, but only those above the requested threshold will be presented in the Hydra viewer tools. Implementation of multi-processor options where SFs support that is also planned. Hydra is being explored for integration into the EMU and VLASS pipelines. The heterogeneous nature of Hydra data, arising from multiple SFs, will add versatility to future radio survey data processing.

\section{Acknowledgements}
M. M. Boyce, S. A. Baum, Y. A. Gordon, D. Leahy, C. O'Dea, and A. N. Vantyghem acknowledge partial support from the NSERC of Canada. S. Riggi acknowledges INAF for financial support through the PRIN TEC programme ``CIRASA.'' Partial support for L. Rudnick came from U.S. National Science Foundation Grant AST17-14205 to the University of Minnesota. M. I. Ramsay acknowledges support from NSERC and the University of Manitoba Faculty of Science Undergraduate Research Award (USRA). C. L. Hale acknowledges support from the Leverhulme Trust through an Early Career Research Fellowship. Y. A. Gordon is supported by US National Science Foundation grant 2009441. H. Andernach benefited from grant CIIC 138/2020 of Universidad de Guanajuato, Mexico. D. Leahy acknowledges support from NSERC 10020476. M. J. Micha{\l}owski acknowledges the support of the National Science Centre, Poland through the SONATA BIS grant 2018/30/E/ST9/00208. S. Safi-Harb acknowledges support from NSERC through the Discovery Grants and the Canada Research Chairs programs, and from the Canadian Space Agency. M. Vaccari acknowledges financial support from the Inter-University Institute for Data Intensive Astronomy (IDIA), a partnership of the University of Cape Town, the University of Pretoria, the University of the Western Cape and the South African Radio Astronomy Observatory, and from the South African Department of Science and Innovation's National Research Foundation under the ISARP RADIOSKY2020 Joint Research Scheme (DSI-NRF Grant Number 113121) and the CSUR HIPPO Project (DSI-NRF Grant Number 121291). E. L. Alexander gratefully acknowledges support from the UK Alan Turing Institute under grant reference EP/V030302/1 and from the UK Science \& Technology Facilities Council (STFC) under grant reference ST/P000649/1. A. S. G. Robotham  acknowledges support via the Australian Research Council Future Fellowship Scheme (FT200100375). H. Tang acknowledges the support from the Shuimu Tsinghua Scholar Program of Tsinghua University.

Hydra is written primarily in Python, with some elements of Cython \citep{behnel_2011} and R, along with their standard libraries. Hydra uses alphashape \citep{bellock_2021}, APLpy \citep{robitaille_2012}, Astropy \citep{astropy_2013, astropy_2018}, Matplotlib \citep{hunter_2007}, NumPy \citep{harris_2020}, and pandas \citep{mckinney_2010,reback_2020} Python libraries commonly used in astronomy. Hydra utilizes click, gzip, Jinja, tarfile, and YAML Python libraries as part of its overall architectural infrastructure. Hydra encapsulates Aegean \citep{hancock_2012,hancock_2018}, Caesar \citep{riggi_2016,riggi_2019}, ProFound \citep{robotham_2018,hale_2019}, PyBDSF \citep{mohan_2015}, and Selavy \citep{whiting_2012} SF software using Docker. The technical diagrams in this paper were created using macOS Preview and Microsoft PowerPoint. We acknowledge the authors of the aforementioned software applications, languages, and libraries.

CIRADA is funded by a grant from the Canada Foundation for Innovation 2017 Innovation Fund (Project 35999) and by the Provinces of Ontario, British Columbia, Alberta, Manitoba and Quebec, in collaboration with the National Research Council of Canada, the US National Radio Astronomy Observatory and Australia’s Commonwealth Scientific and Industrial Research Organisation.

ASKAP is part of the ATNF which is managed by the CSIRO. Operation of ASKAP is funded by the Australian Government with support from the National Collaborative Research Infrastructure Strategy. ASKAP uses the resources of the Pawsey Supercomputing Centre. Establishment of ASKAP, the Murchison Radio-astronomy Observatory and the Pawsey Supercomputing Centre are initiatives of the Australian Government, with support from the Government of Western Australia and the Science and Industry Endowment Fund. We acknowledge the Wajarri Yamatji people as the traditional owners of the Observatory site.

\bibliographystyle{pasa-mnras}
\bibliography{references}

\end{document}